\documentclass[pra,twocolumn,showpacs,nofootinbib]{revtex4}
\usepackage{amssymb,amsmath}
\usepackage{epsfig}
\usepackage{graphicx}
\usepackage{rotating}  
\def\shiftleft#1{#1\llap{#1\hskip 0.04em}}
\def\shiftdown#1{#1\llap{\lower.04ex\hbox{#1}}}
\def\thick#1{\shiftdown{\shiftleft{#1}}}
\def\b#1{\thick{\hbox{$#1$}}}

\begin{document}
\title{The isotopic effects in the
scattering and the kinetics of the atomic cascade

of excited $\mu^- p$ and $\mu^- d$ atoms} 
\author{V.P. Popov}
\email{popov@nucl-th.sinp.msu.ru}
\author{V.N. Pomerantsev}
\email{pomeran@nucl-th.sinp.msu.ru}

\affiliation{Skobeltsyn Institute of Nuclear Physics,
 Lomonosov 
Moscow State University, 
119234 Moscow, Russia}           
\date{\today}    

\begin{abstract} 
The quantum-mechanical calculations of the differential and
integrated cross sections of the elastic scattering,  Stark transitions,
and Coulomb de-excitation at collisions of excited $\mu^- p$ and $\mu^- d$
atoms with hydrogen isotope atoms in the ground state are performed. 
The scattering processes are treated in a unified manner in the  framework
of the close-coupling approach. The used basis includes both open and
closed channels corresponding to all  exotic atom states with principal
quantum numbers from $n=1$  up to $n_{\rm max}=20$. The energy shifts of 
$ns$ states due to electron vacuum polarization and finite nuclear size are
taken into account. The kinetics of atomic cascade of $\mu^- p$ and 
$\mu^- d$ atoms are studied in a wide range of relative target densities
($\varphi = 10^{-8} -1$) within  the improved version of the extended
cascade model, in which the results of the numerical quantum-mechanical
calculations of the cross sections for quantum numbers  and  kinetic
energies of muonic atoms, that are of interest for the detailed cascade
calculations, are used as input data. Initial $(n,l,E)$-distributions of
muonic atoms at the instant of their formation and the target motion are
taken into account explicitly in present cascade calculations. The
comparison of the calculated cross sections, the kinetic energy distributions
of muonic atoms at the instant of their $np\to 1s$ radiative transitions as well as the
absolute and relative $x$-ray yields for both muonic hydrogen and muonic
deuterium reveals the isotopic effects, which, in  principal, may be
observed experimentally. The present results are mainly in very good
agreement with experimental data available in the literature.
\end{abstract}
\maketitle

\section {Introduction}

The muonic and hadronic hydrogen-like atoms ($\mu^- p$, $\mu^- d$, 
$\pi^-p$, $\pi^- d$, etc.) are formed in highly-excited states, when a heavy 
negatively charged particle $(\mu^-, \pi^-, \mathrm{K}^-, etc.)$ is slowed
down  in the gaseous or liquid hydrogen target and captured on the atomic
orbit.  After the exotic atom formation, its initial distributions in 
quantum numbers and kinetic energy are changed during the so-called
atomic cascade through a number of processes: radiative and Stark
transitions, elastic scattering, external Auger effect, Coulomb
de-excitation (CD), weak decay, and strong absorption in the case of
hadronic atoms.

The general features of these exotic atoms are similar to ones of 
ordinary hydrogen atoms, because their level structure is mainly determined
by the static Coulomb interaction. However, due to significant differences
of exotic particle and electron masses, the distance scale in the exotic 
atom case is much smaller, while  the energy scale is much larger in
comparison with the usual atomic scale. These scale effects make possible
to realize the processes of external Auger effect and CD and open
additional opportunities for study of the quantum electrodynamics, weak or
strong interactions (in the case of hadronic atoms), and scattering
processes  in collisions of the exotic and ordinary  hydrogen isotope
atoms. 

The muonic atoms of hydrogen isotopes are the simplest of exotic atoms
and may be considered as ideal objects to study the isotopic effects in the
lightest exotic atoms. Their experimental and theoretical investigations
serve as the best probe to study various collisional processes of these
atoms with ordinary  hydrogen isotope atoms or molecules. A good
understanding of isotopic effects in the scattering and kinetics of the
atomic cascade of muonic hydrogen and deuterium constitutes a very
important foundation for  pionic deuterium and muonic tritium predictions.
Besides, they are of particular interest in order to reach the reliable
theoretical description of the scattering and de-excitation cascade in the
case of hadronic atoms. The last is especially important for the proper
analysis  of the precise spectroscopic experiments with pionic
\cite{Str1}, kaonic \cite{Baz},  and antiprotonic \cite{DG3} hydrogen
isotope atoms, aimed at extraction of strong interaction widths of low
angular-momentum states. 

The experimental data, such as $x$-ray yields, the products of strong
interaction, the shapes of $x$-ray lines, and the kinetic energy
distribution of exotic atoms in their ground states, are mainly related
to the last stage of the atomic cascade. Therefore, the study of the
kinetics of atomic cascade in hydrogen media provides the only test of
both theoretical approaches used to describe the  collisional processes and
cascade models. From the other hand, the realistic theoretical description
of all processes involved in the atomic cascade plays also a key role to
choose optimal experimental conditions and to improve the quality of the
analysis of experimental data. 

The first theoretical study of the atomic cascade in 
exotic hydrogen-like ($\pi^{-}p$ and
$K^{-}p$) atoms has been performed by Leon and Bethe \cite{LB} 
more than fifty years ago. In this paper and 
later in more refined models \cite{BL,TH}, beyond 
radiative transition rates, only collisional rates of the 
Stark mixing and external Auger effect,
calculated respectively in the 
semiclassical and Born approximation {\it at a fixed
kinetic energy} ($\sim 1$ eV), have been 
taken into account to simulate the
atomic cascade. In these cascade calculations 
the scaling factors for the Stark mixing and 
external Auger effect rates were used and, besides, 
the kinetic energy was treated as a fit parameter. 

However, the high-energy fractions
have been found in energy distributions of 
$\mu^-p$, $\mu^-d$ and $\pi^-p$ atoms in the different time-of-flight 
experiments \cite{JCr, Abbott, RPD, Bad}.
In particular, the high-energy components have been found 
in the neutron time-of-flight
experiments (see \cite{Bad} and references therein)
with pionic hydrogen and in diffusion experiments \cite{Abbott, RPD} 
with muonic hydrogen and deuterium atoms. 

The extended standard cascade model (ESCM) \cite{VEMark,CascalII} 
introduces a number of improvements compared to the earlier 
models: for example, the scattering from molecular hydrogen 
at high $n$ is calculated as opposed to the phenomenological 
treatment in other cascade models. Cascade calculations 
in this model include the evolution of the 
kinetic energy during de-excitation cascade.
However, the ESCM predictions could not  
describe a number of experimental data, in which the
kinetic energy distribution of an exotic atom at the instant of its
$x$-ray emission from a specific level in 
muonic \cite{DSCov, DGot4}
and pionic (e.g., see \cite{DGot3} and references therein) atoms
or at the time of the charge-exchange 
reaction (neutron time-of-flight experiment \cite{Bad}) 
is needed for the analysis of the experimental data.

To interpret and analyze these experimental data a more sophisticated 
approach based on the reliable and self-consistent  description of all
essential collisional processes involved in the de-excitation cascade is required. 
The processes of elastic scattering, Stark transitions, and CD appear to
be of significant importance, because of their considerable influence on
the kinetic energy distribution of exotic atoms during the atomic cascade.

 The study of the elastic scattering and Stark mixing cross sections
in the exotic atom - ordinary hydrogen atom collisions 
has been performed within a fully quantum-mechanical adiabatic 
approach~\cite{adiab1,adiab2,adiab3,GPP} about fifteen years ago. 
Later, these cross sections have been also calculated
in the framework of the close-coupling model~\cite{JM} applying
the screening dipole approximation to describe
the exotic atom-hydrogen atom interaction for the low states $n=2-5$. 
However, this approximation can not be justified
for low-energy scattering (see \cite{GPP}). 
Thus, the latter approach as well as various modifications 
of the semiclassical model~\cite{BL,TH,JM} lead to 
unreliable results in the low-energy region where 
only the lowest partial waves are important.
 
The theoretical study of the CD has also a long 
history (e.g., see \cite{BF,PS1, AAA,JM2,PS2,KM,MF}).
Nevertheless, in the most important region $n=2-8$ 
of the principal quantum number, relevant for the kinetics 
of the atomic cascade, 
this process has been regarded as the least studied 
until last decade~\cite{PPForm, KPP, diff, pion}.

The present research continues the cycle
of our papers devoted to the 
detailed theoretical study of the scattering processes and 
the kinetics of the atomic cascade in lightest exotic 
hydrogen-like atoms. 
In particular, in our previous papers 
(e.g., see \cite{PPForm, diff, KPP, pion}), 
the dynamics of collisions
of excited exotic atoms with hydrogen ones has been studied
in the framework of the close-coupling approach. 
         
The present study has at least 
three main goals. First, to obtain in framework of
a unified quantum-mechanical approach  
the cross sections of elastic
scattering, Stark transitions, and CD in collisions
of excited $\mu^- d$ atoms 
with ordinary deuterium atoms, which,
according to our knowledge, have never 
been calculated before. Second, to provide a
comprehensive set of differential and integrated 
cross sections for all collisional transitions 
$nl\to n'l'$  in $\mu^- p$ and $\mu^- d$ atoms
at relative energies needed for the
detailed study of the kinetics of the de-excitation cascade 
in the hydrogen isotope targets. 
Finally, to investigate 
isotopic effects in the scattering
cross sections of excited $\mu^- p$ and $\mu^- d$ 
atoms, calculated on the same theoretical footing, 
and in various characteristics of the
atomic cascade, such as absolute and relative 
$x$-ray yields,  
kinetic energy 
distributions of muonic hydrogen and deuterium 
atoms at the instant 
of their radiative transitions, and
cascade times.

The paper is organized as follows. The brief description of  the
theoretical approach used to describe the scattering of the excited muonic
hydrogen and deuterium atoms in the hydrogen isotope targets  is given in
Sec.~II.  In Sec.~III, we demonstrate the typical examples illustrating the
closed channel effect and the convergence of  calculated cross sections
relative  to the extension of the basis and the number of  partial waves
taken into account.  The isotopic effect in the scattering cross sections
and rates of different CD is also represented in Sec.~III. The kinetics of
the atomic cascade of  $\mu^- p$ and $\mu^- d$ atoms is studied in Sec.~IV.
Here, the various results calculated in the improved version of  the atomic
cascade are presented and compared  with the known experimental data. The
conclusions  are summarized in Sec.~V. 

  Atomic units are used unless otherwise stated.  
  
\section { Cross sections  of the collisional processes}
 
\subsection {Main features of the approach}

To study the scattering processes in collisions of 
$(\mu^- p)_{nl}$ and $(\mu^- d)_{nl}$ atoms in the excited
$(nl)$ states ($n$ and $l$ are the principal quantum
number and orbital angular momentum of the exotic atom) 
with ordinary hydrogen or 
deuterium atoms in the ground state, the 
close-coupling approach is applied. In the framework of this 
approach the processes, 
\begin{equation}
(\mu^- a)_{n l} + (be^-)_{1s} \to (\mu^- a)_{n'l'} + (b e^-)_{1s},
\label{Cd}
\end{equation}
of the elastic scattering ($n'=n, l'=l$), Stark transitions
($n'=n, l' \neq l$), and CD ($n' \leqslant n-1$)
are treated in a unified manner.
Here, $a=b=p$ in the case of the muonic hydrogen atom,
and $a=b=d$ in the case of the muonic deuterium 
atom, respectively.

The close-coupling approach was  
applied earlier by the authors 
for the fully quantum-mechanical description of the
scattering processes (\ref {Cd}) in collisions of excited 
exotic (muonic, pionic, kaonic, and 
antiprotonic)  hydrogen atoms with ordinary hydrogen 
ones~\cite{PPForm, KPP,diff,pion} 
and hydrogen molecules~\cite{PPM}, as well as 
for the collision-induced
absorption or annihilation in the case of hadronic 
atoms~\cite{PPInd,PPInHyp}. 
For the benefit of the reader, the outlines 
of the applied close-coupling 
approach are briefly reminded.

In this paper, as in our previous studies,
the target atom is assumed to be in its ground state 
during the collision.
In order to properly take into account the electronic degrees  of freedom in considered
processes, it is necessary to consider the quantum four-body problem.  However, in our
case we are dealing with a collision of two neutral subsystems  (atoms of the muon and
ordinary hydrogen). A muonic atom can be considered as a point dipole with a small dipole
moment. So, the polarization of atom in the collision leads to rapidly decreasing   Van
der Waals interaction $\sim 1/R^6$.  Such an interaction can affect only elastic
scattering cross section. Due to the smallness of the dipole moment  this effect can not
be significant.  Note that even in the case of alpha decay of heavy nucleus accounting
distortions of the inner electron shell  (due to the interaction of electrons with the
charged alpha particle) has no significant effect on the rate  of the alpha decay, as
shown in the papers~\cite{Karp1, Karp2}.

Thus, at the present stage of the study 
of the considered processes, we are limited to the ``frozen'' electron approximation.
Then the four-body scattering problem under consideration
is reduced to the three-body one and can be 
described by the wave function 
$\Psi_{E}^{JM{\it p}} (\boldsymbol{\rho},\mathbf{R})$ 
of the three-body system $(\mu^- a)_{nl}
 + A$ ($A$ denotes an ordinary 
 hydrogen or deuterium atom). The set of
Jacobi coordinates $(\mathbf{R},
\boldsymbol{\rho})$ of the system
is defined as:
\begin{equation}
\mathbf{R} =
\mathbf{R}_A - \mathbf{R}_{\mu a},\quad
\boldsymbol{\rho} = \mathbf{r}_{\mu } -
\mathbf{r}_{a}.\quad  
 \label{Jc}
\end{equation}
Here, $\mathbf{R}_A$ and  $\mathbf{R}_{\mu a}$ 
are the center-of-mass radius-vectors of hydrogen and exotic atoms;
$\mathbf{r}_a$ and $\mathbf{r}_{\mu}$ are radius-vectors
of the nucleus $a$ and muon in the laboratory system. 
 
 At a fixed total energy $E$ of the system, definite quantum numbers
of the total angular momentum $(J, M)$ and parity 
${\it p} = (-1)^{l+L}$, the exact wave function 
$\Psi_{E}^{JM{\it p}} (\boldsymbol{\rho},\mathbf{R})$
satisfies the time-independent Schr\"{o}dinger equation,
\begin{equation}
(H-E)\Psi_{E}^{JM {\it p}}({\boldsymbol{\rho}}, {\bf R})=0.
\label{Scheq} 
\end{equation}
The Hamiltonian, $H$, after the separation of the 
center-of-mass motion of the total system, can be
written in terms of Jacobi coordinates (\ref{Jc}) as
\begin{equation}
H = -\frac{1}{2M_r}\Delta _{\mathbf{R}} +
h_{\mu}(\boldsymbol{\rho}) 
+V(\boldsymbol{\rho},\mathbf{R}),
 \label{Ham}
\end{equation}
where $M_r=M_{\mu a}M_{t}/(M_{\mu a} + M_{t})$ is the reduced mass 
of the system with $M_{\mu a} = m_{\mu}+m_a$  
and $M_t = m_e+m_b$ ($m_a$, $m_b$, $m_{\mu}$, and  $m_e$ are 
masses of hydrogen isotopes, muon, and electron, respectively). 

The Hamiltonian $h_{\mu}$ of the free exotic atom involves  
Coulomb interaction of the non-relativistic 
two-body problem as well as terms leading to the 
energy shift of $ns$ states relative to the 
degenerate $nl$ states ($l\neq 0$) mainly account for by  
electron vacuum polarization and nuclear finite
size effects. In the case of muonic atoms 
\begin{equation}
\label{hmu}
\langle nlm|h_{\mu}|nlm \rangle = E_{nl}
= -\frac{0.5m_r}{n^2}-\delta_{l0}\varepsilon^{\rm Lamb}_{n\geqslant 2},
\end{equation}
where 
$\langle {\boldsymbol{\rho}}|n l m\rangle=R_{nl}(\rho)Y_{lm}(\hat{\b\rho})$
is the hydrogen-like wave function of muonic atom in the $n l m$ state, 
$m_r~=~m_{\mu}m_a/M_{\mu a}$ is its reduced mass, 
and $\varepsilon^{\rm Lamb}_{n\geqslant 2}$ is the shift of the Coulomb
energy level due to electron vacuum polarization and nuclear finite
size effects. In the present study we use 
$\varepsilon^{\rm Lamb}_2~=~202.08$~meV 
and $\varepsilon^{\rm Lamb}_{2}~=~203.01$~meV~\cite{EB} 
for muonic hydrogen and muonic deuterium, respectively. 
For $ns$ states with $n\geqslant 3$ the energy shift 
due to electron vacuum polarization and nuclear finite
size effects is approximately calculated according to   
$\varepsilon^{\rm Lamb}_n~=~\varepsilon^{\rm Lamb}_{2}(2/n)^3$.  
  
The interaction potential $V(\boldsymbol{\rho},\mathbf{R})$ 
in (\ref{Ham}) is obtained by averaging the sum of two-body Coulomb 
interactions $V_{i j}$ $(i=a, \mu^-$ 
and $j=b, e^-)$ between particles of colliding subsystems 
over the ground-state wave function of the hydrogen 
isotope atom (for details see Appendix A).

The wave function $\Psi_{E}^{JM{\it p}} (\boldsymbol{\rho},\mathbf{R})$
of the three-body system $(\mu^- a)_{nl}+ A$
is expanded on basis states  
\begin{equation}
\langle {\boldsymbol{\rho}}, \mathbf{\hat R}|n l, L:JM\rangle\equiv 
R_{nl}(\rho){\cal Y}_{l L}^{JM}(\hat {\b\rho}, \hat{\bf R}),
 \label{af}
\end{equation}
constructed from non-relativistic hydrogen-like wave functions 
$\langle {\boldsymbol{\rho}}|n l m\rangle$
of the muonic atom and
\begin{equation}
{\cal Y}_{l L}^{JM} (\hat {\b\rho}, \hat{\bf R})\equiv
i^{l+L}\sum_{m \lambda}\langle l m L \lambda |JM \rangle
Y_{lm}(\hat{\b\rho}) Y_{L \lambda}(\mathbf{\hat R}).
 \label{ang}
\end{equation} 

The substitution of the expansion for $\Psi_{E}^{J M {\it p}}$,
\begin{equation}
\Psi_{E}^{J M {\it p}}( {\boldsymbol{\rho}}, \mathbf{R})=
\frac{1}{R} \sum_{nl L}G_{nlL}^{EJ {\it p}}(R)
\langle{\boldsymbol{\rho}}, \mathbf{\hat R}|nl,L:JM\rangle, 
\label{twf}
\end{equation}
into the Schr\"{o}dinger equation (\ref{Scheq}),
leads to the set of the close-coupling second-order differential 
equations for the radial channel wave functions $G_{nlL}^{EJ {\it p}}(R)$:
\begin{align}
\left(\frac{d^2}{dR^2} + k^2_{nl} -
\frac{L(L+1)}{R^2}\right)G^{EJ {\it p}}_{nlL}(R) \nonumber
\\= 2M_r
\sum_{n'l'L'}W^{J {\it p}}_{n'l'L', nlL}(R)\,G^{EJ {\it p}
}_{n'l'L'}(R), \label{CCE}
\end{align}
where
\begin{equation}
k^{2}_{nl}=2M_{r}(E_{\rm cm}+E_{n_i l_i}-E_{nl})
\label{wn}
\end{equation}
specify the channel wave numbers; $E_{\rm cm}$,
$E_{n_i l_i}$, and $E_{nl}$ are respectively the relative motion 
energy and bound energies of the exotic atom in the entrance 
and current channels. The basis used in the present study
includes exotic atom states corresponding to both open 
($k^{2}_{nl}>0$) and closed ($k^{2}_{nl}<0$, \rm{Im}~$k_{nl} >0$)
channels.

The explicit analytical expression for a matrix  
\begin{equation}
W^{J {\it p}}_{n'l'L',nlL}(R)=\langle nl,L:JM|
V(\boldsymbol{\rho},\mathbf{R})|n'l',L':JM\rangle 
\label{potW}
\end{equation}
of the interaction potential, $V(\boldsymbol{\rho},\mathbf{R})$,
between the asymptotic initial $|nlL; J\rangle$ and final $|n'l'L'; J\rangle$
channels is obtained by integrating over 
$\boldsymbol{\rho}$ and $\mathbf{\hat R}$
without any additional approximations. 
As a result, the matrix elements (\ref{potW}) 
are reduced to the finite sum of
multipole interactions $W_{t;n'l',nl}(R)$ 
(allowed by the angular momentum
and parity selection rules)
with the asymptotic behavior:
\begin{equation}
\label{aspotW0}
W_{t;n'l',nl}(R)\propto R^{t} \,\,\,(R\to 0)
\end{equation}
and 
 \begin{equation}
W_{t;n'l',nl}(R) \propto \exp(-2R)/R^{t+1}\,\,\,(R\to \infty),
\label{aspotW1} 
\end{equation} 
where $t$ is a multipolarity, which changes from 
$t_{\rm min} = |l-l'|$ up to $t_{\rm max} = l+l'$ 
(for details see Appendix A).
  
 The scattering problem (\ref{CCE}-\ref{potW}) was solved
 numerically by applying the propagator matrix method,
 developed earlier (for details see \cite{PPForm}),   
at each fixed value of the relative energy, total angular
momentum  and parity. In the frame of this method, instead of 
radial channel functions $G_{nlL}^{EJ {\it p}}(R)$, we find their ratios
in two nearest points $R_i$ and $R_{i+1}$ 
($R_{i+1}=R_i + h$; $h$ is the constant integration step), starting
from $R_1=0$. Then, 
using the two-point matching conditions,
the $K$  matrix, defined in the subspace
of open channels ($k^{2}>0$), and thereby 
the $T$ matrix ($T=-2iK(I-iK)^{-1}$) are calculated.
The method
allows to avoid numerical problems with both the exponentially growing
and exponentially damping functions 
and significantly improves the accuracy of the calculated
$T$ matrix and correspondingly cross sections. 
 
The degeneracy-averaged partial-wave differential cross sections 
for processes (\ref{Cd}) are given by
\begin{equation}   
\frac{d \sigma^{J}_{n l \to n'l'}}{d\Omega} = 
\frac{1}{2l+1}\frac{k_{n'l'}}{k_{nl}}\sum_{m, m'}
|f_{nlm\rightarrow n'l'm'}^{J}(\Omega)|^2. 
\label{sigma dif}                   
\end{equation}
Here, the partial-wave on-shell scattering amplitudes, 
$f_{nlm\rightarrow n'l'm'}^{J}(\Omega)$, for the transition 
$i\rightarrow f$ ($i$ and $f$  denote the quantum numbers
of initial  
$|n\,l\,m\rangle$ and final $|n'\,l'\,m'\rangle$ states of
exotic atom, respectively) 
are determined by the transition matrix $T^J_{nlL\to n'l'L'}$: 
\begin{align}
&f_{i\rightarrow f}^{J}(\Omega)=  
\frac{i\sqrt{\pi}}{\sqrt{k_{nl}k_{n'l'}}} \sum_{LL'\lambda'}
i^{L'-L}\sqrt{2L+1}\langle lmL0 |J m\rangle\nonumber\\ 
&\times 
T^J_{nlL\to n'l'L'}
\langle l'm'L'\lambda'|J m\rangle Y_{L'\lambda'}(\Omega), 
\label{ampl} 
\end{align} 
where the sums go over all values $(L, L', \lambda') $, 
satisfying angular momentum and parity selection rules; 
$\Omega=(\theta, \phi)$ is the center-of-mass 
solid angle. 

The partial-wave integrated cross sections 
for $nl \rightarrow n'l'$ transitions are given by
\begin{equation}
\sigma^{J}_{n l \to n'l'} =  
\frac{\pi}{k_{nl}^2}\frac{2J+1}{2l+1} \sum_{L
L'}|T^J_{nlL\to n'l'L'}|^2. 
\label{sigmaJ}  
\end{equation} 
 
Then the total differential, 
\begin{equation}
\label{totdif}
\frac{d \sigma_{n l \to n'l'}}{d\Omega} = \sum_{J=0}^{J_{\rm max}}
\frac{d \sigma^{J}_{n l \to n'l'}}{d\Omega}, 
\end{equation}
and integrated, 
\begin{equation}  
\sigma_{nl \to n'l'}=\sum_{J=0}^{J_{\rm max}} \sigma^J_{nl \to n'l'},
\label{totsig}
\end{equation} 
cross sections are obtained by summing 
(\ref{sigma dif}) and (\ref{sigmaJ}) 
over partial waves from $J=0$ up to  $J_{\rm max}$ 
until an accuracy better than 0.01\% was reached for 
all collisional energies. 
 
The quantum-mechanical approach described above was applied
to produce the systematic calculations of 
differential and integrated cross sections 
(\ref{sigma dif}, \ref{sigmaJ} - \ref{totsig}) for  
processes (\ref{Cd}) in collisions of excited
$\mu^- p$ and $\mu^- d$
with ordinary $H$ and $D$ atoms, respectively.
These calculations were performed using the basis,
which includes all open and 
closed channels, associated with the
principal quantum number values from $n=1$
up to $n_{\rm max} = 20$. In the case of the 
$(\mu^- p)_{2s}$ and $(\mu^- d)_{2s}$, for a collision energy below
$2p$ threshold,
the used basis was even extended up to $n_{\rm max} = 30$.
Since the energies of $ns$ and
$n l\, (l\geqslant1)$ sub-levels are split 
(mainly due to electron vacuum
polarization), the cross sections with
initial $ns$ and $n l$ ($l\geqslant 1$) states 
were calculated separately for relative energies 
referring to $ns$ and $n p$ thresholds.
The results of our systematic calculations   
are extremely extensive and were used as input data 
for the detailed description
of the atomic cascade kinetics. 
   
Here, we present only
the key points illustrating both the convergence of
various cross sections (elastic scattering, Stark transitions, 
and CD)
with the extension of the basis and the isotopic effect
in the scattering of excited muonic hydrogen and deuterium. 
To demonstrate some of our results, the 
$l$-averaged cross sections of  
$n\to n'$ $(n'\leqslant n)$ transitions, obtained     
by summing  (\ref{totsig}) over $l'$ and averaging 
with the statistical weight $(2l+1)/(n^2-1)$ over $l\geqslant1$ 
are defined as follows
\begin{equation}
\sigma_{n\to n'}^{\rm{av}}(E) =
\frac{\pi}{k_{nl}^2}\frac{1}{n^2-1}\sum_{l\geqslant 1,\,l\,'}(2l+1)
\sigma_{nl \to n'l'}.
\label{avcrsec}
\end{equation}

\subsection{The effect of closed channels in the scattering of excited muonic hydrogen atoms}

In real applications of the close-coupling approach 
the used basis is always truncated. All the
basis states  associated with open channels are 
usually taken into account. The reliability of 
results obtained with this basis is unclear without the
further study of the effect of closed channels.
There are no simple theoretical representations or estimations, 
which allow to choose the optimal basis. 
Hence, the scattering problem has to be 
investigated numerically.
This study is also necessary from the viewpoint of practical
applications, to produce calculations 
within the reasonable computer time.

\begin{figure}[!ht]
\centerline{\includegraphics[width=0.47\textwidth,keepaspectratio]
{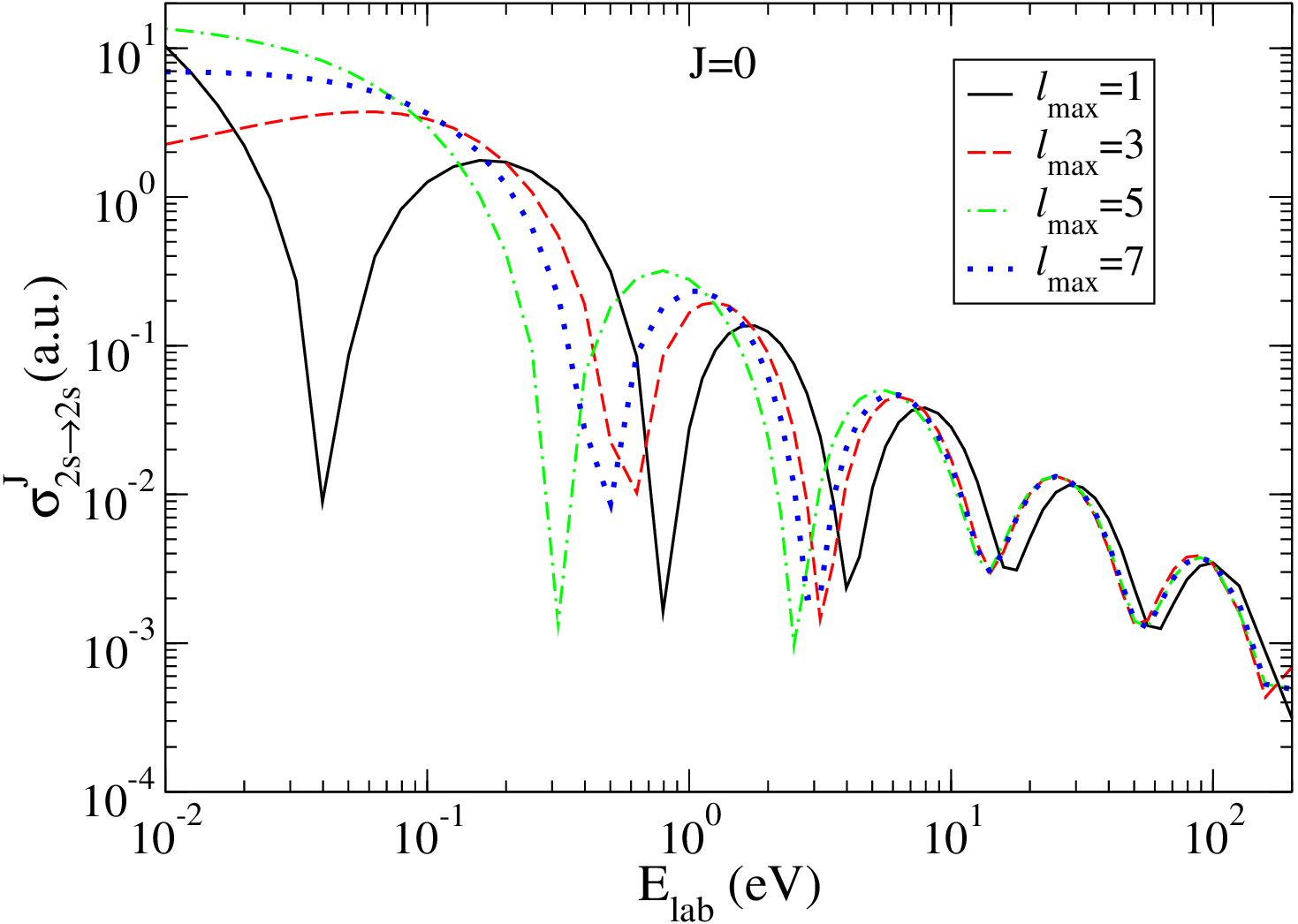}}
 \caption{(Color online) The partial-wave
 elastic cross section for $(\mu^- p)_{2s} + H$ collision
 vs. the laboratory kinetic energy $E_{\rm lab}$  
 calculated for four variants of the 
 basis corresponding to $n _{\rm max}= 8$ and different values
 $l_{\rm max}$ of the internal orbital momentum.}
 \label{part_el_l}
       \end{figure}

\begin{figure}[!ht]
\centerline{\includegraphics[width=0.47\textwidth,keepaspectratio]
{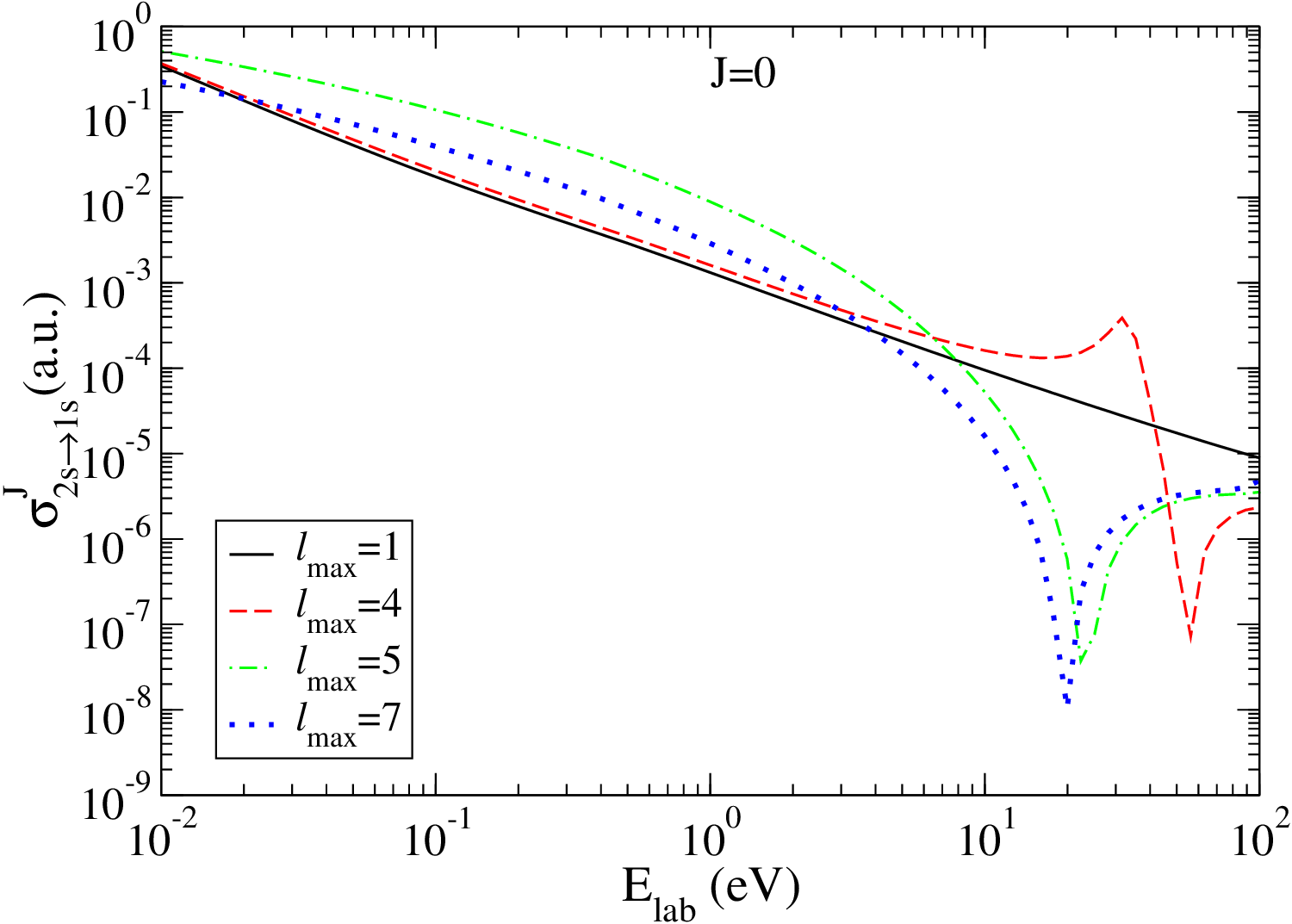}}
 \caption{(Color online) The same as Fig.~\ref{part_el_l} 
 for the partial-wave cross section of the Coulomb 
 de-excitation $2s\to 1s$.}
 \label{part_cd_l}
       \end{figure}

\begin{figure}[!ht]
\centerline{\includegraphics[width=0.47\textwidth,keepaspectratio]
{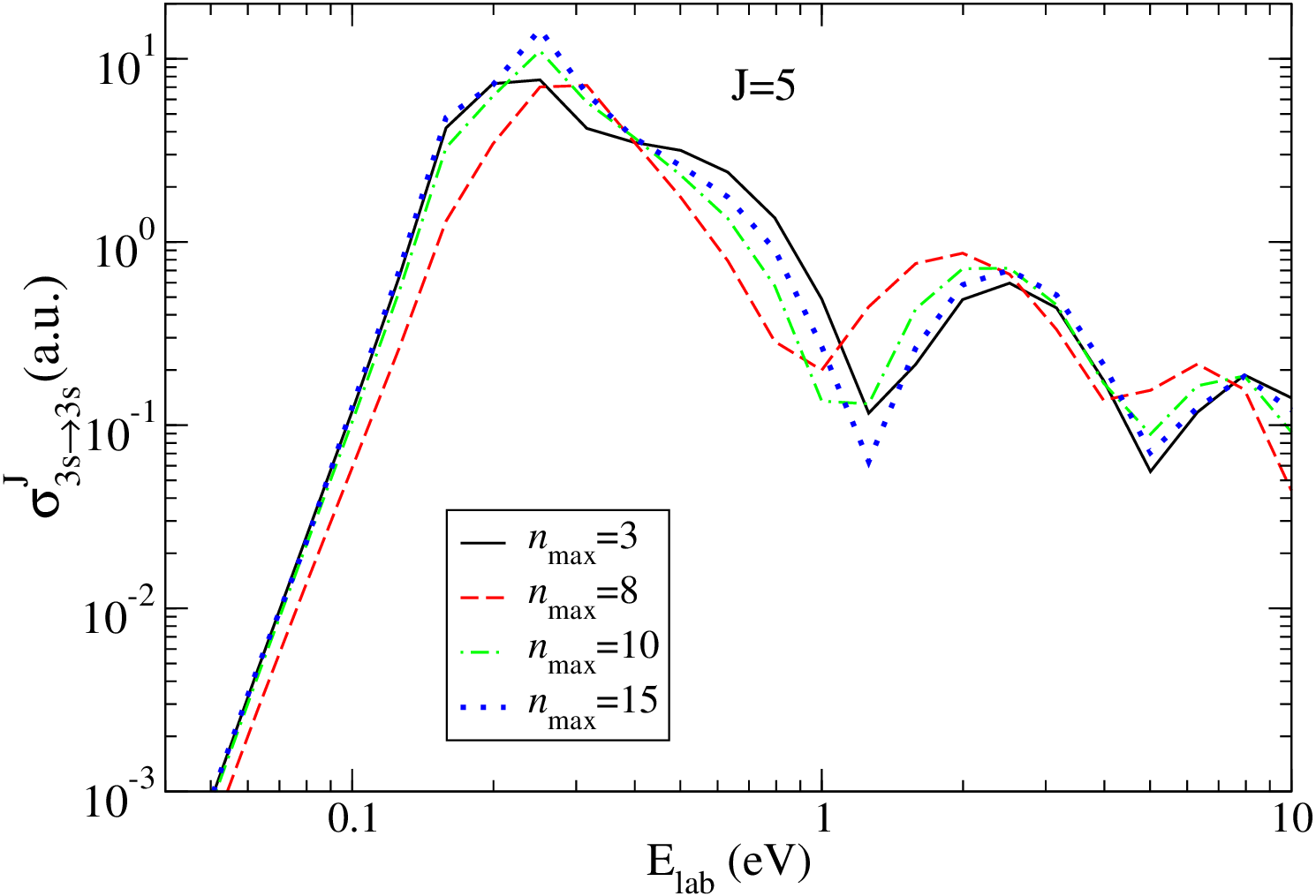}}
\caption{(Color online) The energy dependence of the partial-wave
elastic cross section for $(\mu^- p)_{3s} + H$ collision 
calculated with the basis corresponding to 
different values $n _{\rm max}$.}
 \label{part_el_n}
        \end{figure}

\begin{figure}[!ht]
\centerline{\includegraphics[width=0.47\textwidth,keepaspectratio]
{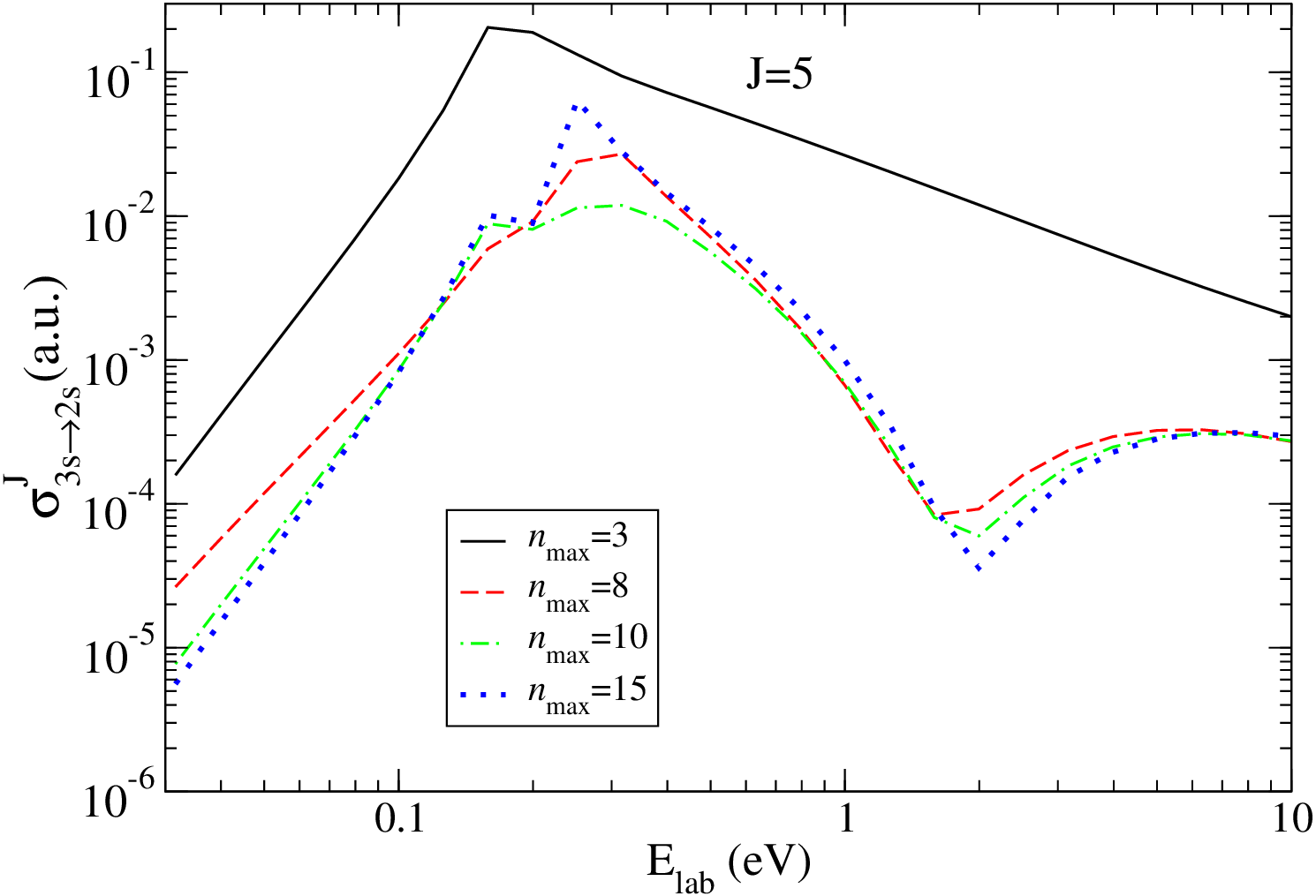}}
 \caption{(Color online) The same as Fig.~\ref{part_el_n} 
 for the partial-wave cross section of the Coulomb 
 de-excitation $3s\to 2s$.}
 \label{part_cd_n}
        \end{figure}

\begin{figure}[!ht]
\centerline{\includegraphics[width=0.47\textwidth,keepaspectratio]
{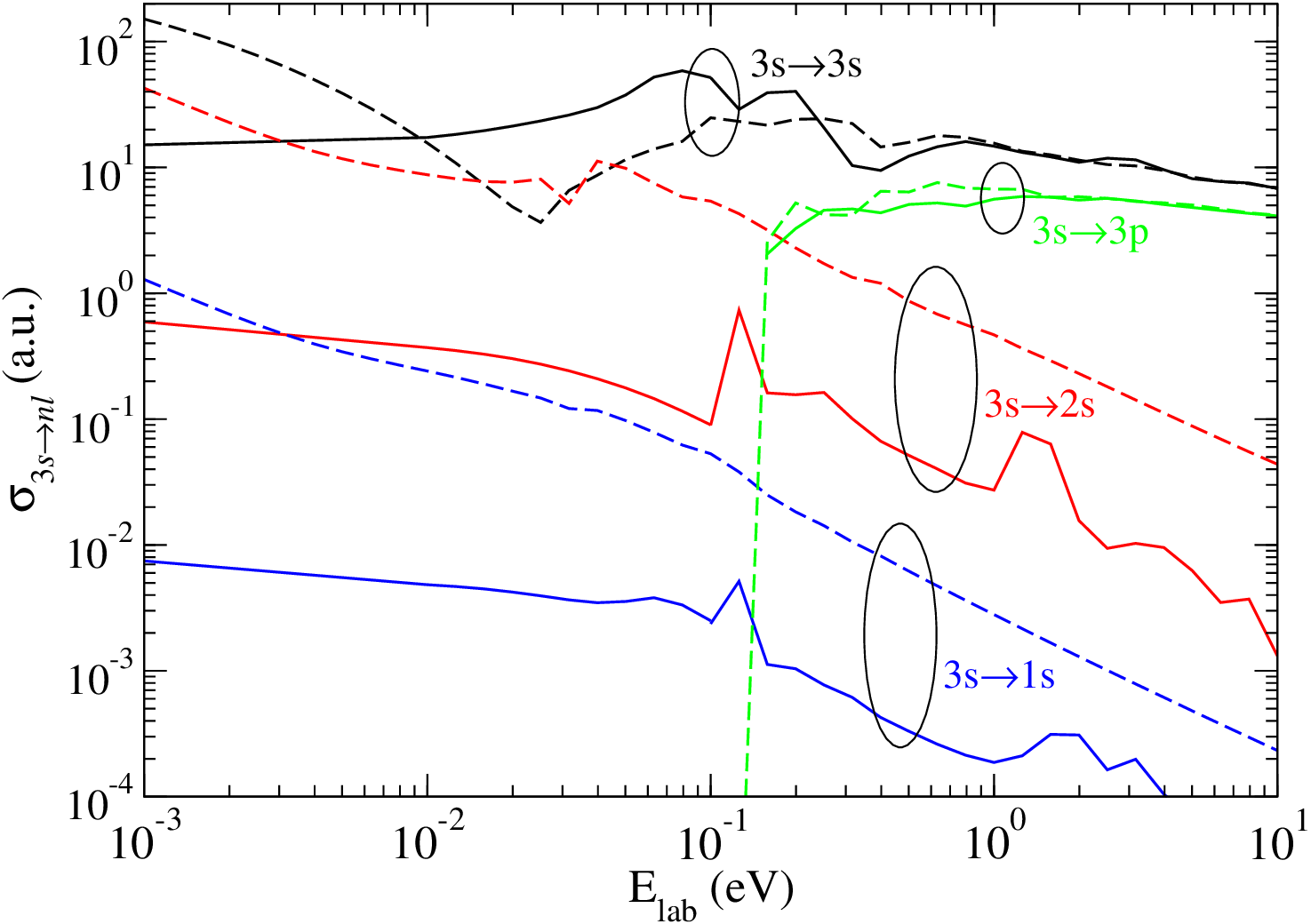}}
 \caption{(Color online) The energy dependence of the integrated
 cross sections for $(\mu^- p)_{3s} + H$ collisions
calculated with different basis sets: 
$n _{\rm max}= 3$ (dashed lines) and
 $n _{\rm max}= 20$ (solid lines).}
 \label{crsec_3s_n}
        \end{figure}

The extension of the basis set by adding 
muonic atom states associated with closed 
channels alters the effective interaction at short 
distances and, as a result, should affect the 
low-energy scattering of exotic atoms in the 
states with small values of the principal quantum number $n$.
To study the closed channel 
effect and convergence of the close-coupling approach
we have performed a lot of calculations with large basis sets.
The key features of the closed channel effect 
in the scattering of the excited muonic hydrogen
are demonstrated in Figs.~\ref{part_el_l}~-~\ref{crsec_3s_n}. 
 
Figure~\ref{part_el_l} shows the dependence of the                
partial-wave cross section ($J=0$) of the
elastic $2s-2s$ scattering on $E_{\rm lab}$ which is the laboratory kinetic energy of muonic atom in
its collision with target at rest. The calculations were performed 
for various basis sets corresponding to   
$n_{\rm max} = 8$ and different values of the internal
angular momentum 
$l \leqslant l_{\rm max}$ ($l_{\rm max}=1, 3, 5, 7$).
At the laboratory kinetic energy $E_{\rm lab}\lesssim 0.4$~eV 
(below the $2p$ threshold)
there are  only two open channel 
associated with $1s$ and $2s$ states of the muonic  
hydrogen. All the other channels are closed: the weakly closed
$2p$ channel and strongly closed channels 
for $n > 2$ (here the energy gap between the open and
closed channels is more than a few 
hundred electron volts). At collision 
energies above the $2p$ threshold the $2p$ channel 
becomes also opened. Thus, the maximum number 
of open channels at energies above the $2p$ threshold 
is three. In the case $J=0$ ($L=l$ and parity ${\it p}=1$), 
the number of channels equals to the number of basis states. 
Therefore,
for the basis with $n_{\rm max}=8$ and $l_{\rm max}=1$ or $l_{\rm max}=7$
one has a 15-channel or 36-channel 
scattering problem, respectively.       

The effective interaction in the system 
is very strong and leads to the formation of a number of bound
states below $2s$ threshold. 
This is confirmed by oscillations of the elastic 
cross section $2s-2s$ in Fig.~\ref{part_el_l}
in accordance with Levinson's theorem. 
 The extension of the basis increases the effective 
interaction (attraction at short distances) and as a 
consequence all maxima and minima 
in the cross section are shifted to lower energies. 

Figure~\ref{part_cd_l} shows the energy dependence of the                
partial-wave cross section ($J=0$) of the
CD $2s\to1s$.  
The calculations were performed for four variants of the basis   
corresponding to   
$n_{\rm max} = 8$ and different values of the internal
angular momentum $l \leqslant l_{\rm max}$ 
($l_{\rm max}=1, 4, 5, 7$). Here the number of open and closed
channels is determined by both the collision energy and the
value of $l_{\rm max}$ as in the case of the elastic scattering
$2s - 2s$. 
          
As it is seen in Figs.~\ref{part_el_l} and \ref{part_cd_l}, 
the  effect of closed channels is very strong.
The cross sections of the elastic 
scattering $2s-2s$ and CD $2s\to 1s$ are
essentially modified with the extension of the basis. 
The convergence is very slow.
 The lowest $l$ orbitals are very important, but, as a rule, 
 do not lead to a reliable result.
Cross sections calculated by taking into account only $ns$ and 
$np$ states differ substantially from those obtained 
with the further extension of the basis by including the 
states with large values of the internal orbital angular 
momentum $(l \geqslant 2)$.
In particular, the extension of the basis results in the 
resonance behavior of both elastic $2s-2s$ and Coulomb
$2s\to1s$ cross sections above $E_{\rm lab}\approx 20$~eV. 
The position and shape of this resonance essentially
change with the extension of the basis (see Fig.~\ref{part_cd_l}).    
 The results obtained with the basis sets corresponding to
$l_{\rm max}=6$ and 7 are 
indistinguishable for both elastic scattering and CD cross sections.

 The convergence of the partial-wave ($J=5$) cross sections  
of the elastic $3s-3s$ scattering and the
CD $3s\to 2s$ of the muonic hydrogen with 
the extension of the basis (by increasing $n_{\rm max}$)
is demonstrated in Figs.~\ref{part_el_n} and \ref{part_cd_n}, 
respectively. Calculations were done for different basis sets 
corresponding to muonic hydrogen states with 
$n_{\rm max}$=3, 8, 10, 15, 20 and all values of the 
internal orbital angular momentum at a given $n_{\rm max}$.

The closed channel effect is very noticeable in the case
of the elastic scattering (see Fig.~\ref{part_el_n}). 
The partial-wave cross section  
of the elastic $3s-3s$ scattering changes 
with the extension of the basis from $n _{\rm max}= 3$ 
to $n _{\rm max}= 8$ about a few times
near its values calculated with the minimal basis 
($n _{\rm max}= 3$), however, its energy behavior 
is conserved. Further enlargement of basis set leads to the convergence 
of the calculated cross section as shown in Fig.~\ref{part_el_n}
(see curves corresponding to $n _{\rm max}= 10$ 
and $n _{\rm max}= 15$). Note that the results of the calculation 
with a basis set $n _{\rm max}= 20$ practically coincide 
with those obtained with the basis $n _{\rm max}= 15$ 
and not shown in Fig.~\ref{part_el_n}.

In the case of the CD 
(see Fig.~\ref{part_cd_n}) the effect is much stronger.       
The extension of the basis leads to a significant reduction
of the cross section of the CD $3s\to 2s$ 
(more than one order of magnitude) and, as in the case
of the Coulomb $2s\to 1s$ de-excitation  (see Fig.~\ref{part_cd_l}),
its energy dependence changes as well. As in the case of the elastic
scattering $3s-3s$, the convergence of the partial-wave CD $3s\to 2s$
cross section is achieved with the basis set corresponding to
$n _{\rm max}= 15$ and further extension of the basis set to
$n _{\rm max}= 20$ leads to variation of the cross section less
than 1\%.   

 \begin{table}[ht]
\caption{The maximum number of closed channels in calculations 
of elastic $3s-3s$ scattering and CD $3s\to 2s$ for different 
basis sets and $J$ values.}
\begin{center}
\begin{tabular}{c||c|c|c|c} \hline \hline
$\; n_{\rm max} \;$&\;\; 3 \;\;&\;\;  8 \;\; & \;\;12 \;\;&\;\; 20 \;\;\\ \hline
$\; J=5 \;$ &\; \;0 \;\;&\;\; 106 \;\; &\;\; 298 \;\; &\;\; 754 \;\;\\ \hline
$\; J=10 \;$ &\;\; 0 \;\; &\;\; 110 \;\.&\;\;  353 \;\; &\;\; 959 \;\; \\  \hline \hline
\end{tabular}
\end{center}
\label{chnum}
\end{table}

In the considered examples of the elastic $3s-3s$ scattering 
and the CD $3s\to 2s$ at the collision energy above $3p$ threshold 
($E_{\rm lab}\gtrsim 0.125$~eV) one has 10 open channels. 
The number of the closed channels essentially depends on
both $n _{\rm max}$ and $J$ values.
It is important to note, that with increasing $J$ the number 
of closed channels increases significantly (see Table~1).
 
Figure~\ref{crsec_3s_n} shows the energy dependence of  integrated
 cross sections for $(\mu^- p)_{3s} + H$ collisions.
 Here we present numerical results for the elastic scattering
 $3s - 3s$, Stark transition $3s \to 3p$, and
  CD $3s\to2s$ and 
 $3s\to1s$. Calculations were performed for
 two variants of the used basis: the minimum basis set  
($n_{\rm max}=3$) and the maximum basis set including all the
muonic atom states associated with $n \leqslant 20$. 
The comparison of the results obtained in these two variants 
shows, that the closed channels strongly affect
all cross sections at collision energies below a few electron volts. 
In particular, the cross section of the elastic scattering $3s - 3s$ changes 
 its energy dependence at low energies with the extension of the basis. 
 A much stronger effect is observed in the case of the
 CD. The cross sections of $3s\to2s$ and 
 $3s\to1s$ CD are reduced more than 
 an order of magnitude with the extension of the basis at 
 all energies under consideration.

Our investigation shows, that 
the convergence 
of cross sections with the extension of the basis
is achieved very slowly in the lowest partial waves and 
at low-energy collisions due to a strong coupling
between closed and open channels. The closed channel effect 
is especially significant in the low-energy scattering of exotic atoms in
low-lying excited states ($n=2-5$), because the basis of the 
open channels is very poor to produce the reliable 
results (e.g., see Fig. 4 in \cite{PPForm}). 
  
The present as well as our previous studies~\cite{PPForm}
show, that the energy dependence of the
cross sections at low-energy collisions is very important for the 
kinetics of the atomic cascade. Therefore, the cross sections of different
processes in the low-energy
region should be calculated using the extended basis set,
which includes a lot of closed channels. All our further calculations are 
performed using a basis set of $n_{\rm max} =  20$, 
the use of which allows to achieve the convergence of all cross 
sections in the whole energy range. Used basis set contains 
hundreds or even thousands of states depending on the values 
of the total angular momentum of the system. According to the present 
study, in the case of the muonic deuterium the convergence is achieved
much faster than in the case of muonic hydrogen. Nevertheless, all calculations 
for both muonic hydrogen and muonic deuterium were performed 
with the same basis set corresponding to $n_{\rm max} =  20$. 
 
\subsection{$J$-dependence of different cross sections}

 \begin{figure}[!ht]
\centerline{\includegraphics[width=0.47\textwidth,keepaspectratio]
{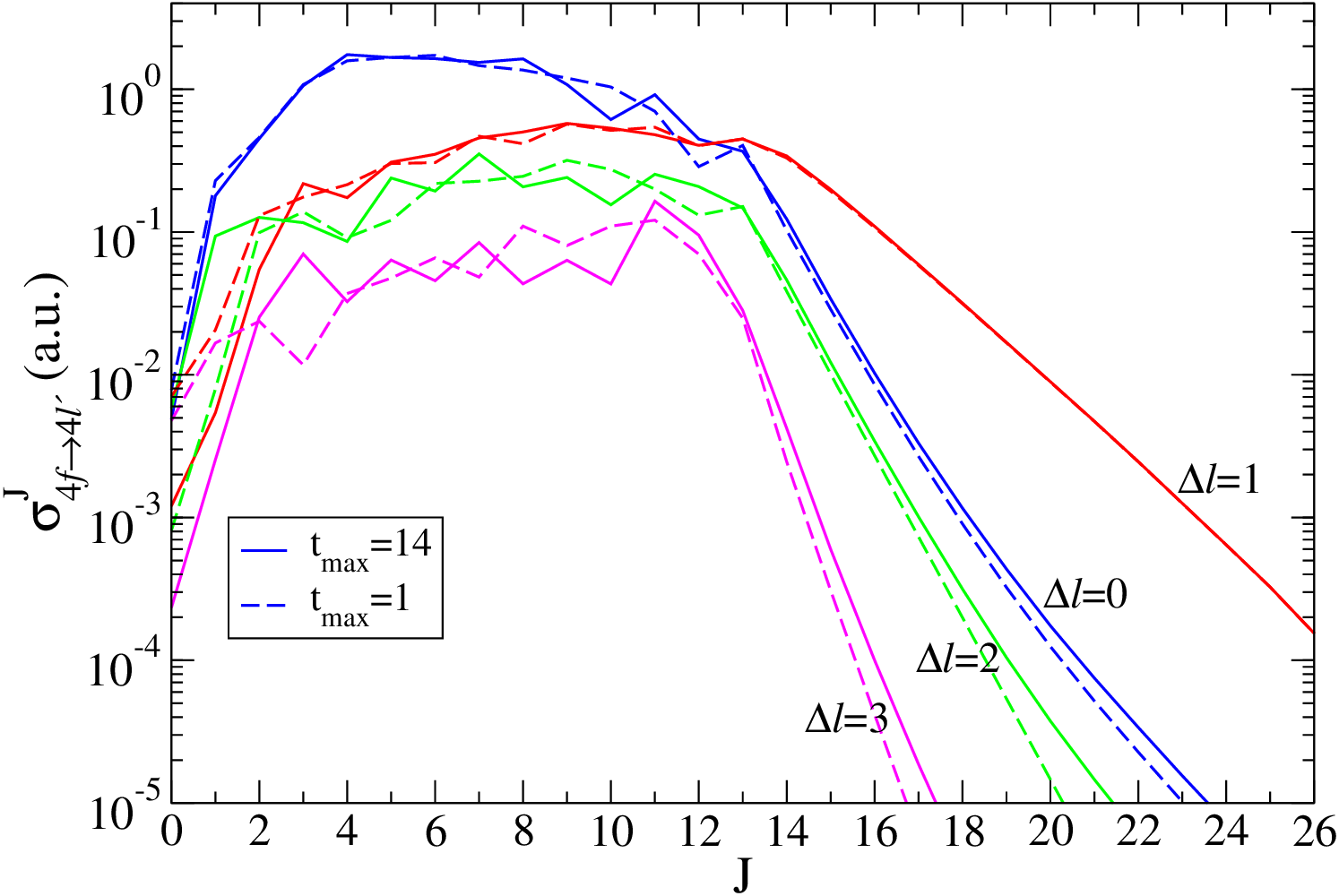}}
 \caption{(Color online) The partial-wave
 elastic $4f-4f$ and Stark $4f\to 4l$ $(l=0 - 2)$ cross sections 
 for $(\mu^- p)_{4f} + H$ collision vs. the total 
 angular momentum $J$ at the laboratory kinetic energy 
 $E_{lab}=$~1~eV. The calculations were performed 
 with the basis set including
 all the exotic atom states with principal quantum numbers 
 $n\leqslant 20$ ($\Delta l = \left|l_i - l_f\right|$)
 for two variants of interaction potentials: $t_{\rm max}=14$
 (solid lines) and  $t_{\rm max}=1$ (dashed lines).}
 \label{part_el_t}
 \end{figure}
Figure~\ref{part_el_t} shows the $J$-dependence of 
cross sections for the elastic $4f - 4f$
scattering and Stark $4f\to 4l$ $(l=0-2)$ 
transitions calculated at the laboratory kinetic energy
$E_{\rm lab}=1$~eV.
In these calculations the basis
includes all the exotic atom states with principal quantum numbers
from $n=1$ up to  20 (note, that all channels 
associated with $n\geqslant 5$
are closed). The behavior of partial-wave cross sections in the range 
of the total angular momentum $J < J^{*}$
($J^{*}\approx 13$ at $E_{\rm lab}=1$~eV) is quite similar. 
It is noteworthy, that the value of $J^*$ grows with the increase
in the principal quantum number (in the entrance channel)
and a collision energy. 
 
The partial-wave cross section of the elastic
scattering is, on average, significantly larger (about one 
order of magnitude) than 
the partial-wave cross sections of Stark transitions 
($\Delta l = \left|l_i - l_f\right| \geqslant 1$). 

At higher values of the total angular momentum $J~>~J^*$ 
all the partial-wave cross sections rapidly decrease. The rate of
this decrease essentially depends on the
$\Delta l $ value.
The partial-wave cross sections of Stark transitions 
with the $\Delta l =1$ decrease much slower 
than the partial-wave cross sections of the elastic
scattering ($\Delta l =0$) and Stark transitions 
with $\Delta l \geqslant 2$. 
Such a behavior of partial-wave cross sections is 
explained in the case of transitions with $\Delta l =1$
by the direct dipole ($t=1$) coupling
(see (\ref{aspotW1})) between entrance and final channels,
whereas in the $\Delta l \neq 1$ case 
the corresponding cross sections are mainly determined
by the effective dipole coupling through intermediate states.
 
This conclusion is confirmed in Fig.~\ref{part_el_t},
where the same cross sections were
calculated for two variants of interaction potentials.
In the first variant the matrix of interaction potentials includes
only monopole and dipole terms ($t_{\rm max}=1$), whereas
in the second variant all the allowed multipoles
up to $t_{\rm max}=14$ were taken into account.    
For transitions with $\Delta l =1$  
the direct dipole interaction ($t=1$) 
determines the coupling between initial
and final channels and very accurately reproduces  
the $J$-dependence of Stark cross sections with $\Delta l =1$.
Here the higher multipoles with $t\geqslant 2$ practically
do not change this cross section (see Fig.~\ref{part_el_t}). 
In the case of transitions with $\Delta l \geqslant 2$ the
direct interaction with $t\geqslant 2$ gives a noticeable
contribution only at small $J$, whereas the
effective dipole coupling largely determines both the value of the
corresponding cross section and its $J$-dependence at large $J$
values, as shown in Fig.~\ref{part_el_t}. 

The fast decrease of the partial-wave cross sections 
with the increase in the total angular 
momentum at $J~>~J^*$ 
is a general feature, which is also revealed
in partial-wave cross sections 
of both elastic scattering (e.g., $4s~-~4s, 4p~-~4p$, 
and $4d~-~4d$) 
and Stark transitions with a fixed $\Delta l$ for
the given $n$.  

\begin{figure}[!ht]
\centerline{\includegraphics[width=0.47\textwidth,keepaspectratio]
{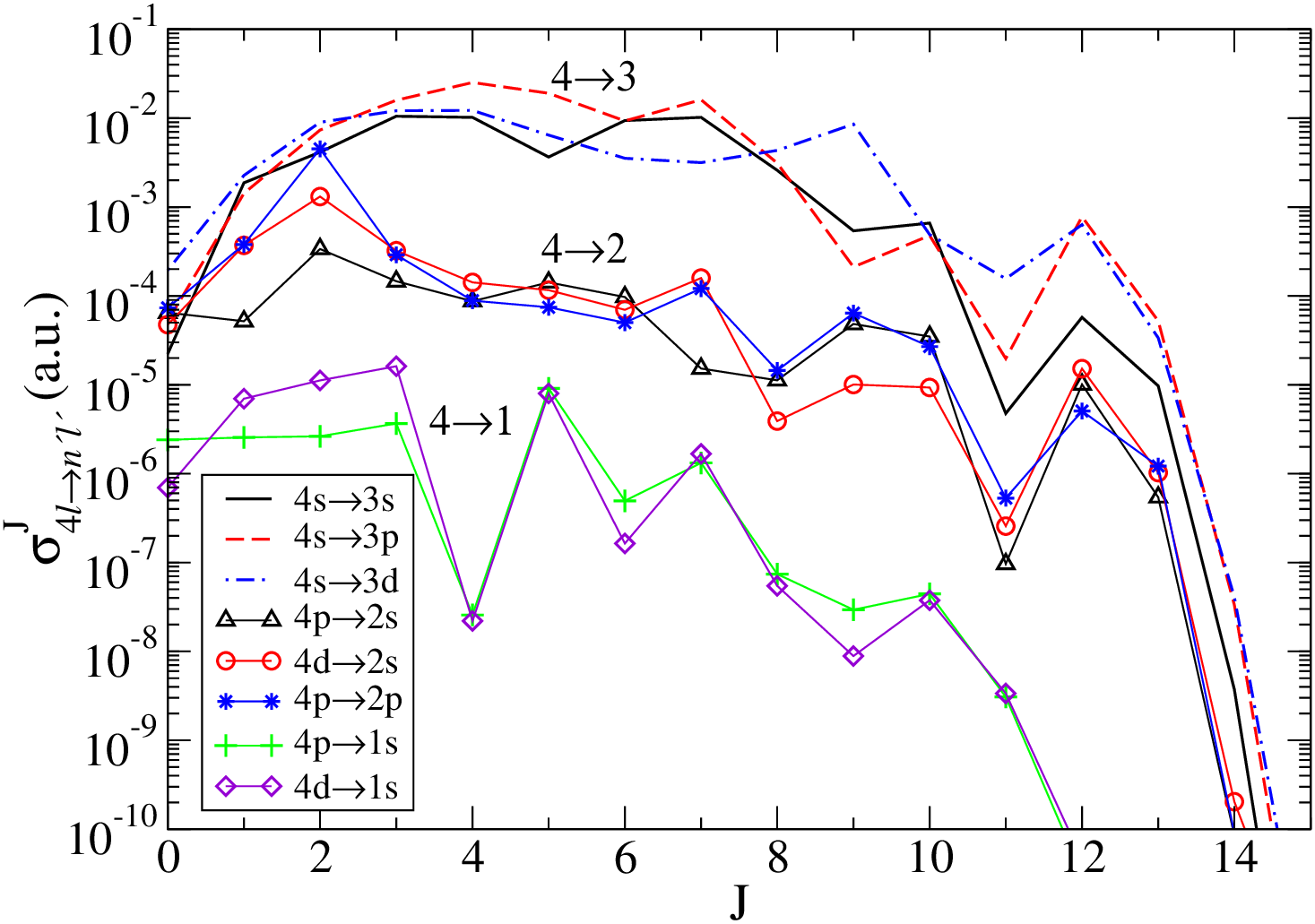}}
 \caption{(Color online) The partial-wave 
 CD cross sections for 
$(\mu^- p)_{4l} + H$ collisions vs. the total 
 angular momentum $J$ calculated at 
 the laboratory kinetic energy $E_{\rm lab}=$~1~eV
 with the basis including all the exotic atom states 
 with $n\leqslant 20$.}
 \label{part_cd_e}
 \end{figure}

Figure~\ref{part_cd_e} shows the $J$-dependence 
of partial-wave cross sections for different Coulomb
transitions ($\Delta n =n_i - n_f =1-3$) in 
$(\mu^- p)_{4l} + H$ collisions. Calculations 
were performed with the basis including all the exotic
atom states  with $n \leqslant 20$
at the laboratory kinetic energy $E_{\rm lab}=1$~eV.
The present study shows that, in contrast to the convergence 
with respect to $J_{\rm max}$ observed above in 
Fig.~\ref{part_el_t},  
the partial-wave cross sections of the CD 
(see Fig.~\ref{part_cd_e}) very sharply decrease at 
$J_{\rm max}\gtrsim 12$ for  $4\to3 $, $4\to 2$, and 
$4\to 1$ transitions independently from $\Delta l$. 
Such a behavior of CD cross sections 
is explained by the affect of the 
centrifugal barrier, which prevents partial waves with 
$J_{\rm max} \gtrsim 12$ (at a given energy $E_{lab}=1$~eV)
from penetrating into the interaction range, where the process
of the CD mainly occurs. 

The fast decrease of partial-wave cross sections 
(see Figs. (\ref{part_el_t} -\ref{part_cd_e})) 
with the increase in the total angular momentum in the range $J~>~J^*$ 
is a general property, manifested also in partial-wave cross sections 
of elastic scattering, Stark transitions, and CD 
at all the collision energies and for all the excited states of an exotic atom.

The comparison of the results shown in
Fig.~\ref{part_el_t} and Fig.~\ref{part_cd_e}
allows to explain 
the different $J$-dependence of partial-wave cross sections in
Fig.~\ref{part_el_t} at $J > J^*$. In this range of 
$J$-values the coupling between open
channels corresponding to the CD and   
closed channels can be 
neglected and calculations with the minimum basis set
($n_{{\rm max}}=4$ in the considered example) 
allow to obtain results which are in correspondence 
with the extended basis case.
The cross sections of Stark transitions with $\Delta l =1$
are largely determined by the direct dipole coupling 
and survive for much larger $J$ values in comparison with
processes corresponding to $\Delta l \neq 1$. These observations allow
significantly to reduce the dimension of the system
of close-coupling equations and as a consequence the
computer time. 
        
\subsection{Isotopic effect in the scattering
of excited $\mu^- p$ and $\mu^- d$ atoms}

As far as we know, the calculations of 
scattering cross sections for $(\mu^- d)_{n} + D$ collisions 
in the framework of a fully quantum-mechanical 
approach have not been done so far and the isotopic 
effect in the scattering of the muonic atoms of hydrogen 
and deuterium has not been investigated either.
 In the present study the cross sections of elastic
 scattering, Stark transitions and CD
 for both muonic hydrogen and deuterium were calculated
 within the same fully quantum-mechanical approach. 
 Calculations were performed using 
 basis states corresponding to $n_{{\rm max}}=20$
and $l \leqslant l_{{\rm max}}=n_{{\rm max}}-1$.    
 This allows to study the isotopic effect
 in the scattering of the excited muonic hydrogen 
 and deuterium on the same footing.
 
\begin{figure}[!ht]
\centerline{\includegraphics[width=0.47\textwidth,keepaspectratio]
{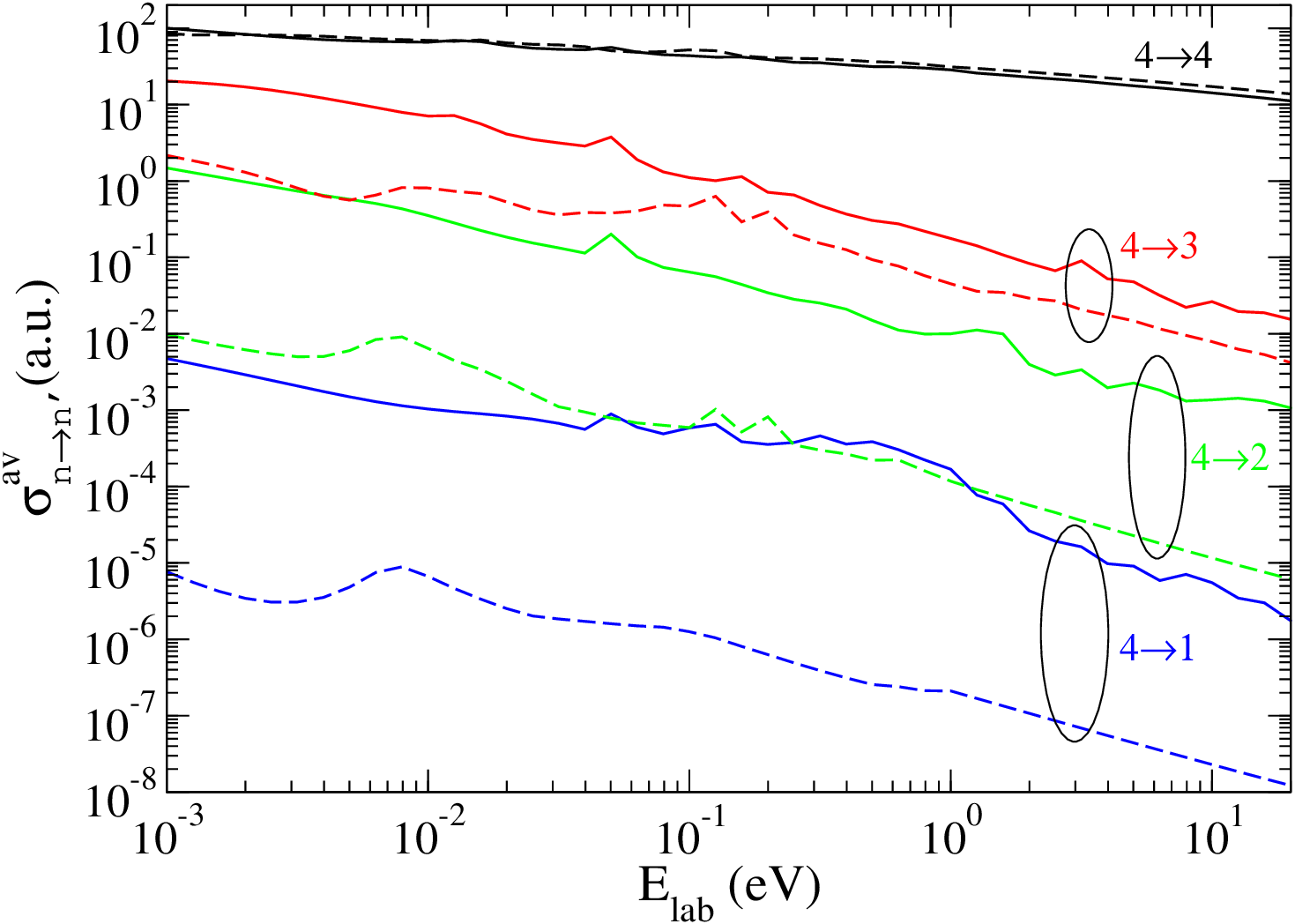}}
\caption{(Color online) The $l$-averaged 
cross sections for $(\mu^- p)_{n} + H \to(\mu^- p)_{n'} + H$ (solid lines) and 
$(\mu^- d)_{n} + D\to(\mu^- d)_{n'} + D$ (dashed lines)
collisions ($n=4$) vs. the laboratory kinetic energy $E_{\rm lab}$. }
\label{crsec_n4_av}
        \end{figure}

\begin{figure}[!ht]
\centerline{\includegraphics[width=0.47\textwidth,keepaspectratio]
{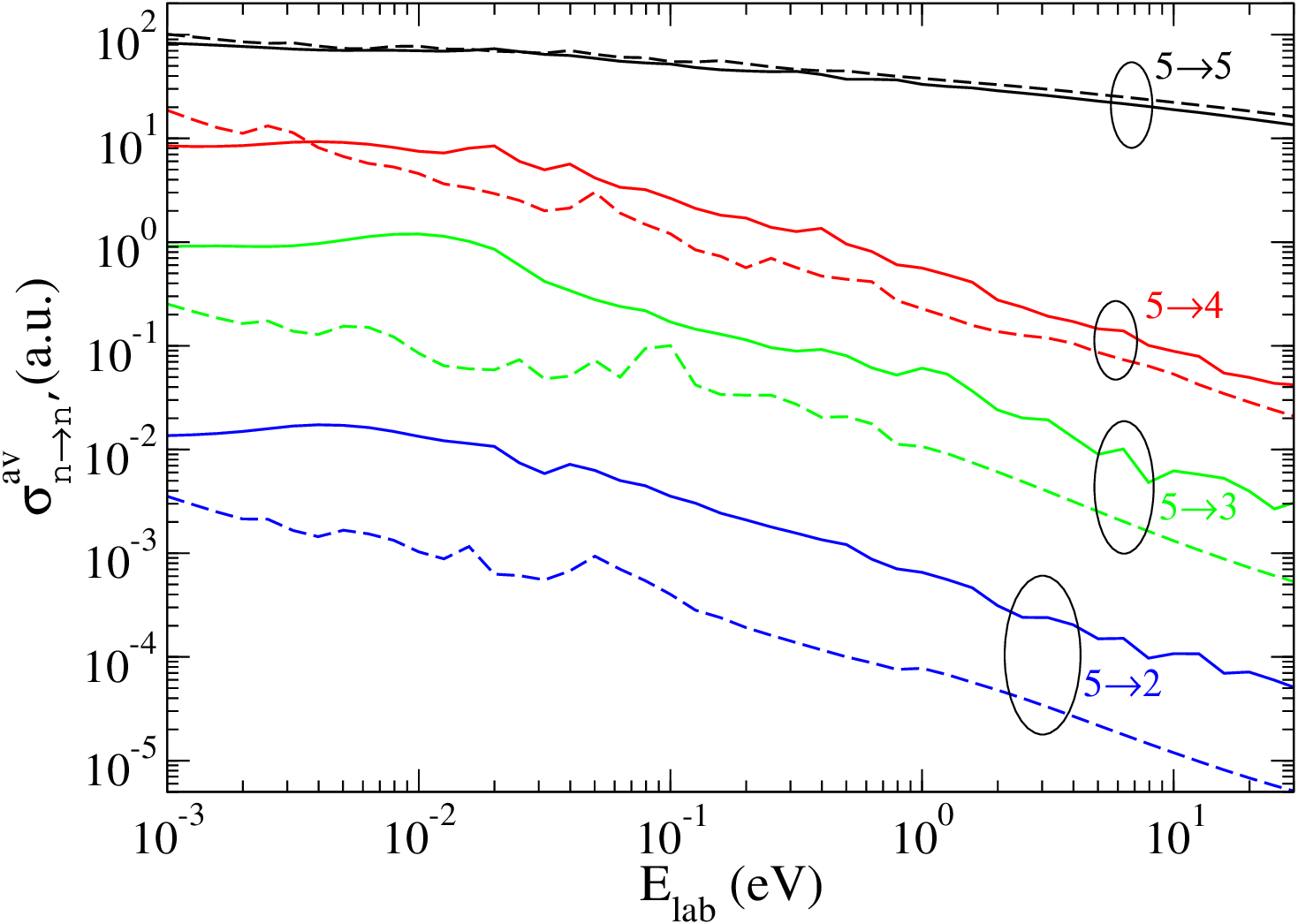}}
 \caption{(Color online) The same as in Fig.~\ref{crsec_n4_av}
 for $n=5$.}
 \label{crsec_n5_av}
        \end{figure}

\begin{figure}[!ht]
\centerline{\includegraphics[width=0.47\textwidth,keepaspectratio]
{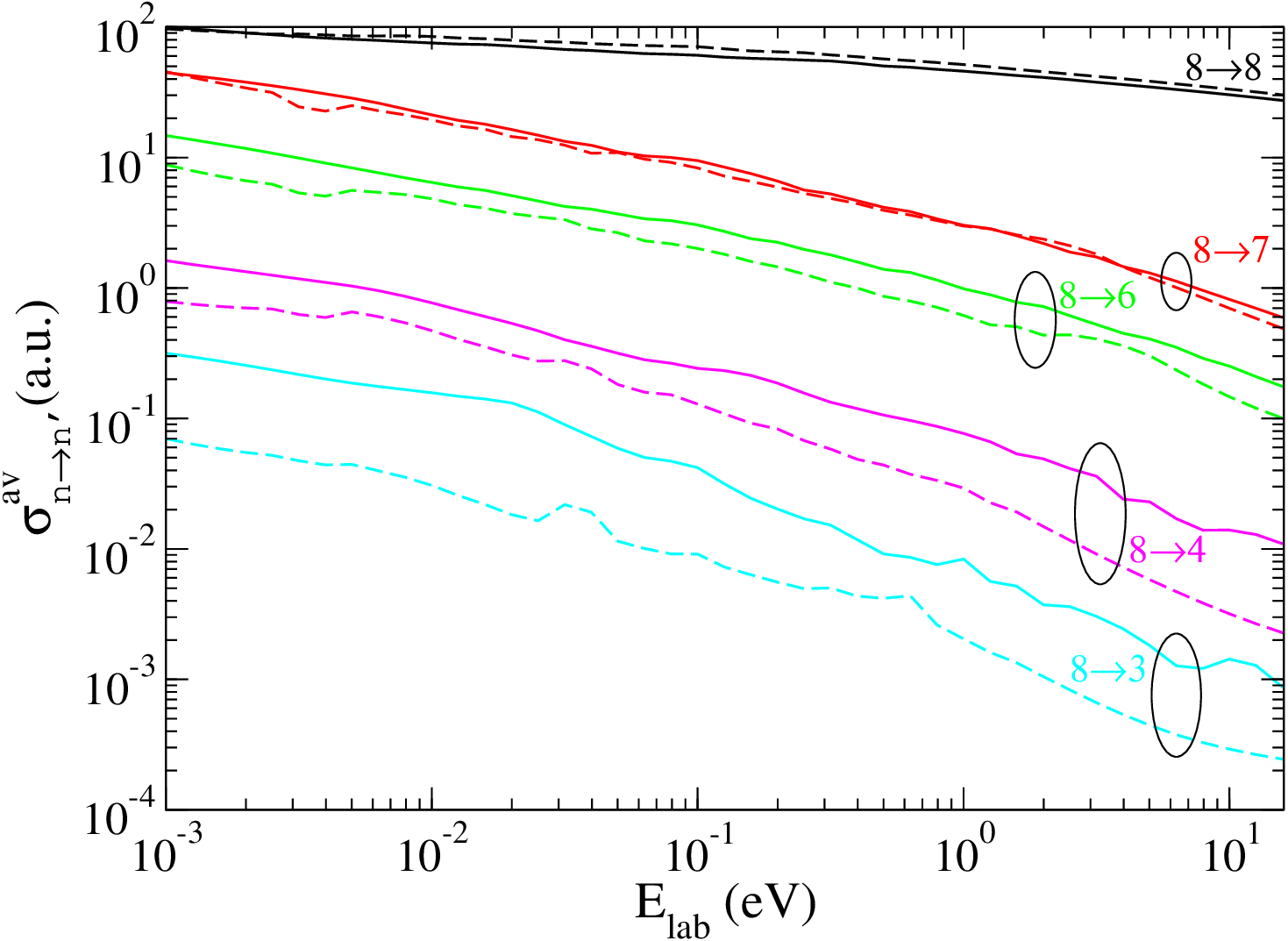}}
 \caption{(Color online) The same as in Fig.~\ref{crsec_n4_av}
 for $n=8$.}
 \label{crsec_n8_av}
        \end{figure}

 Figures~\ref{crsec_n4_av} - \ref{crsec_n8_av} show the energy
 dependence of $l$-averaged cross sections for  
 $(\mu^- p)_{n} + H \to (\mu^- p)_{n'} + H$ and 
$(\mu^- d)_{n} + D \to (\mu^- d)_{n'} + D$
collisions ($n'\leqslant n$) with values of the
 principal quantum number $n=4, 5$, and $8$. 
The comparison of the results obtained for the muonic 
hydrogen and deuterium
reveals, that the $l$-averaged cross sections 
without changing the principal quantum number
differ not more than (10 -15)\%. 
At all collision energies these cross sections in the case 
of the $(\mu^- d)_n +D$ scattering are slightly more than 
in the $(\mu^- p)_n + H$ one. 
                      
  Unlike the $l$-averaged cross sections of
Stark transitions, a much stronger isotopic effect is observed 
in $l$-averaged cross sections of the CD      
($n\to n'\leqslant n-1$). 
For a given $n$ value and fixed $\Delta n = n-n'$ 
the cross section of the CD  
in the case of $(\mu^- d)_{n} + D$ collision 
is always less than in the case of 
$(\mu^- p)_{n} + H$ collision and this difference grows 
essentially with increasing $\Delta n$, as shown 
in Figs.~\ref{crsec_n4_av} - \ref{crsec_n8_av}. 
On the other hand, the isotopic effect decreases with
the increase in the principal quantum number of
the initial state. In particular, the $l$-averaged cross sections
of the CD $8\to7$ of the muonic hydrogen and
deuterium are almost indistinguishable on eye, while in the case of
the $l$-averaged cross sections
of CD $5\to4$ and especially $4\to3$ 
the isotopic effect is very strong.       

The suppression of the cross sections of CD 
in the case of the muonic deuterium
as compared with the muonic hydrogen is 
mainly explained by the weakening of the effective
coupling, 
$W^{J {\it p}}_{n'l'L', nlL}(R)G^{EJ {\it p}}_{n'l'L'}(R)$ 
(see \ref{CCE}), between open channels as well as 
between open and closed channels.   
At a fixed energy the ratio of the absolute values of 
wave numbers of the scattering problem is determined 
mainly by the reduced masses of the system. In 
the scattering of $\mu d$ and $\mu p$ atoms it is equal to
$|k_{nl}(\mu d)/k_{nl}(\mu p)| \approx 1.4$. 
Therefore, in the case of the muonic deuterium the wave functions 
of closed channels are exponentially damped at smaller distances, 
and the wave functions of open channels respectively oscillate 
much faster. Both these factors lead to a significant 
weakening of the effective coupling between different channels and, 
as a consequence, a significant suppression of 
CD cross sections. 

The interplay of various processes in the kinetics of the atomic
cascade is determined by their rates. The rates of
collisional processes in the laboratory frame are defined as follows
\begin{equation}
\label{rate}
\lambda_{nl \to n'l'}(E_{\rm{lab}}) = 
\varphi N_{\rm{LHD}} \sigma_{nl \to n'l'}(E_{\rm{lab}})
\sqrt{\frac{2E_{\rm{lab}}}{M_{\mu a}}},
\end{equation}
where $\varphi = N/N_{\rm{LHD}}$ is the relative density of the target
in the units of the liquid hydrogen density, 
$N_{\rm{LHD}} = 4.25~\times~10^{22}~{\rm atoms/cm}^3$
and $E_{\rm lab}$ is a laboratory kinetic energy of a muonic atom.
 While $\mu$-decay and radiative transition rates do not depend on
the kinetic energy of the muonic atom, target density, and 
temperature, the rates of collisional processes essentially
depend on these physical parameters.          

\begin{figure}[!ht]
\centerline{\includegraphics[width=0.47\textwidth,keepaspectratio]
{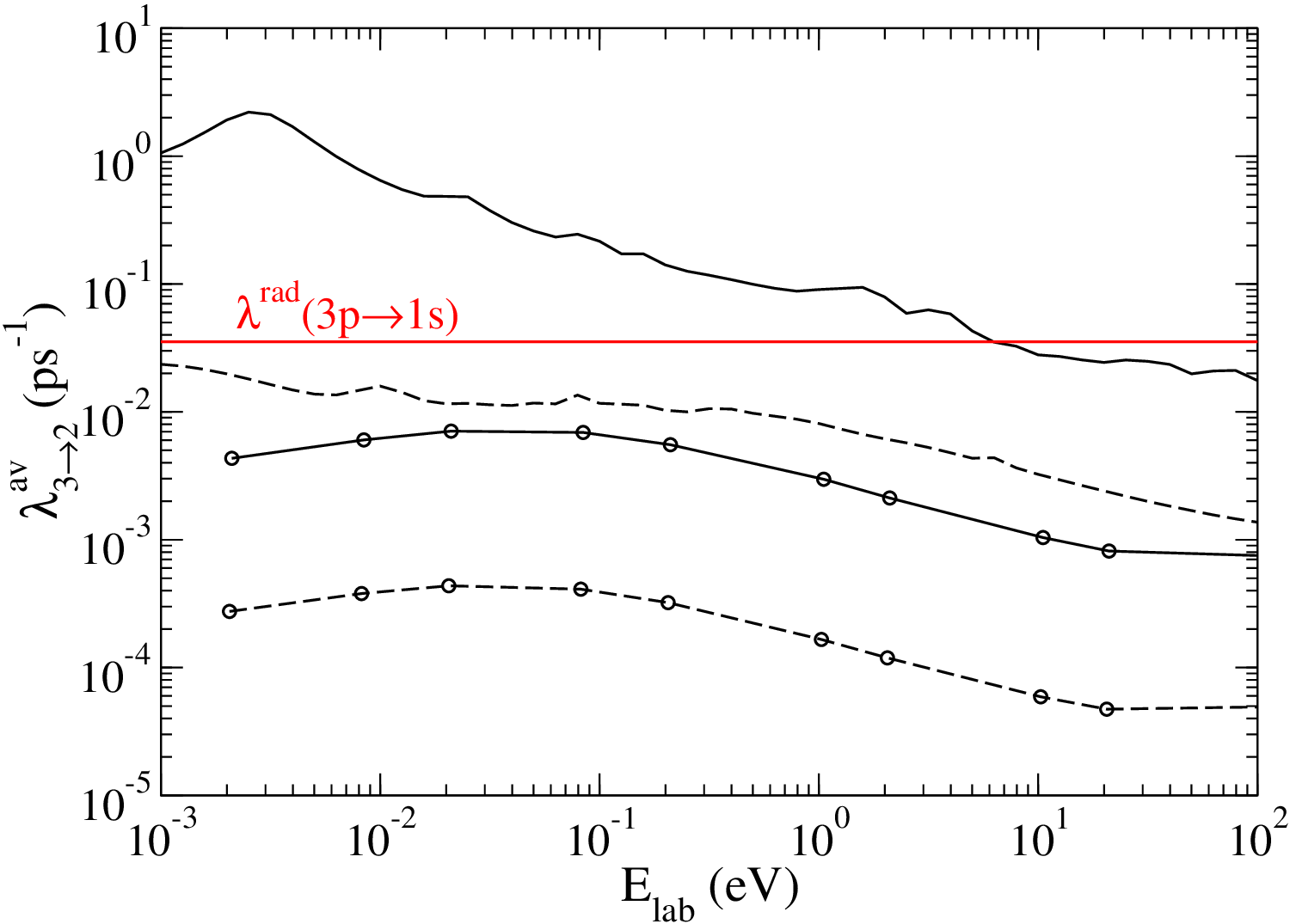}}
 \caption{(Color online) The $l$-averaged 
 Coulomb $3\to2$ de-excitation rates calculated at the 
 liquid hydrogen density ($\varphi=1$) for muonic hydrogen (solid line) 
 and deuterium (dashed line) atoms
 vs. the laboratory kinetic energy $E_{\rm lab}$. 
 The curves with open circles are from \cite{KM}.
 The horizontal line shows the rate of the radiative
 $3p\to1s$ transition in both muonic hydrogen and deuterium
 (see the text).}
    \label{rates_n3_av_p}
    \end{figure}

\begin{figure}[!ht]
\centerline{\includegraphics[width=0.47\textwidth,keepaspectratio]
{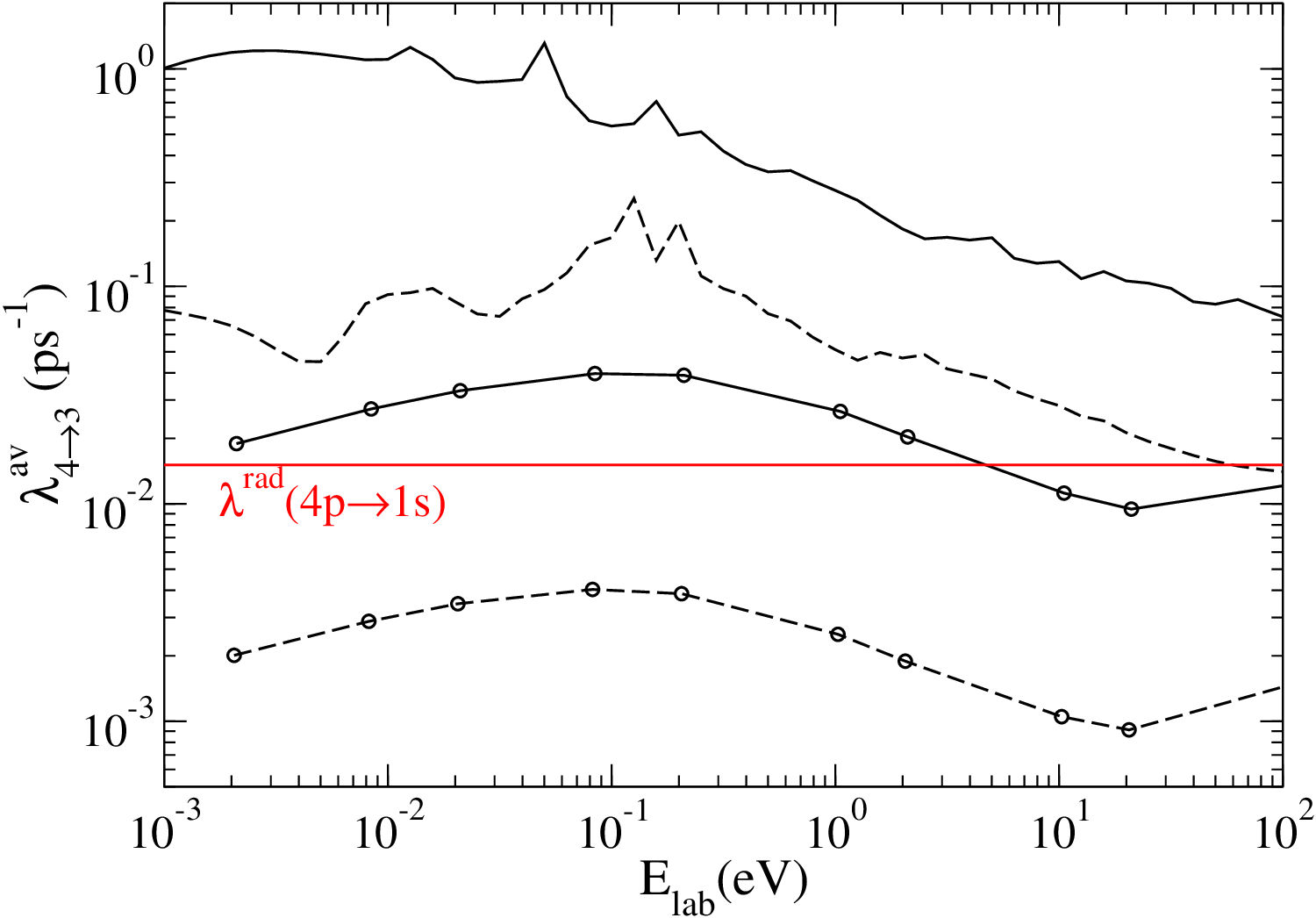}}
 \caption{(Color online) The same as in Fig.~\ref{rates_n3_av_p}
 for Coulomb $4\to3$ de-excitation rates vs. the laboratory kinetic 
 energy $E_{\rm lab}$. The curves with open circles are
 from \cite{KM}.
 The horizontal line shows the rate of the radiative
 $4p\to1s$ transition in both muonic hydrogen and deuterium.}
 \label{rates_n4_av_p}
    \end{figure}

The energy dependence of $l$-averaged CD 
rates 
\begin{equation}
\label{avrate}
\lambda_{n \to n'}^{\rm{av}}(E_{\rm{lab}}) = 
\varphi N_{\rm{LHD}} \sigma_{n \to n'}^{\rm{av}}(E_{\rm{lab}})
\sqrt{\frac{2E_{\rm{lab}}}{M_{\mu a}}},
\end{equation}
for $(3~\to~2)$ and
$(4~\to~3)$ transitions in muonic hydrogen and deuterium calculated 
at a liquid hydrogen density ($\varphi=1$) are shown 
in Figs.~\ref{rates_n3_av_p} and \ref{rates_n4_av_p} 
together with the rates $\lambda ^{rad}(np\to 1s)$
of radiative transitions $(3p~\to~1s)$ and $(4p~\to~1s)$, respectively.  
A small difference between the
radiative rates of $(np\to 1s)$ transitions in muonic
hydrogen and deuterium due to their reduced masses is not visible
at the scale of Figs.~\ref{rates_n3_av_p} and \ref{rates_n4_av_p}.  
     
 Here the
 isotopic effect revealed in cross sections 
 (see Figs.~\ref{crsec_n4_av} - \ref{crsec_n8_av})
 is additionally enhanced  by the trivial kinematic factor associated 
 with the difference of exotic atom velocities 
 at a fixed energy in the laboratory system
 (due to the different masses 
 of muonic hydrogen and deuterium atoms). 
The isotopic effect is very strong and 
significantly changes the rates of 
CD especially in the low-energy region. 
In particular, this effect results 
in decreasing the rate of the 
CD $(~5\to~4)$ more than four times
(see Fig.~\ref{crsec_n5_av}), about an
order of magnitude the 
rate of the CD $(4~\to~3) $,
and even much more the rate of the CD $(3~\to~2)$
in the case of the $(\mu^- d)$ atom, as compared with
the case of the $(\mu^- p)$ atom. 
Comparing  the rates of CD $(3~\to~2)$ and 
$(4~\to~3)$ respectively with
radiative transition rates
$(3p~\to~1s)$ and $(4p~\to~1s)$, 
we can conclude 
that the acceleration of the muonic hydrogen due to CD 
should be more significant than in the
muonic deuterium case. This isotopic effect has to lead to 
the much lesser Doppler broadening of the $K$ lines 
in the case of the muonic deuterium in comparison 
with the case of the muonic hydrogen. 
This effect must be especially noticeable at the relative 
density above $\varphi \approx 10^{-3}$. 

The comparison of the present results with those
obtained in the framework of the asymptotic theory 
of non-adiabatic transitions (see \cite{KM} and references
therein) is given 
in Figs.~\ref{rates_n3_av_p} and \ref{rates_n4_av_p}.
It is seen, that the rates of the CD from paper \cite{KM} 
only qualitatively reproduce the isotopic effect, but they are too small
to explain the experimental data (e.g., see \cite{Bad, DSCov}).
It should be noted that all the previous calculations
of the CD were realized within semiclassical
or adiabatic approaches with a two-level approximation
which is not suitable for the treatment of 
the deeply inelastic process such as CD (e.g., see \cite{KPP,AIg}). 
 
The observed isotopic effect in both cross sections and 
the rates of the CD allows to predict a similar effect in 
the pionic deuterium and the muonic tritium.  
This effect should lead to a strong weakening of the 
high-energy components in the energy spectrum 
of the pionic deuterium and especially in the case 
of the muonic tritium compared to the pionic and muonic 
hydrogen atoms, respectively. In particular, the predicted 
isotopic effect allows for the first time to explain 
the significant suppression of the contributions of high-energy 
components originating from CD 
transitions ($4\to3$) and ($5\to3$) in shift/width 
experiments~\cite{Str2} with the pionic deuterium.          
              
\section {Kinetics of the atomic cascade}

The life history of the exotic atom after its formation is 
determined by the complex interplay of collisional 
and non-collisional processes.
In these processes the initial 
distributions on quantum numbers 
$(n, l)$ and the laboratory kinetic energy $E$ 
(at the instant of the exotic atom formation) 
change during the de-excitation cascade
until the exotic atom arrives in the ground state or  
the weak decay (absorption or annihilation in the case
of hadronic atoms) of the exotic particle occurs. 

The theoretical predictions of various characteristics of the
atomic cascade represent 
an idealized view of what one might expect in experimental
studies, because they are free from standard 
experimental problems associated with the energy resolution, 
efficiency of the registration, statistic, and so on. 
Besides, the model
representations used in the analysis of experimental
data may also lead to additional uncertainties. 
Hence, the reliable theoretical predictions 
may be very useful both for choosing the optimal conditions
of experiments and improving the results extracted 
from the analysis of experimental data.

 The evolution of the kinetic energy  and 
$(n,l)$ distributions of exotic atoms plays a significant role
in the competition of various cascade processes and
should be followed up during the whole atomic cascade. 
To realize this, the comprehensive sets of differential
and integrated cross sections of collisional processes, 
involved in the atomic cascade, were calculated 
(in a framework of the present 
quantum-mechanical approach) in the 
wide range of quantum numbers ($n\leq 8$) and kinetic energies
of muonic atoms, that are of interest to reliable cascade
calculations. Besides, the realistic model simulating the kinetics of the
atomic cascade was developed.
It is especially important from the viewpoint of various
precise experiments with 
muonic~\cite{RPD,FKott,RPohl,PDH,Ludh,DSCov, Natur,Mud,DGot4} 
and hadronic ~\cite{HSchr,DGott, DG3,DG1,DG2, 
Str1, Str2, DG4,DGot3,Baz, Carg} hydrogen atoms.

For the collisions of highly excited exotic atoms ($n~>~8$)
with molecular hydrogen the results of the classical-trajectory
Monte Carlo calculations \cite{JM2} were used. 
It is important to note that the isotopic effect revealed 
in the cross sections for scattering of muonic atoms in states with 
$n < 9$, is virtually nonexistent in their scattering in highly excited 
states with $n>9$. The only source of the isotopic effect in 
the kinetics of atomic cascade here is a trivial kinematic 
factor of the transition from the cross sections of processes 
to their rates.

\subsubsection{Cascade processes}

In the present paper we use as
the improved version of the ESCM originally 
developed in \cite{VEMark}
and later essentially improved in the paper \cite{CascalII}
and in recent papers \cite{JPP,PPForm,PPKin}. The improvements
were mainly achieved due to new theoretical results for
the cross sections of the collisional processes and 
initial distributions of muonic atoms in the quantum numbers and 
the laboratory kinetic energy at the instant of their formation.
The ESCM includes: radiative transitions, CD, 
external Auger effect, and  
Stark transitions, as well as the elastic scattering
and $\mu^-$ - decay. 

The most known cascade process, beyond the
$\mu^-$-decay, is the radiative 
$n_i l_i \to n_f l_f$  de-excitation described by 
electro-dipole transitions, 
in which the initial and final values of the internal 
orbital angular momentum of exotic atoms satisfy 
the selection rule $\Delta l = \left|l_{i}-l_{f}\right|~=~1$. 
The rates of radiative transitions are calculated
with very high accuracy and together with the $\mu^-$ - decay 
almost entirely determine the de-excitation cascade 
at very low target densities $\varphi \lesssim 10^{-7}$.
In this density range the information about
initial  $n$-, $l$-, and $E$- distributions 
is practically conserved up to the end 
of the cascade (e.g., see \cite{PPForm}). 

At higher densities the collisional processes 
become very important, because their rates are proportional
to the target density and, as a rule, have a strong
energy dependence. 
 The external Auger effect, elastic scattering,
Stark transitions, and CD are the main 
collisional processes determining the data observed 
or extracted from various experiments. 

The external Auger effect does not change the kinetic 
energy of the exotic atom, because the transition energy 
is practically carried away by the target electron.
The Auger transitions with the minimal $\Delta n = n_{i}-n_{f}$
needed for the ionization of the target are the most
probable. The rates of the external Auger effect reach their 
maximum at $\Delta n = 1$ (the transition $7\to 6$ in 
the muonic atom case) and they
rapidly decrease both for $n < 7$ and $\Delta n > 1$.
The cross sections of the external Auger effect were calculated 
in the semiclassical approximation and used through the whole cascade

The elastic scattering, Stark transitions and 
CD significantly change 
both the kinetic energy and quantum numbers 
of exotic atoms.  
The Stark transitions affect the population of $nl$ 
sub-levels and together with the elastic scattering,  
as a rule, decelerate exotic atoms, 
thus changing the interplay of different de-excitation
processes during the atomic cascade.
The process of the CD  plays a 
very important role in the acceleration 
of exotic atoms.
In this process the energy of the transition $n\to n'< n$ is 
converted into the kinetic energy of the relative motion
and shared between colliding 
objects (exotic and target atoms). 

The competition of the external Auger effect and the CD 
significantly depends on both the initial state of the muonic atom 
and its kinetic energy. In addition, their interplay differs for muonic 
hydrogen and deuterium atoms, due to the isotopic effect mainly 
in the CD of these muonic atoms.    
 \begin{figure}[!ht]
\centerline{\includegraphics[width=0.47\textwidth,keepaspectratio]
{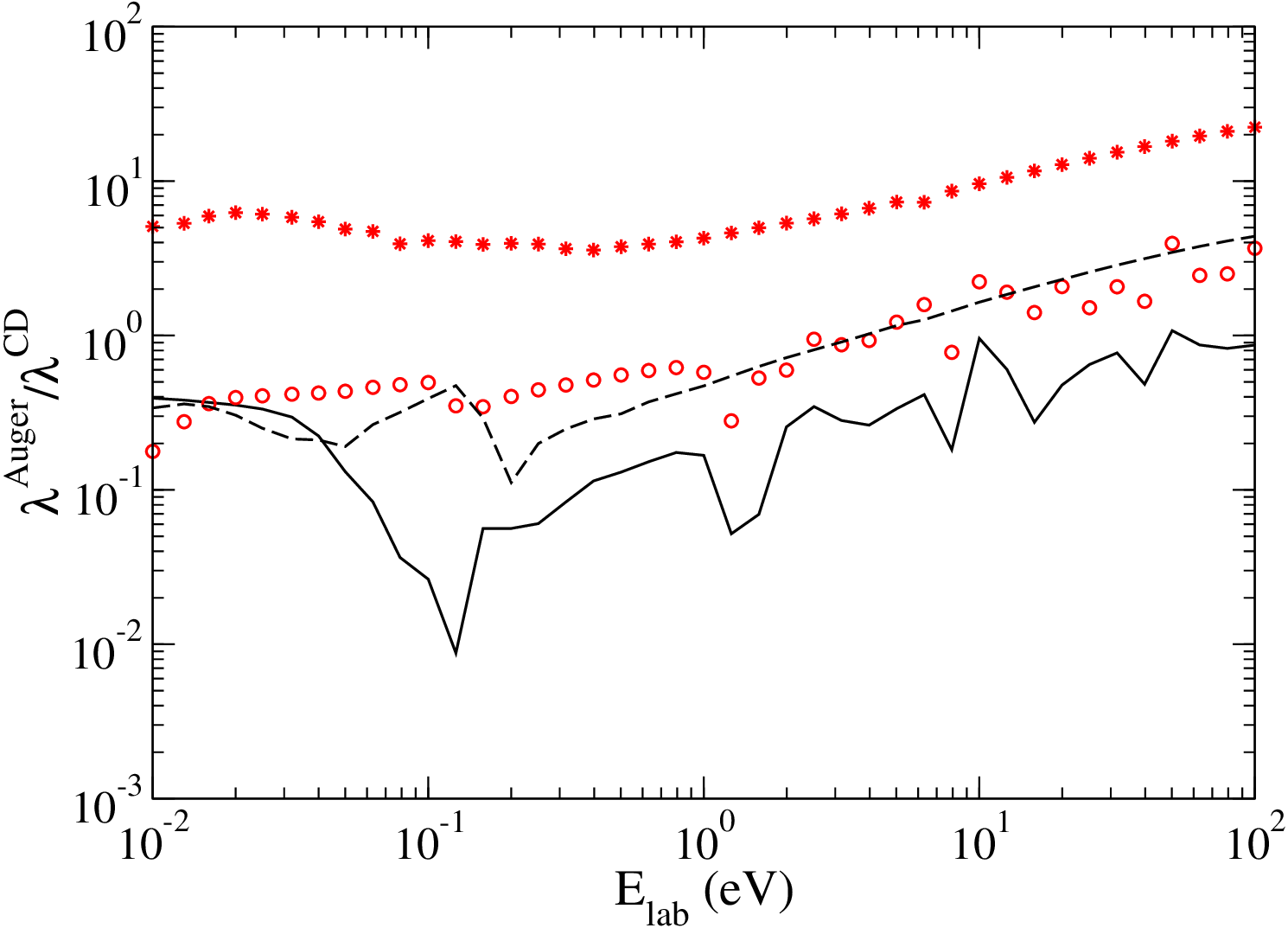}}
 \caption{(Color online) 
 The ratio of the external Auger effect and CD rates, 
 $\lambda^{Auger}/\lambda^{CD}$, vs. the laboratory
 kinetic energy calculated for $\mu^- p$ and $\mu^- d$ 
 atoms in the $3s$ state (solid and dashed lines, respectively)
 and in the $3p$ state (open circles and asterisk, respectively).}
 \label{ratio_Aug-CD_n3}
    \end{figure} 
The energy dependence of the ratio of the external Auger effect
and CD rates calculated for $\mu^- p$ and $\mu^- d$ atoms in 
$3s, 3p$ states and in $5s, 5p$ states is shown in 
Figs.~\ref{ratio_Aug-CD_n3} and \ref{ratio_Aug-CD_n5}, respectively.
As can be seen, this ratio has a nontrivial energy dependence which
essentially depends on the exotic atom state. The competition 
between these processes significantly affects the energy 
distribution of muonic atoms, in particular, appears in 
the spectra of muonic atoms at the time of radiative transitions
(e.g., see Figs.~\ref{mup_2-1} and \ref{mud_2-1} below).
 \begin{figure}[!ht]
\centerline{\includegraphics[width=0.47\textwidth,keepaspectratio]
{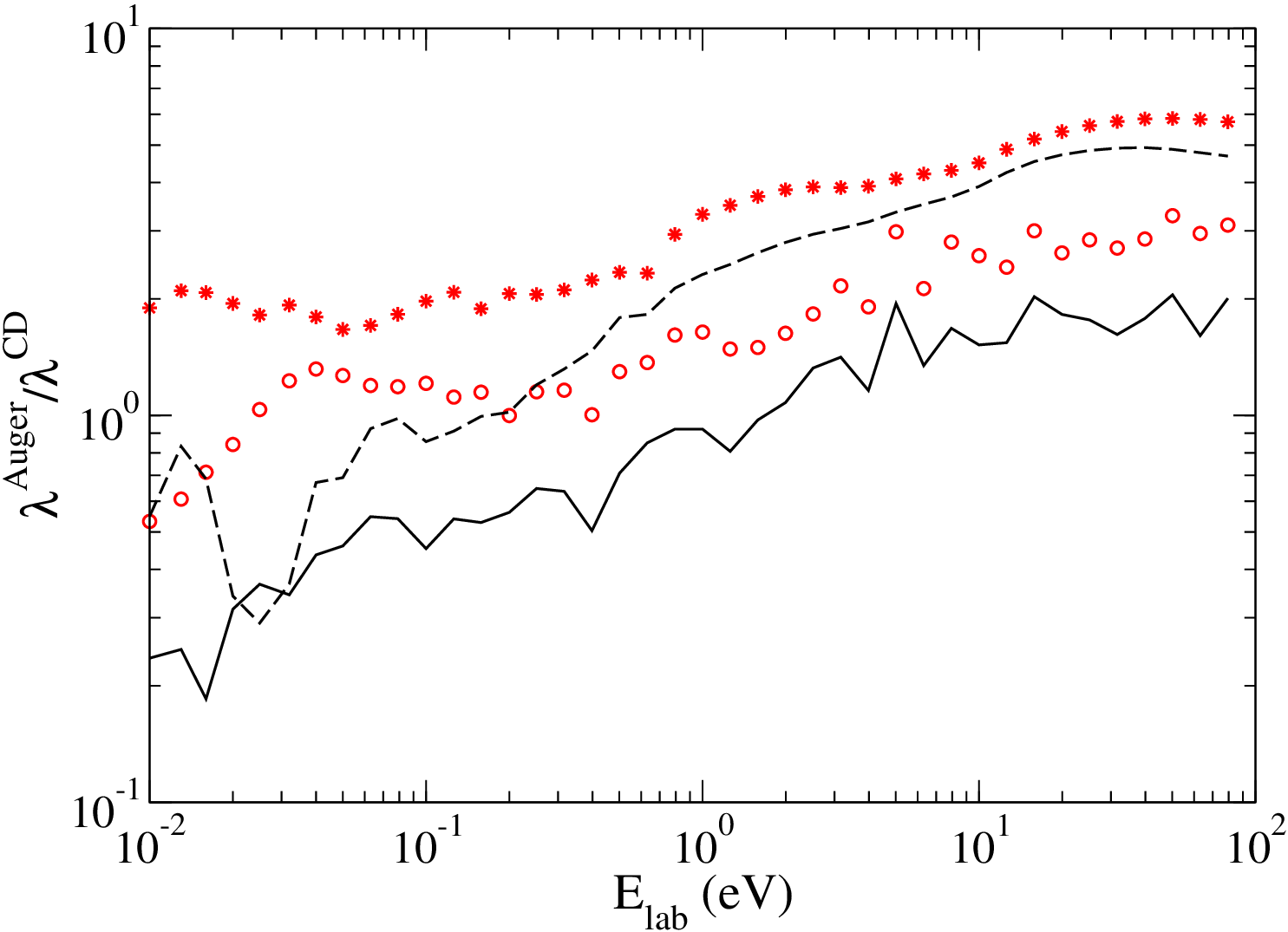}}
 \caption{(Color online) The same as in Fig.~\ref{ratio_Aug-CD_n3}
 for $\mu^- p$ and $\mu^- d$ atoms in the 
 $5s$ state (solid and dashed lines, respectively)
 and in the $5p$ state.}
 \label{ratio_Aug-CD_n5}
    \end{figure}
    
According to our knowledge, 
a fully quantum mechanical calculations of the 
cross sections of the charge-exchange process and the external Auger 
reaction with the formation and subsequent decomposition of 
the intermediate molecular complex is currently lacking. 
Estimates of the cross sections (rates) of these 
processes (e.g., see \cite{MF,KM, KM2} and references therein) 
allow us to conclude that their role in the kinetics of atomic 
cascade is insignificant in comparison with the direct 
Coulomb de-excitation and Auger ionization of the target atom.   
Thus, the above mentioned processes are not taken into account
in the present version of the ESCM.    
    
\subsubsection{Initial distributions and the thermal motion
of the target}

The simplest model of 
the exotic atom formation, usually used in cascade calculations, 
is based on the assumption, that the exotic atoms 
are formed in states with a
fixed principal quantum number $n=n_0 \approx \sqrt{m_{r}}$ 
($\approx 14$ for muonic hydrogen) and the statistical 
$l$ distribution  $(2l+1)/n_{0}^2$ $(l \leqslant n_0 -1)$ of 
$n_{0} l$ sub-levels (e.g., see \cite{CascalII} and references therein). 
This simple representation of the exotic atom
formation is not confirmed by more 
elaborated studies \cite{KP,KPF,JC}, in which
the molecular structure of the target was taken into account. 
According to these 
studies, the muonic atoms are formed in states with a broad
distribution in the principal quantum number,
the maximum of which is 
shifted to lower $n\cong 11 < n_0$. Besides, the populations 
of $nl$ sub-levels
have a non-statistical distribution over $l$ for each value of the
principal quantum number.
  
In the present cascade model the
initial $n$-, $l$-, and $E$- distributions are assumed to be 
factorized as in our previous 
papers \cite{PPForm,GPP,PPKin}. 
The probability densities of 
initial $n$- and $l$- distributions 
(denoted by $f_n$ and $f_l$, respectively) as well as 
of the initial kinetic energy distribution
$\omega_{\rm{in}}(E)$ were assumed to be 
the same for both muonic hydrogen
and deuterium atoms. 

Here we use:

-- the Gaussian $n$-distribution
 \begin{equation}
 f_n \propto {\rm e}^{-\alpha_n (n-n_0)^2},
 \label{Gn}
 \end{equation} 
centered at $n_0=11$ ($\alpha_n=0.5$); 

-- the modified statistical 
$l$-distribution 
  \begin{equation}
 f_l  \propto (2l+1){\rm e}^{-\alpha_l (2l+1)}
 \label{Fl}
 \end{equation}
 ($\alpha_l = 0.08$) for each value of the
principal quantum number $n$; 

-- the two-exponential $E$-distribution 
\begin{equation}
\omega_{\rm{in}}(E)=\frac{\varkappa}{E_1}\exp\left(-\frac{E}{E_1}\right)
+\frac{1-\varkappa}{E_2}
\exp \left(-\frac{E}{E_2}\right),
\label{InitE}
\end{equation} 
where parameters $E_1=0.469$ eV, $E_2=4.822$ eV, 
and $\varkappa =0.805$ have been determined in \cite{PPForm} 
by fitting experimental 
data \cite{RPD} obtained
at the target pressure $p_{{\rm H}_2}=0.0625$ hPa
and room temperature. 
In the present study
these initial distributions were used 
for both muonic 
hydrogen and deuterium atoms at all target densities
and temperatures. 

\subsubsection{The thermal motion of the target and 
kinematics of the binary collision}

The thermal motion of the target is very important and
should be taken into account in the realistic description
of the kinetics of the atomic cascade. 
There are, at least, three main reasons to do it in a proper way.
First, the cross sections (or rates) of collisional processes
have a strong energy dependence especially at low energies
(e.g., see Figs.~\ref{crsec_n4_av} - \ref{rates_n4_av_p}).
Second, during the atomic cascade the kinetic 
energy of the exotic atom  
may be comparable with the energy of the thermal
target motion. Finally, it is necessary for the exact
knowledge about possible collisional processes 
and their correct description nearby
$ns - nl\,(l\geqslant 1)$ thresholds. 

In the present research, 
the exact relations between kinematic characteristics
in the laboratory and center-of-mass systems 
{\it before and after the binary collision} were obtained 
and used in cascade calculations. 
In contrast to the case of the target at rest 
considered in the literature, 
the energy $E_{cm}$ of the relative motion 
{\it before the collision}
depends on both energies 
of colliding subsystems and azimuthal angle $\chi$
of the collision in the laboratory system in the following way
\begin{equation}
E_{cm}=\frac{1}{M}(M_{t}E+M_{\mu a}E_{t}
-2\sqrt{M_{t}M_{\mu a}EE_{t}}\cos{\chi}).
\label{en1}
\end{equation}
 Here, $M_{\mu a}$ and $M_{t}$ are masses of 
 exotic and target atoms, $M=M_{t}+M_{\mu a}$;
 $E$ and $E_{t}$ are kinetic energies of exotic and target
 atoms in the laboratory system before the collision. In the 
 present cascade calculations  we assume 
 the uniform distribution of the azimuthal angle $\chi$. 
 Besides,
 the thermal motion of the target is described by a Maxwell distribution,
\begin{equation}                            
F(E_t)=\frac{2}{\sqrt{\pi}}\sqrt{\frac{E_t}{
(kT)^3}}\exp\left (-\frac{E_t}{kT}\right), 
\label{Mw}                                            
\end{equation}
where $T$ is the target temperature and $k$ is the Boltzmann constant.
  
 To obtain the kinetic energy 
 of the muonic atom in the laboratory system
{\it after the collision}, we apply the three-dimension geometry and the
 conservation laws of the energy and momentum. As a result, 
 the following formula for the kinetic
 energy of the muonic atom in the laboratory system after the collision 
 was derived
 \begin{align}
&E'=\frac{1}{M}\big\{M_tE'_{\rm cm}+M_{\mu a}(E+E_t -E_{\rm cm}) +
\nonumber\\
 &\left[M_{\mu a}(E_{\rm cm}-E_t)+M_t(E-E_{\rm cm})
 \right]\sqrt{E'_{\rm cm}/E_{\rm cm}} \cos \theta\big\}.
 \label{elf}
 \end{align}  
 Here, $\theta$ is the scattering angle in the center-of-mass system, 
 $E'_{\rm cm}=E_{\rm cm}+\Delta_{i\,f}$ is the kinetic energy of the relative
 motion after the collision, with 
 $\Delta_{i\,f}=E_{n_i\, l_i}-E_{n_f\, l_f}$ being
 the difference of the bound energies of muonic atom states before
 and after the collision. It is noteworthy, that in the case of the 
 external Auger effect the released energy is mainly carried
 away by the electron and in cascade calculations
 we neglect the recoil energy obtained by the muonic atom,
 hence, $E=E'$. 

\subsubsection{Simulation of the atomic cascade}

In the represented model of the atomic cascade 
the life history of each muonic atom can be traced from 
the moment of its formation until the transition to the ground state
or $\mu^-$- decay. A multi-step sequence
randomly selected events, which simulate 
the kinetics of the atomic cascade, 
can be schematically described as follows:

-- At the very beginning of the cascade the initial values of quantum
numbers $(n_i, l_i)$ and the laboratory kinetic energy $E$ of 
the muonic atom
are randomly chosen according to their 
distributions Eqs.~(\ref{Gn})-(\ref{InitE});

-- Further, the kinetic energy $E_t$ of the target 
(at a fixed target temperature) in accordance with Eq.~(\ref{Mw})
and the azimuthal angle $\chi$
of the collision (in the laboratory system) are randomly chosen
and the relative motion energy $E_{\rm cm}$ (before the collision) 
is calculated according to Eq.~(\ref{en1});
 
-- The rates of collisional
processes for  a given $E_{cm}$ together
with the rates of radiative transitions and $\mu$-decay are used
to calculate their total and relative rates (probabilities) 
at chosen above values 
of $n_i, l_i, E_{\rm cm}$ for all possible final states;

-- In accordance with these relative rates, the event type 
and the quantum numbers $(n_f, l_f)$  of the final state 
are randomly chosen. 
In the case of collisional process, the scattering angle is 
randomly chosen in accordance with the angular distribution 
for a given process.

-- As a result of these random events, 
one gets the type of the process, quantum numbers $(n_f, l_f)$ 
of the final state, and the new value of the laboratory 
kinetic energy $E'$ of the
muonic atom after the collision calculated according to Eq.~(\ref{elf}). 
These new values $(n_f, \,l_f,\,E')$ are used
as initial ones and the simulation of the atomic cascade continues
until $\mu^-$-decay occurs or the ground state 
of the muonic atom is reached. 

The present cascade calculations for muonic 
hydrogen and deuterium atoms were performed in the 
relative density range,
covering eight orders of magnitude ($\varphi = 10^{-8} - 1$). 
To obtain good statistics, the destiny of the $10^7$
muonic atoms have been analyzed in cascade
calculations for each value of the target density
and temperature. The various characteristics of the
de-excitation cascade in muonic hydrogen and deuterium 
were calculated: the absolute and relative
$x$-ray yields of different 
$K_i$ ($i =\alpha, \beta, \gamma$, etc.)
lines, the integrated kinetic energy
distributions and mean energies of muonic atoms 
at the instant of their radiative 
$np\to 1s$ transitions, cascade times, and so on.
Hereafter, the generally accepted designations 
$K\alpha$,  $K\beta$, $K\gamma$, and $K\delta$
are used for $K_i$ lines corresponding to the
$2p\to 1s, 3p\to 1s$, $4p\to 1s$, and $5p\to 1s$ radiative transitions.
Their energies are 1.90~keV,
2.25~keV, 2.37~keV, and 2.43~keV in $\mu^- p$ atom and 
accordingly 2.00~keV, 2.37~keV, 2.50~keV, and 2.56~keV 
in $\mu^- d$ atom.
 
Some of our results are compared with
the available experimental data on   
the density dependence of the relative $x$-ray yields of 
$K$ lines \cite{HA2, Breg, Laus,Laus1}, 
intensity ratio of $K\alpha$ and 
$K\beta$ lines \cite{HA2,Laus,Laus1},
and cascade times \cite{RPD, Ludh,Mud}.  
       
\subsection{Absolute and relative $x$-ray yields of $K$ lines}

The absolute $x$-ray yield $Y_i$ 
 of the specific $K_i$ line
 is determined as the intensity
 of this line per one formed muonic atom. 
Besides,  $Y_{\geqslant \delta}$ is a summary absolute
$x$-ray yield of all $K$ lines corresponding to
radiative transitions $np\to 1s$ ($n \geqslant 5$), and 
$Y_{\rm tot}=\sum_{i}Y_i$ ($i = \alpha, \beta,$ etc.) is a total
absolute yield of all $K$ lines.
    
\begin{figure}[!ht]
\centerline{\includegraphics[width=0.47\textwidth,keepaspectratio]
{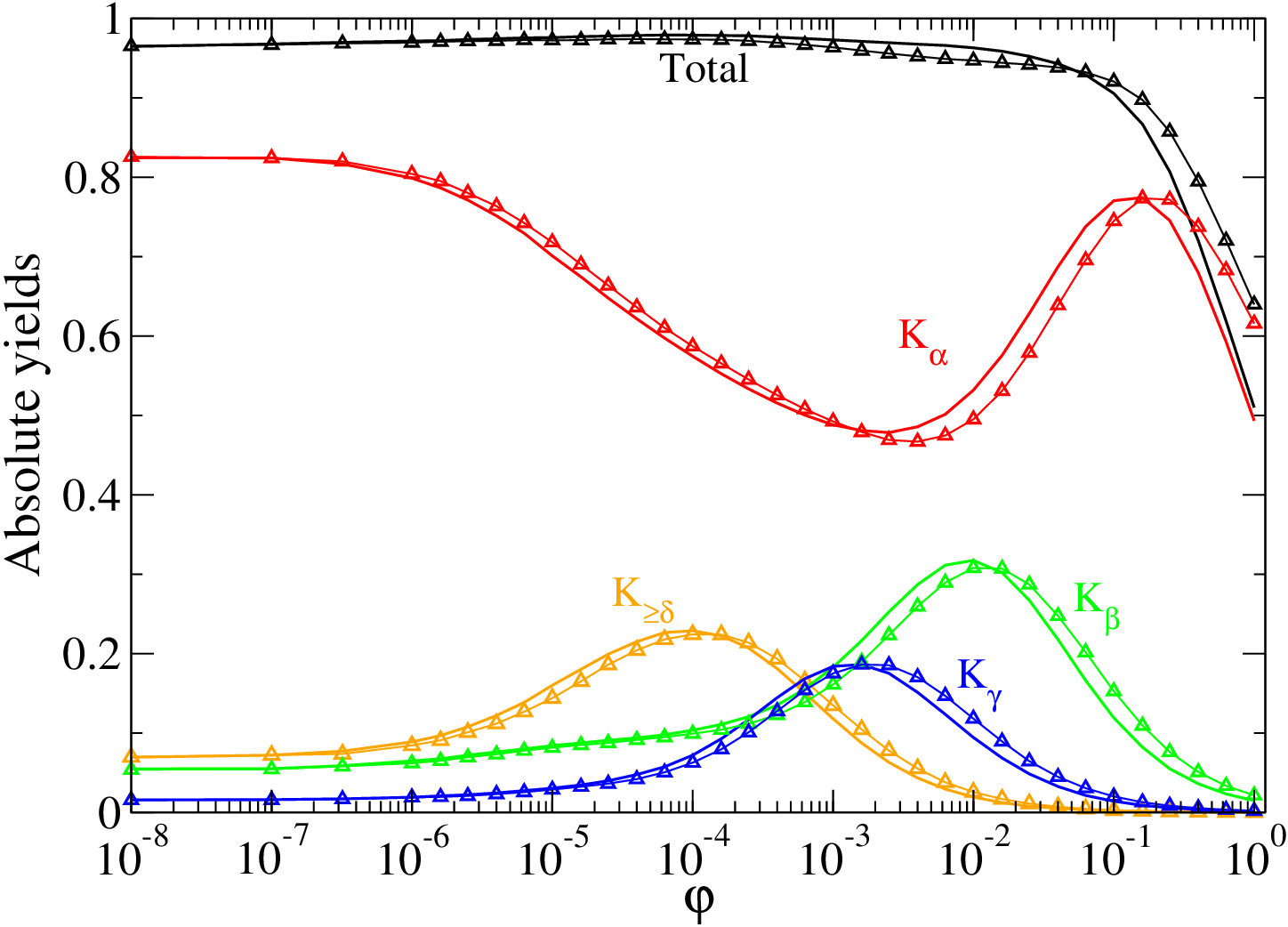}}
\caption{(Color online) The density dependence of absolute 
 yields for $K\alpha$, $K\beta$,
 $K\gamma$, and $K_{\geqslant \delta}$ lines,  
 and $Y_{\rm tot}$
 calculated for $\mu^- p$ (lines) and $\mu^- d$ 
(lines with triangles) atoms at target temperature T=30~K.}
\label{abs_y}
 \end{figure}

 The calculated absolute $x$-ray
 yields of $K\alpha$,  $K\beta$, and $K\gamma$ lines 
 as well as the summary absolute yield of $K_{\geqslant \delta}$
 lines and the total absolute yield  
 $Y_{\rm tot}$ for $\mu^- p$ and $\mu^- d$ atoms 
 are shown in Fig.~\ref{abs_y}.
The calculations were performed in the relative hydrogen density
range $\varphi =10^{-8} - 1$ at target temperature $T=30$~K.
  
 As a whole, the absolute $K$ line yields as well as 
 the total absolute yields have a similar dependence 
 on the density both for $\mu^- p$ and for $\mu^- d$ atoms.
 Their density dependencies reveal a few general features 
 discussed below.
 
At the lowest relative densities $\varphi \lesssim 10^{-7}$, 
the de-excitation cascade in $\mu^- p$ and $\mu^- d$ atoms 
are largely determined by the initial $n$- $l$- distributions and the
rates of radiative transitions and muon decay. The rate of the muon
decay is applied to be equal to 
$\lambda_{\mu}=4.54 \cdot 10^5$~$s^{-1}$, 
that corresponds to the free muon lifetime 
$\tau _{\mu} = 2.2\cdot 10^{-6}$~s.
Here the radiative transitions dominate over other processes, 
thus leading to the population of circular states of the muonic atom
and mainly to the radiative transition $2p\rightarrow 1s$ at the
end of the cascade. At these densities the
$np$ $(n \geqslant 3)$ states are populated much weaker
during the cascade mainly because of 
initial conditions defined in Eqs.~(\ref{Gn}) and (\ref{Fl}). 
We see the corresponding picture in Fig.~\ref{abs_y} both for 
$\mu^- p$ and $\mu^- d$ atoms. In particular,
at $\varphi = 10^{-8}$ the present cascade calculations 
predict for $\mu^- p$ and $\mu^- d$ atoms practically 
the same absolute yields: $Y_{\alpha} = 0.825$, 
$Y_{\beta} = 0.054$, $Y_{\gamma} = 0.016$, and 
$Y_{\geqslant \delta} = 0.070$ (here the total absolute
yield amounts to $Y_{\rm tot} = 0.965)$.
At this density about 3.5\% of all formed muonic 
atoms undergo the $\mu$-decay: 2\% 
of $\mu$-decay occur in the $2s$ state (both above
and below $2s-2p$ threshold) and accordingly 1.5\% of muons
decay in states with $n \geqslant 3$.
 
The absolute $Y_{\alpha}$ and 
$Y_{\rm tot}$ yields calculated 
previously (see Fig. 6 in \cite {CascalII})
for muonic hydrogen at the density $\varphi = 10^{-8}$
do not agree with the present predictions and 
are equal to  about $0.95$ and $1$, respectively. 
Besides,
the so-called radiative mode of the arrival population of the
$2s$ state (see Eq. 21 in \cite{PPForm}) calculated
with these absolute yields~\cite {CascalII} is obtained to be 
$\simeq 0.7\%$, 
that is almost three times less than the present prediction 2\%. 
All these differences can be
explained only by the effect of initial distributions on quantum
numbers $n$ and $l$ ($n_i = 14$ and the $l_i$ distribution 
is statistical) applied in \cite {CascalII}. 
 
 In the wide density range from $\varphi =10^{-8}$ up to 
$\varphi \approx 2.5 \cdot 10^{-3}$ the density dependence of
the total absolute is very weak. 
Contrary to this, the absolute yields of individual  
$K_i$ lines have a more complicated density
dependence. 
With the increase in density above $\varphi =10^{-7}$ the
role of collisional processes in highly excited states ($n\geqslant 5$)
is steadily enhanced, that leads to the
decrease in $K\alpha$ line yield and simultaneously to 
the increase in the absolute yields of the other 
$K_i$ ($ i =\beta, \gamma$, etc.) lines, conserving 
the total absolute yield $Y_{\rm tot}$ practically 
unchangeable in the limit of a few percent.

The absolute $K\alpha$ yield reach their minimum values
$Y_{\alpha}=0.478$ at $\varphi \simeq 2.5 \cdot 10^{-3}$ 
for $\mu^- p$ atom and $Y_{\alpha}=0.465$ 
at $\varphi \simeq 3.5 \cdot 10^{-3}$ for $\mu^- d$ atom. 
 At higher densities the absolute $K\alpha$ 
yield rapidly grows and achieves 
its maximum value 0.78 at $\varphi = 0.13$ for $\mu^- p$ atom
and at $\varphi = 0.18$ for $\mu^- d$ atom, respectively. 
The enhancement of the absolute 
$K\alpha$ line yield observed in this density region
is mainly explained by both the external Auger and 
CD $3\to 2$ of muonic
atoms.

Further density increase leads to a significant reduction of both
$Y_{\alpha}$ and $Y_{\rm tot}$. 
At $\varphi = 1$ the present calculations predict: $Y_{\alpha}=0.493$ and 
$Y_{\rm tot}=0.510$ (for $\mu^- p$ atom) and $Y_{\alpha}=0.616$ and 
$Y_{\rm tot}=0.640$ (for $\mu^- d$ atom). 
Therefore, the isotopic effect predicted at the liquid 
hydrogen density for $Y_{\alpha}$ and $Y_{\rm tot}$ is about 25\%.
It would be very useful to check these predictions in measurements 
of absolute $x$-ray yields at this density for
both muonic hydrogen and muonic deuterium atoms. 

A sharp decrease in the absolute yields of the $Y_{\alpha}$ and 
$Y_{\rm tot}$ is explained by the formation (during the de-excitation 
cascade) of a muonic atom $(\mu^{-}p)_ {2s}$ (or $(\mu^{-}d)_ {2s}$) 
with kinetic energy below the $2p$ threshold and its subsequent 
CD $2s\to1s$.
In the density region $\varphi = 0.1 - 1$ both the population of
the $2s$ state below the $2p$ threshold and the rate of the
CD $2s\to 1s$ in the case of the $\mu^- p$ atom
is on average about a few times more than corresponding values in the 
$\mu^- d$ atom case.
 
 In the present study the
 collision-induced radiative decay of the $2s$ state below
 $2p$ threshold was not considered. This process 
 will be studied in details in a separate paper
 together with the other decay modes of the $2s$ state. 
 Note, that this process leads to the increase in $Y_{\alpha}$ 
 and correspondingly in $Y_{\rm tot}$.
      
 The density dependence of absolute $K$ line yields 
 $(n~\geqslant~3)$ reveals quite similar behavior with the increase
 in density, which may be
conventionally divided into three stages and correspondingly
three density regions.  
In the first region,
the absolute yields slowly increase (about two times) with the density 
compared to their initial values at the
lowest density  ($\varphi =10^{-8}$). The corresponding density 
regions are different
for various  $K_{\geqslant \beta}$ lines: 
$\varphi \lesssim  8\cdot 10^{-5}$ for $K\beta$ line,
 $\varphi \lesssim  2\cdot 10^{-5}$ for  $K\gamma$ line, and 
 $\varphi \lesssim  6\cdot 10^{-6}$ for  $K\delta$ line, etc. 
 In the second region, 
 the absolute yields of $K_{\geqslant \beta}$ lines grow 
 very rapidly achieving
 their maximum values at different densities 
 (e.g., $\varphi =3\cdot 10^{-4}$ for
 $K\delta$ line, $\varphi =1.2\cdot 10^{-3}$ for
 $K\gamma$ line, and  $\varphi =1.3\cdot 10^{-2}$ for
 $K\beta$ line). Finally, in the third region, the absolute yields of
 $K_{\geqslant \beta}$ lines practically go to zero with the further 
 density increase, excluding the $K \beta$ line. Thus, 
 the absolute yields of $K_{\geqslant \beta}$ lines form 
 the family of the similar curves with maxima, 
 which values increase about
two times, whereas $n$ is decreasing.

The physical reason of the maximum formation in absolute
yields of all $K$ lines ($K\alpha$, $K\beta$, 
 $K\gamma$, etc.)
 has a similar nature --- the interplay between the rates of radiative
 transitions, from one side, and the rates of the external Auger 
 effect and CD, from the other side. 
 All maxima are largely formed
 by the increase in rates of external 
 Auger and CD transitions from the above lying states.
 At the same time, the absolute yields of 
 $K_{\geqslant \beta}$ lines decrease in the third region,
 because the summary rate of the external Auger effect and Coulomb 
 de-excitation from states with $n \geqslant 3$ becomes comparable
 and even much more than
 the rate of the corresponding radiative $np \to 1s$ transition.

The $\mu^- d$ absolute yields slightly differ from $\mu^- p$ ones
at densities below $10^{-3}$. At higher densities the isotopic effect
increases and becomes much stronger in the absolute yields of  
$K\gamma$, $K\beta$, and especially $K\alpha$ lines.

It is noteworthy, that the significant contribution to the 
total absolute yield comes from $K_{\geqslant \delta}$ lines 
in a wide density range $\varphi \lesssim 3\cdot 10^{-3}$. Their
summary yield is comparable with $Y\beta $ at density 
$\varphi \lesssim 10^{-6}$ and even about two times more
at $\varphi \lesssim 10^{-4}$. This result should be 
taken into account in the analysis of experimental data.

\begin{figure}[!ht]
\centerline{\includegraphics[width=0.47\textwidth,keepaspectratio]
{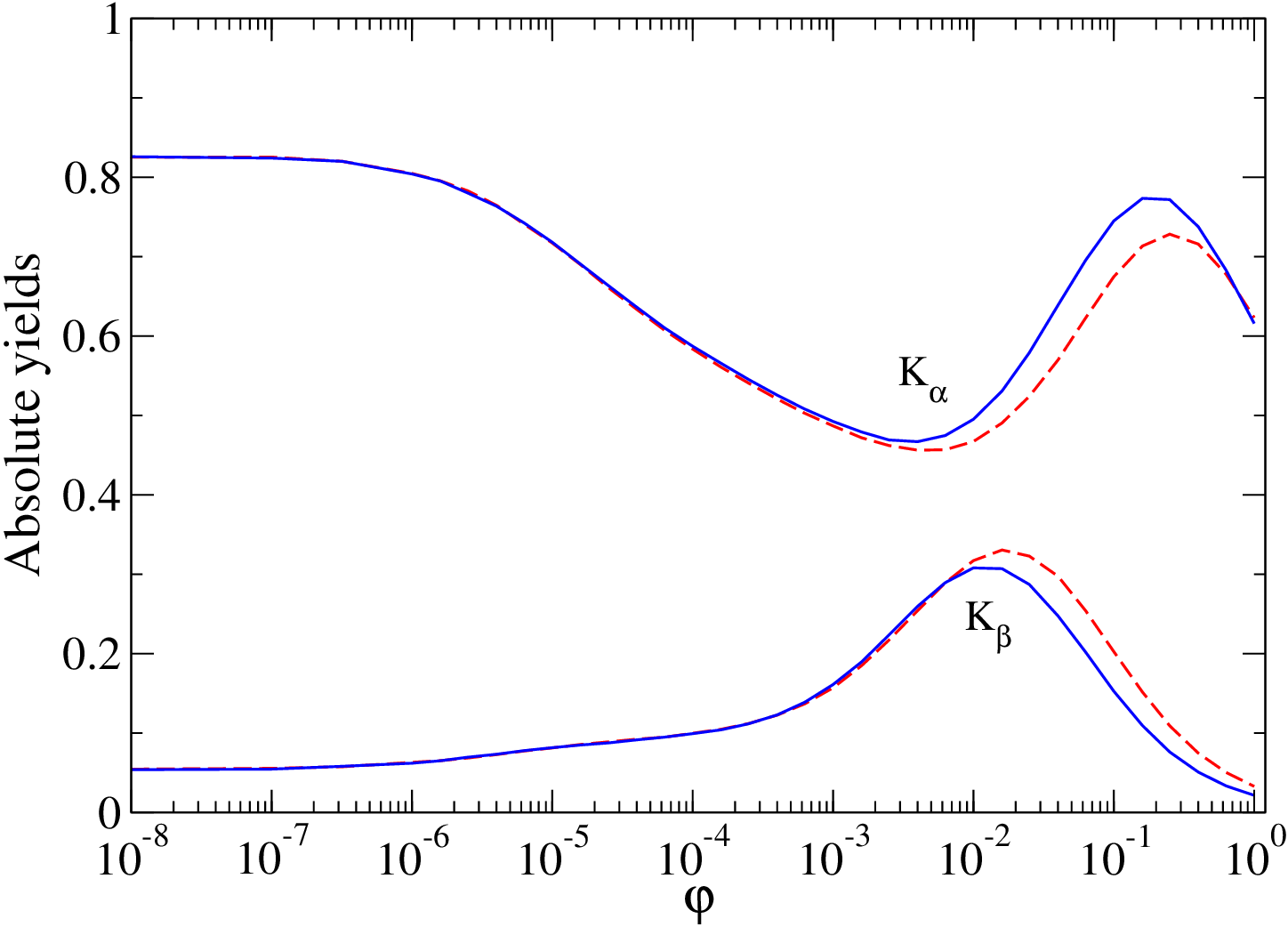}}
\caption{(Color online) The temperature effect in the density dependence 
of the absolute $x$-ray yields for $K\alpha$ and $K\beta$ lines in muonic 
deuterium calculated at target temperature T=30~K (solid lines) and 
T=300~K (dashed lines).}
\label{abs_y_mud-temp}
 \end{figure}
  The exact kinematics of binary collisions taking into account
 the target motion (see Eqs.~(\ref{en1})-(\ref{elf})) is 
 used explicitly in
 the present cascade model. This leads to effects, 
 which could be observed in
 precision experiments produced at densities above 
 $\approx10^{-2}$ and essentially different target temperature.
 To illustrate this effect,  
 the absolute $x$-ray yields of $K\alpha$ and $K\beta$  lines 
for the $\mu^- d$ atom were calculated 
at two values of the target temperature 
$T=30$~K  and $T=300$~K (see Fig.~ \ref{abs_y_mud-temp}).  
One can see, that the absolute yields of 
both $K\alpha$
and $K\beta$ lines differ essentially 
at the density above $\varphi \gtrsim10^{-2}$ 
with the increase in temperature from 
$T=30$~K  to $T=300$~K. 
 
The observed temperature effect is explained
by the strong energy dependence of the CD rate
shown in Figs.~\ref{rates_n3_av_p}. 
The increase in target
temperature  from T=30~K to T=300~K and accordingly the mean
kinetic energy of the target on one order of magnitude makes 
the process of the CD $3\to2$
less effective (at a fixed density) than the radiative 
transition $3p\to 1s$. As a result, at the density above 
$\varphi \approx 10^{-2}$ the absolute yields 
of the $K\beta$ line calculated at T=300~K
increase, whereas the absolute yields
of the $K\alpha$ line decrease in comparison with their values
calculated at the target temperature T=30~K.

The discussed above regularities in absolute $x$-ray yields 
of different $K$ lines should also take
place in their relative
yields defined by the ratio $Y_{i}/Y_{\rm tot}$. 
The present results for $\mu^- p$ and $\mu^- d$ atoms  
are compared with available experimental 
data \cite{HA2, Breg, Laus,Laus1}.

  \begin{figure}[!ht]
\centerline{\includegraphics[width=0.47\textwidth,keepaspectratio]
{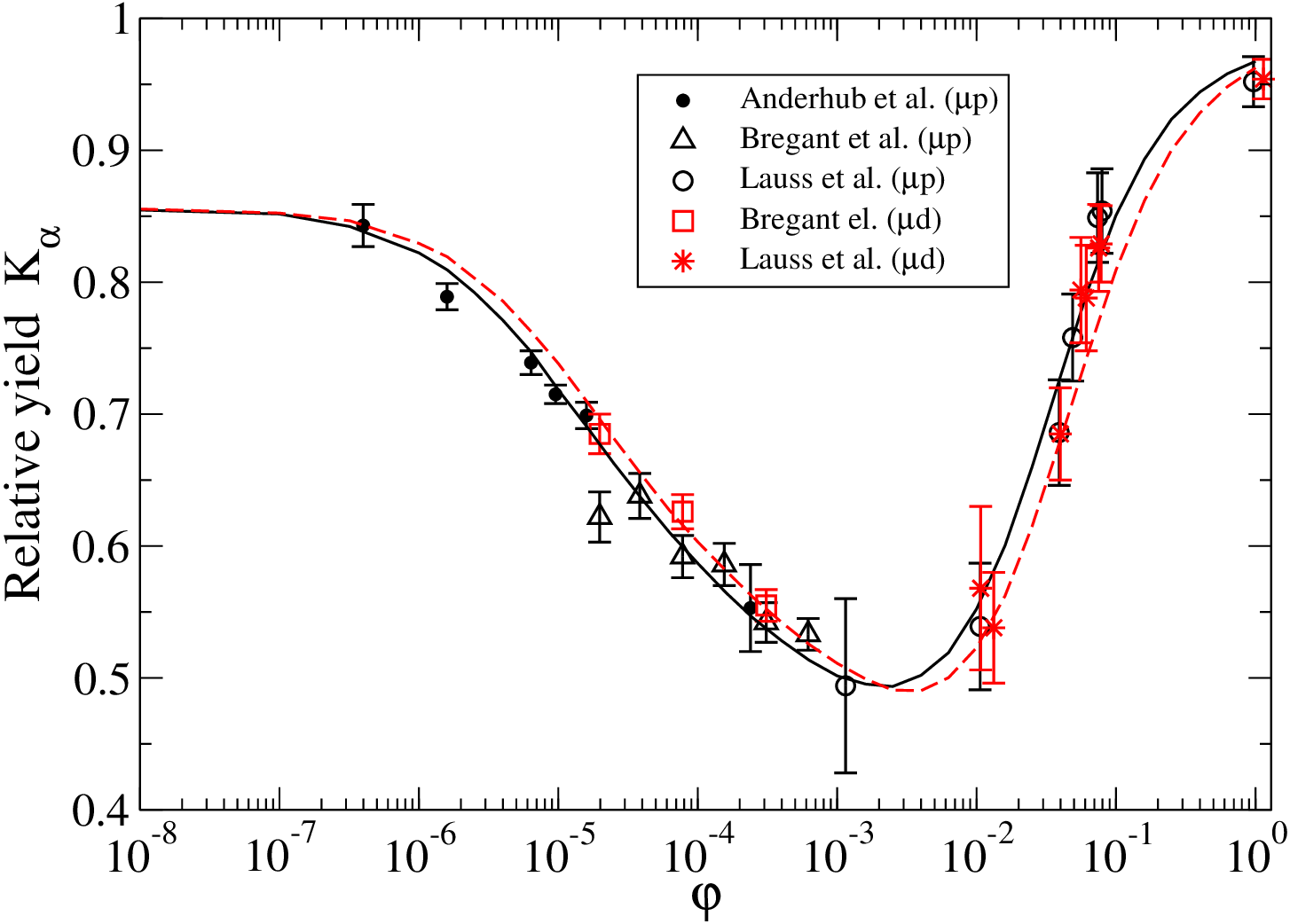}}
\caption{(Color online) The density dependence 
of the relative $K\alpha$ line
$x$-ray yields
for $\mu^- p$ (solid line) and $\mu^- d$ (dashed line) atoms. 
The experimental data are from
\cite{HA2, Breg, Laus,Laus1}.}
\label{rel_a_pd}
\end{figure} 
  The density dependence of the relative 
 $K \alpha$ line $x$-ray yields  for
muonic hydrogen and deuterium is shown in Fig.~\ref{rel_a_pd}. 
The calculations
were done at the target temperature $T=30~K$ that corresponds 
to experimental conditions \cite{Laus,Laus1}. It is noteworthy, 
that the temperature dependence of both absolute 
and relative $x$-ray yields is very weak at the density range 
below $\varphi \approx 10^{-3}$, where experimental data \cite{HA2, Breg}
have been obtained. As a whole, the theoretical
values of relative $x$-ray yields of $K \alpha$ line are in
very good agreement with experimental 
data \cite{HA2, Breg, Laus,Laus1}, obtained in the wide density range
from $\varphi =3.98\times10^{-7}$ up to 1.14 
(in the case of $\mu d$ atom). 
In our opinion, 
the observed disagreements between theoretical and experimental
results for $\mu^- p$ in a few values of density 
(e.g., at $\varphi \simeq 1.6\cdot 10^{-6}$ and 
$\varphi \simeq 2\cdot10^{-5}$) 
can not be explained by the possible     
theoretical uncertainties in the present study. In particular, 
the experimental value of the relative $K \alpha$ line $x$-ray yield
at $\varphi \simeq 2\cdot 10^{-5}$ (0.622 \cite{Breg}) 
is obviously much lesser than
the value 0.699 obtained in \cite{HA2} at a very close value of 
density $\varphi \simeq 1.6\cdot 10^{-5}$,
which, in turn, is in excellent agreement with the present result 0.690.

The comparison of the relative $x$-ray yields of $K\alpha$ line
calculated for muonic hydrogen and deuterium reveals the
isotopic effect observed above in absolute yields (see Fig.~\ref{abs_y}). 
At density below $\sim 10^{-3}$, the relative yields 
of $K\alpha$ line in the case of $\mu^- d$ atoms are slightly
larger than for $\mu^- p$ atom case, that is in correspondence
with experimental data~\cite{Breg}. At densities higher than
$\varphi \approx 3\cdot 10^{-3}$ we observe the opposite picture, viz.
the relative $x$-ray yields of $K\alpha$ line 
in the case of $\mu^- d$ atoms are lesser than in $\mu^- p$ atom case.
Unfortunately, the accuracy of experimental data~\cite{Laus,Laus1}
does not allow to confirm our predictions in this density range.
Besides, there are not any experimental data in a very interesting
density range $0.1\lesssim \varphi < 1$.    
  
\begin{figure}[!ht]
\centerline{\includegraphics[width=0.47\textwidth,keepaspectratio]
{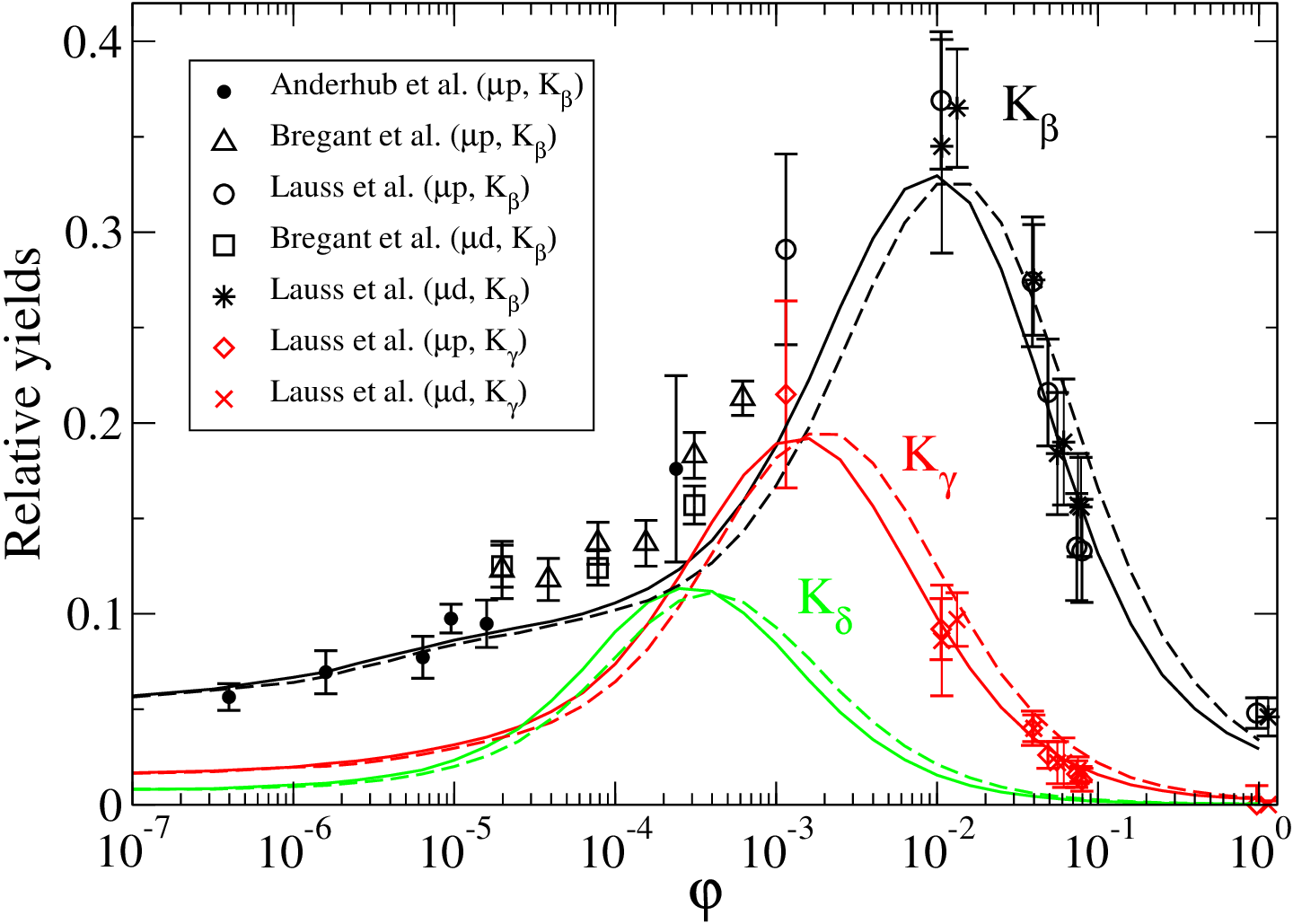}}
\caption{(Color online) The density dependence of relative 
$x$-ray yields of $K\beta$, $K\gamma$, and $K\delta$ lines
in muonic hydrogen (solid lines) and deuterium (dashed lines). 
The experimental data are from \cite{HA2, Breg, Laus,Laus1}.}
\label{rel_b_pd}
\end{figure}
In Fig.~\ref{rel_b_pd} the density dependencies of the relative 
$x$-ray yields of $K \beta$, $K \gamma$, and $K \delta$ lines 
 from muonic hydrogen and deuterium are shown. 
 The calculations
were done at target temperature $T=30~K$.
The agreement between the present
theoretical results and experimental data~\cite{HA2, Breg, Laus,Laus1} 
is rather good practically at all densities under consideration. 
In our opinion, the
observable disagreements between the theoretical and experimental 
relative yields of the $K \beta$ line in the density range 
$\varphi \approx 3\cdot 10^{-5} - 3\cdot10^{-3}$ may be explained
only by experimental reasons: the separation of    
$K \beta$, $K \gamma$, and $K (\geqslant \delta)$ lines, 
the efficiency of their registration and so on. 
It would be very useful to check out these predictions 
by remeasuring relative yields in this density range 
with a much better accuracy.

\begin{figure}[!ht]
\centerline{\includegraphics[width=0.47\textwidth,keepaspectratio]
{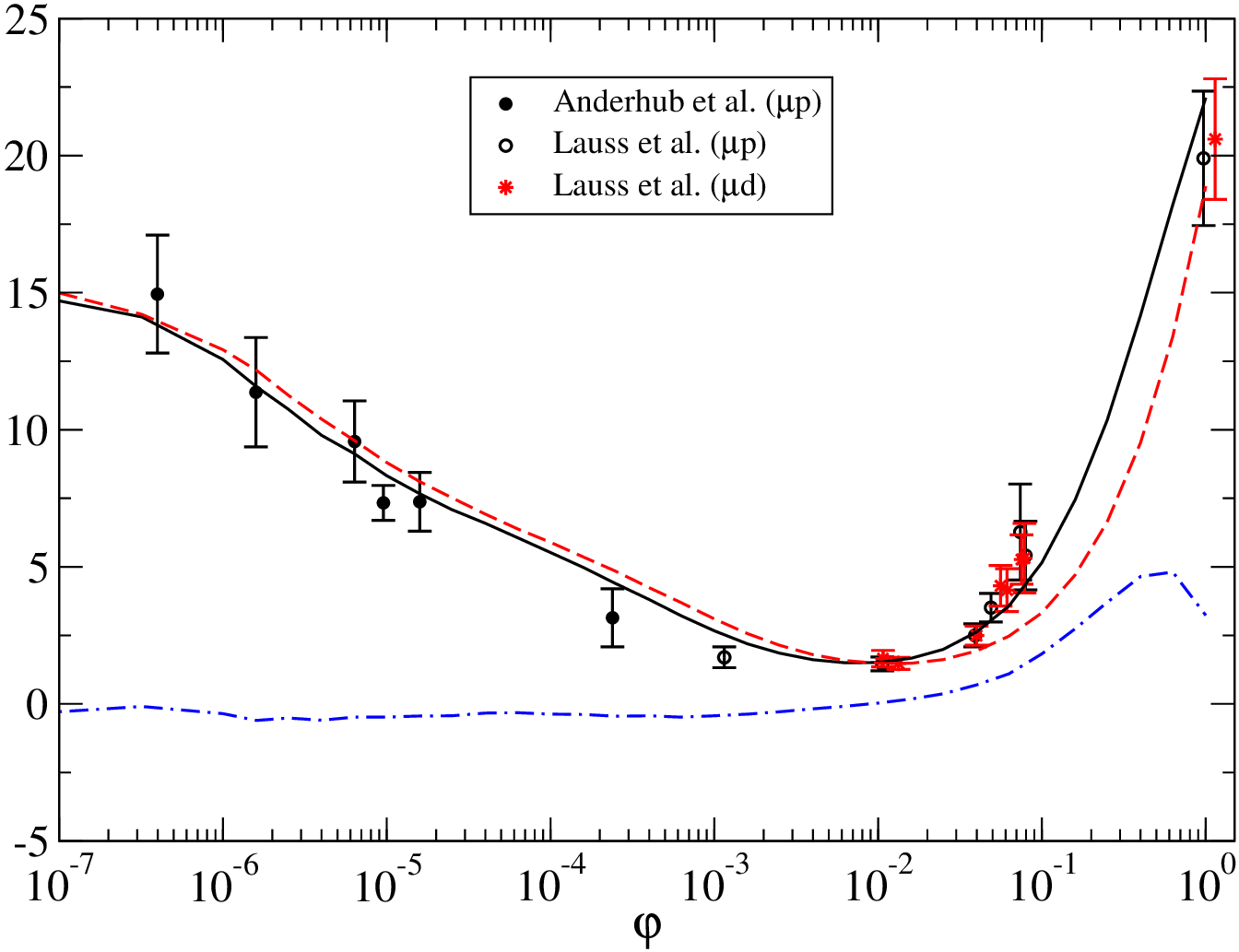}}
\caption{(Color online) The density dependence of the 
$Y_{\alpha} / Y_{\beta}$ ratio 
for muonic hydrogen (solid line) and 
deuterium  (dashed line), as well as
their differences (dashed-dotted line).
The experimental data are from
\cite{HA2,Laus,Laus1}.}
\label{rat_Ka_Kb}
\end{figure}
Figure~\ref{rat_Ka_Kb} shows our results for the density dependence of 
$Y_{\alpha} / Y_{\beta}$ ratios for muonic hydrogen, $R_{\mu p}$, and 
deuterium, $R_{\mu d}$, as well as their differences, 
$(R_{\mu p}-R_{\mu d})$. This difference serves as a direct 
demonstration of the isotopic effect.
Comparing $R_{\mu p}$ with $R_{\mu d}$ at densities below $10^{-2}$,
one can see that the ratio $R_{\mu d}$ is slightly larger than the 
$\mu p$ one. At densities  
$5\cdot 10^{-2} \lesssim \varphi \lesssim 0.8 \cdot 10^{-2}$
we predict the density dependence of $R_{\mu d}$, which
is systematically lesser than experimental data~\cite{Laus1}.
However, at liquid hydrogen density both theoretical prediction and
experimental result are in an excellent agreement.

Our results have a natural explanation. In the case of the
muonic deuterium the rate of the CD
$3\to 2$ is suppressed more than one order of magnitude 
in comparison with the muonic hydrogen 
(see Fig.~\ref{rates_n3_av_p}). This leads to
the increase in $Y_{\beta}$ and correspondingly to the
decrease in $Y_{\alpha}$ in the case of the $(\mu^- d)$ atom as compared
with $(\mu^- p)$ one in the density range above $\approx 10^{-2}$.
It is noteworthy, that differences $(R_{\mu p}-R_{\mu d})$ 
grow at densities above $10^{-2}$ and reach their maximum
at the density about $7\cdot 10^{-1}$. 

The present study shows that additional measurements of the yields
with much better accuracy in the density region between 
$3\cdot 10^{-5}$ and $3\cdot 10^{-3}$
as well as at densities above $\approx5\cdot 10^{-2}$ 
are extremely important. 

\subsection{Kinetic energy distribution}

In addition to the natural line width and the resolution
of the apparatus (response function),
the experimental profile (line shape) 
of the $np\to1s$ $x$-ray lines in muonic and hadronic hydrogen 
atoms is determined  by 
the Doppler broadening due to the motion of the exotic atom 
at the instant  of its radiative
transition. Therefore, the kinetic energy of exotic atoms, 
resulting from the complex interplay of various cascade 
processes, preceding the $x$-ray emission,
is needed in the analysis of precision spectroscopic 
experiments. 
The kinetic energy distribution of the exotic atom changes during
the de-excitation cascade and very sensitive to the rates of 
collisional processes involved in the atomic cascade. 
Thus, it may be considered as one of
the refined tests of both the theoretical
approaches, applied to describe various scattering processes,
and the models of kinetics of the atomic cascade.

The calculations of the kinetic energy distributions of exotic 
atoms at the time of their radiative  $np\to 1s$ transitions 
in muonic and pionic hydrogen atoms have been performed 
previously (e.g., see \cite{JPP,PPKin}),
to improve the description of the Doppler broadening in 
measured profiles of $x$-ray lines: 
for the $\mu^- p$ atom\cite{DSCov,DGot4}
and for the $\pi^- p$ atom (see \cite{DGot3} and references
therein).

In order to take into account the ${K_i}$ line broadening, which
originates from the motion of exotic atoms, the probability density
$\omega_{K_i} (E; \varphi , T)$ of their energy distribution 
should be calculated for the given target density and temperature 
and be used in the analysis of the experimental line shape.
Here we represent the distribution function 
of the probability, 
\begin{equation}
\label{IntE}
I_{K_i}(E; \varphi , T)=\int _{0}^{E}\omega_{K_i}(E'; \varphi , T)dE',
\end{equation}
that the muonic atom has a kinetic energy 
less than $E$ at the instant of the ${K_i}$ line emission. 

\begin{figure}[!ht]
\centerline{\includegraphics[width=\columnwidth,keepaspectratio]
{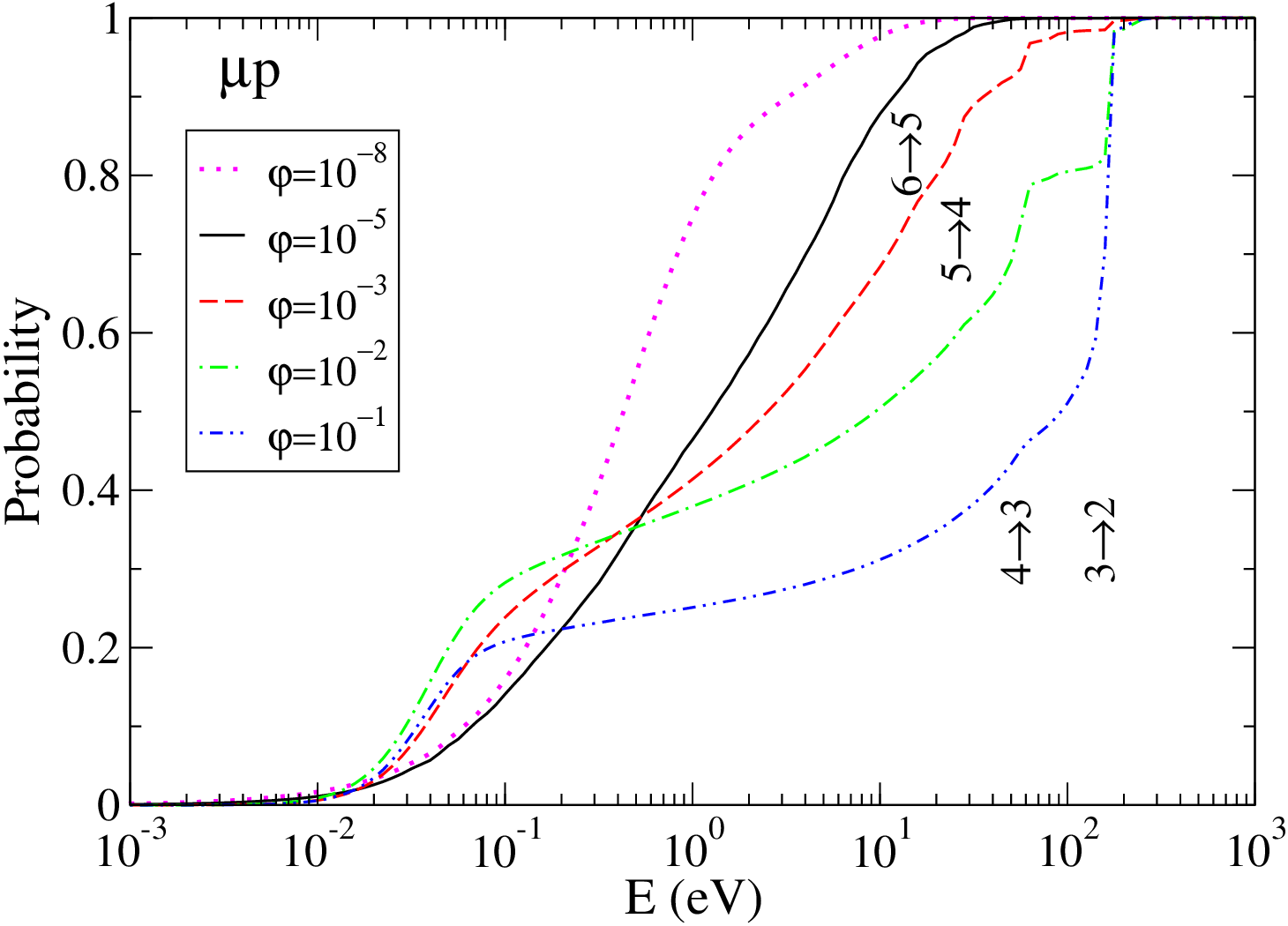}}
\caption{(Color online) The   kinetic 
energy distribution of the $\mu^- p$ atom at the
instant of the radiative transition $2p\rightarrow 1s$  
calculated at various relative target densities $\varphi$ 
(target temperature $T~=~300$~K). The kinetic 
energies, which muonic atoms acquire after CD 
transitions $n\to n'$, are shown by arrows.} 
\label{mup_2-1}
\end{figure}

\begin{figure}[!ht]
\centerline{\includegraphics[width=\columnwidth,keepaspectratio]
{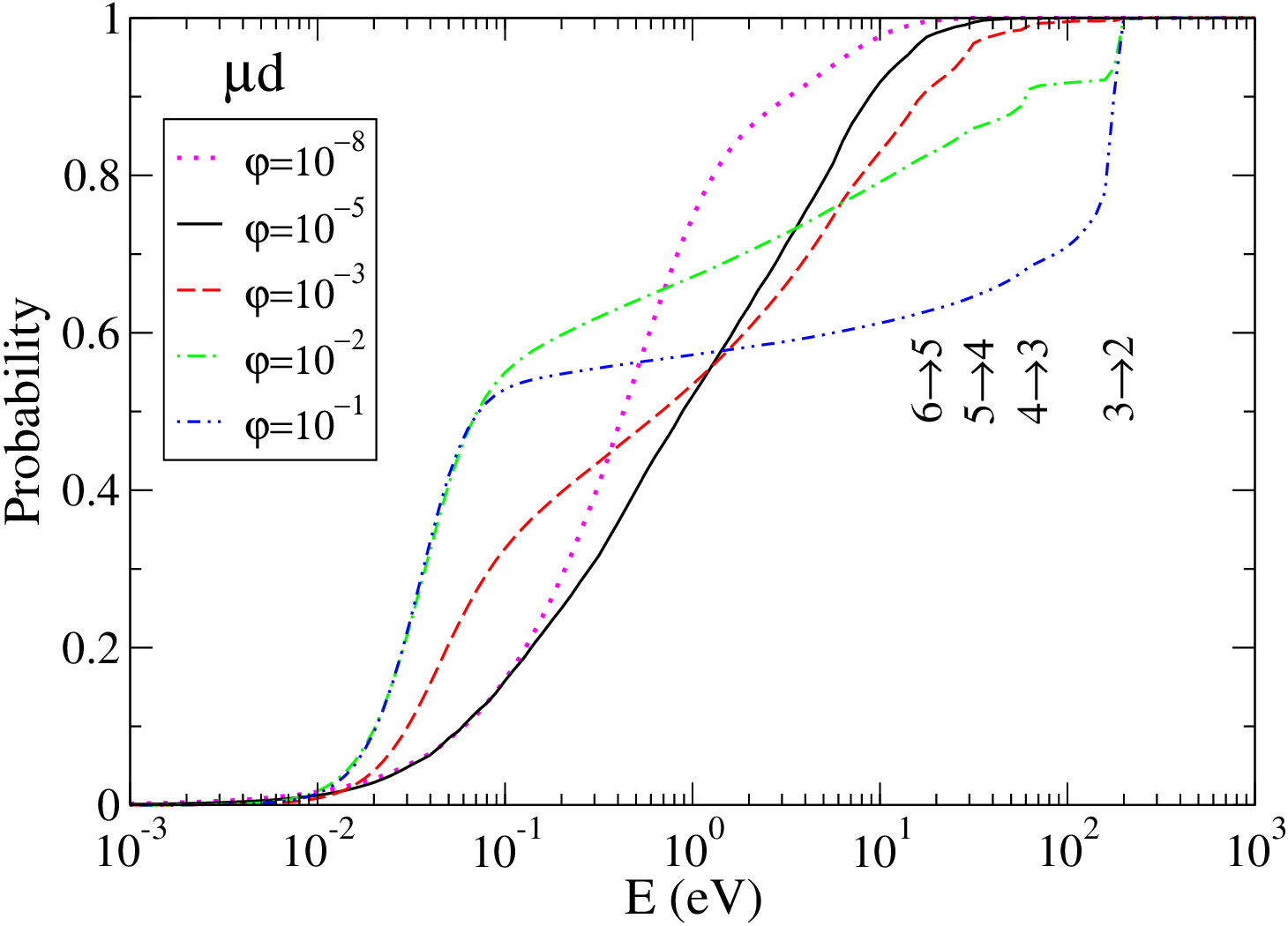}}
\caption{(Color online) The same as in Fig.~\ref{mup_2-1} 
for $\mu^- d$ atom at the instant of the 
radiative $2p\rightarrow 1s$ transition.} 
\label{mud_2-1}
\end{figure}

The energy distribution at low $n$ can be determined by measuring
 the Doppler broadening of $K$ $x$-ray lines, since high-energy 
 components lead to a significant
 Doppler broadening, which is especially pronounced for $K\alpha$
 line.
The calculated distribution functions 
$I_{K\alpha}(E; \varphi, T)$ of $\mu^- p$ and $\mu^- d$ 
atoms are shown in Figs.~\ref{mup_2-1} 
and \ref{mud_2-1}, respectively.
The calculations were performed at target 
temperature $T=300$ K and a few values of the relative target density: 
$\varphi = 10^{-8}, 10^{-5}, 10^{-3}, 10^{-2}$, and $10^{-1}$.
The energy distributions of muonic atoms 
in Figs.~\ref{mup_2-1} and \ref{mud_2-1} calculated at a relative density  
$\varphi = 10^{-8}$ coincide with the initial energy distribution of 
muonic atoms at the time of their formation (see Eq.~(\ref{InitE})).
Here we will use the arbitrary choice of 2~eV as the boundary 
between low-energy and high-energy fractions in the energy distributions
of muonic atoms.

With the increase in density, the role of Coulomb 
transitions grows, and the energy distribution of muonic 
atoms becomes more energetic.   
In particular, in the case of muonic hydrogen, as shown 
in Fig.~\ref{mup_2-1}, the low-energy fraction of the muonic hydrogen
with kinetic energy less than $2$~eV 
decreases drastically  
with the relative density increase 
from $\varphi =10^{-8}$ to $\varphi =10^{-1}$. 
The probability of this fraction is about 0.87 at density $\simeq 10^{-8}$ 
(in correspondence with Eq.~(\ref{InitE})) and strongly 
decreases with the density growth to: about 0.58 at $\varphi =10^{-5}$, 
0.48 at $\varphi =10^{-3}$, 0.41 at $\varphi =10^{-2}$, 
and 0.26 at $\varphi =10^{-1}$.
At the same time, the probability of the high-energy fraction
with kinetic energy $E > 2$~eV 
becomes much more pronounced with the 
increase in density and reveals noticeable contributions of
preceding Coulomb transitions, e.g., $7\to6$, 
$6\to5$, $5\to4$, $4\to3$, and $3\to2$.
 
In the case of the muonic deuterium (see Fig.~\ref{mud_2-1}),
the evolution of the kinetic energy distribution 
(in the same density range)
 is quite different. 
Here, the present cascade calculations 
predict practically the similar change of the probability 
of the low-energy fraction  
from 0.87 to 0.64  in the density range $\varphi = 10^{-8}-10^{-5}$. 
However, at higher densities ($\varphi >10^{-5}$), the contributions of 
Coulomb transitions $6\to5$, $5\to4$, $4\to3$, and $3\to2$ are much 
weaker in comparison with the muonic hydrogen case. As a result,
the probability of the low-energy fraction is about 
0.60 at $\varphi =10^{-3}$ and $\varphi =10^{-1}$
and constitutes even 0.70 at $\varphi =10^{-2}$. 
The strong isotopic effect
observed in the kinetic energy distributions of muonic
hydrogen and deuterium atoms is mainly explained by
the strong isotopic effect in the cross sections of Coulomb 
de-excitation discussed above in the previous section. 

 As shown in Figs.~\ref{mup_2-1} and \ref{mud_2-1}, 
the calculated kinetic 
energy distributions have distinctive high-energy 
components arising from various Coulomb
 transitions with $\Delta n \geqslant 1$, preceding the radiative
de-excitation $2p\to 1s$. In particular, the contributions of Coulomb
de-excitations $6\to5$ (at 14.6 eV for $\mu^-p$ 
and 15.8 eV for $\mu^- d$), $5\to4$ (at 26.9 eV for $\mu^-p$ 
and 29.1 eV for $\mu^- d$), and $4\to3$ (at 58.2 eV for $\mu^-p$ 
and 62.9 eV for $\mu^- d$) are revealed
in the wide density range $\varphi >10^{-5}$. They survive until
the de-excitation of muonic atoms to the $2p$ state due to 
successive Auger and radiative transitions between above lying
states. With the density
increase, the contributions of individual CD
transitions become much less pronounced because of the 
deceleration through the elastic 
scattering and Stark mixing. Nevertheless, the contribution of the
CD transition $4\to3$ is even noticeable at target 
density $10^{-1}$. It is noteworthy, that the contribution of the
CD transition $3\to2$ (at 166.2 eV for $\mu^- p$ 
and 179.9 eV for $\mu^- d$) 
with the density increase from $\varphi =10^{-3}$  
to $\varphi =10^{-1}$ grows from a few percent up to
55\% for $(\mu^- p)_{2p}$ and only to about 30\% for $(\mu^- d)_{2p}$. 

\begin{figure}[!ht] 
\centerline{\includegraphics[width=\columnwidth,keepaspectratio]
{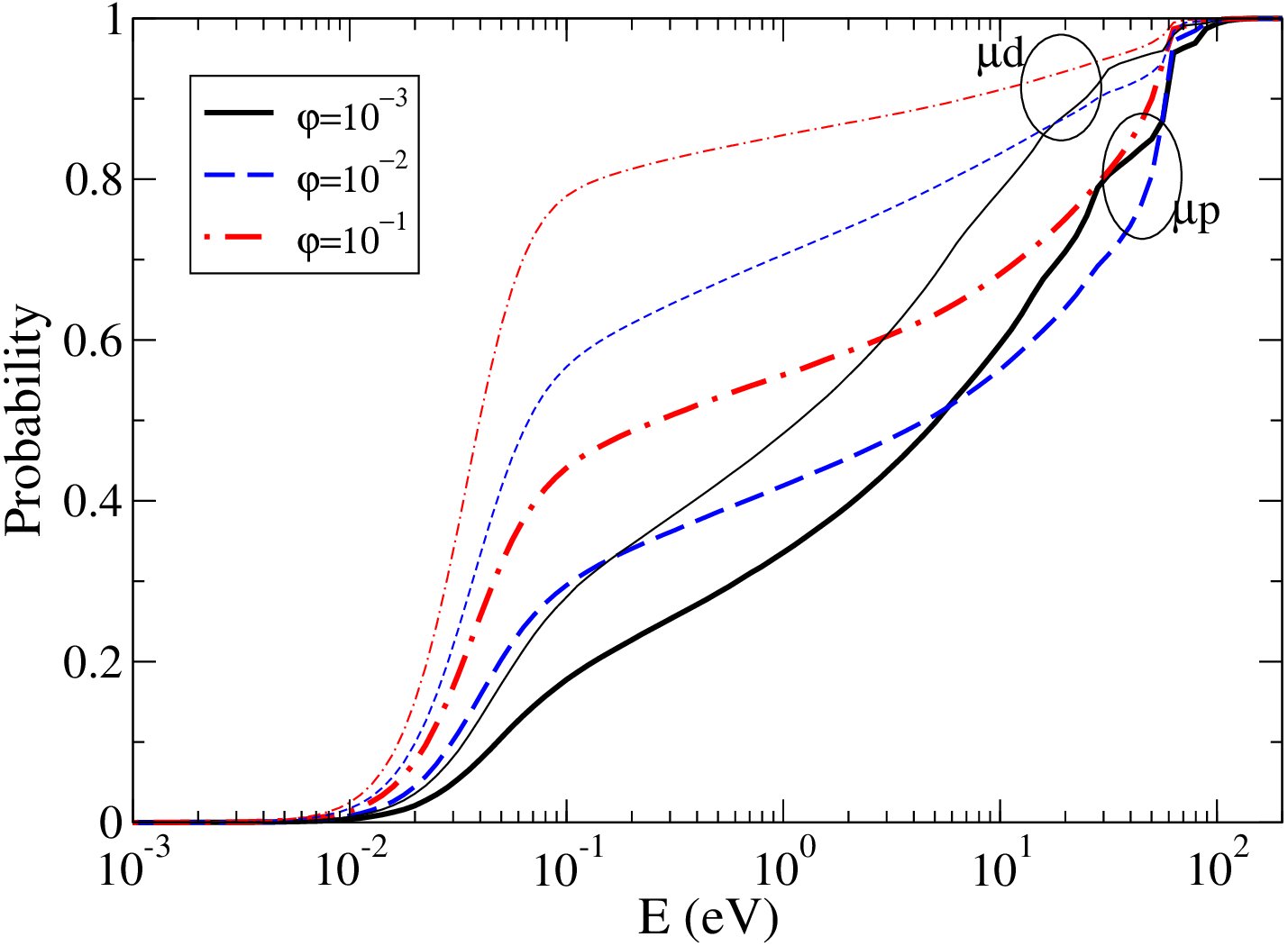}}
\caption{(Color online) The kinetic  
energy distributions of $\mu^- p$ (thick lines) 
and $\mu^- d$ (thin lines) atoms  at the 
instant of their radiative $3p\rightarrow 1s$ transition 
calculated at a few values of the relative target density $\varphi$
(target temperature $T=300$~K).} 
\label{mu_p_D_3}
\end{figure}
The energy distributions of $\mu^- p$ and $\mu^- d$ atoms, 
calculated at the instant of their radiative $3p\rightarrow 1s$ 
transition are shown in Fig.~\ref{mu_p_D_3}.
The calculations were performed for three values of
the relative target density: 
$\varphi = 0.001$, $\varphi = 0.01$, and $\varphi = 0.1$.
The target temperature is chosen to be $T=300$~K.

Here the energy distributions of of $\mu^- p$ and $\mu^- d$ atoms
calculated at the same density differ significantly. The energy 
distributions of $(\mu^- p)_{3p}$ contain quite strong contributions
from preceding $(5\to 4), (5\to 3)$, and especially $(4\to 3)$ 
CD. For example, the Coulomb de-excitation
$(4\to 3)$ at $\varphi = 0.01$ makes up to about 25\%
in the case of the muonic hydrogen, whereas in the case
of the muonic deuterium it contributes only about 7\%.
The predicted probabilities of low-energy fractions ($E\leqslant 2$~eV)
at densities $\varphi = 0.001$, $\varphi = 0.01$, and $\varphi = 0.1$
have also a very strong differences: 40\%, 47\%, and 58\% for 
$(\mu^- p)_{3p}$ and correspondingly 55\%, 74\%, and 87\% for
$(\mu^- d)_{3p}$.

The strong isotopic effect discovered in the present study 
in the scattering and the kinetics of the atomic cascade 
of muonic hydrogen and deuterium atoms allows to explain
the true reason, why the kinetic energy distribution of the
$(\pi^- p)_{3p}$ (at the instant of the radiative $(3p\to 1s)$
transition), scaled to the $(\pi^- d)_{3p}$ case, 
is unable adequately to reproduce the Doppler induced 
width~\cite{Str2}. There are also other reasons that are 
not the subject of this study, which does not allow to justify 
the procedure of the scaling of the kinetic energy distribution. 
In the case of the pionic deuterium the rates of the absorption
from $ns$ states and collision-induced absorption~\cite{PPInd}
as well as the rates of radiative transitions are larger than 
in the case of pionic hydrogen. As a result of these processes, 
the probability of the CD in the case 
of the pionic deuterium, in addition to the isotopic effect
discussed above, becomes also lesser during the de-excitation 
cascade than in the pionic hydrogen case.  

\begin{figure}[!ht]
 \centerline{
 \includegraphics[width=0.47\textwidth,keepaspectratio]
 {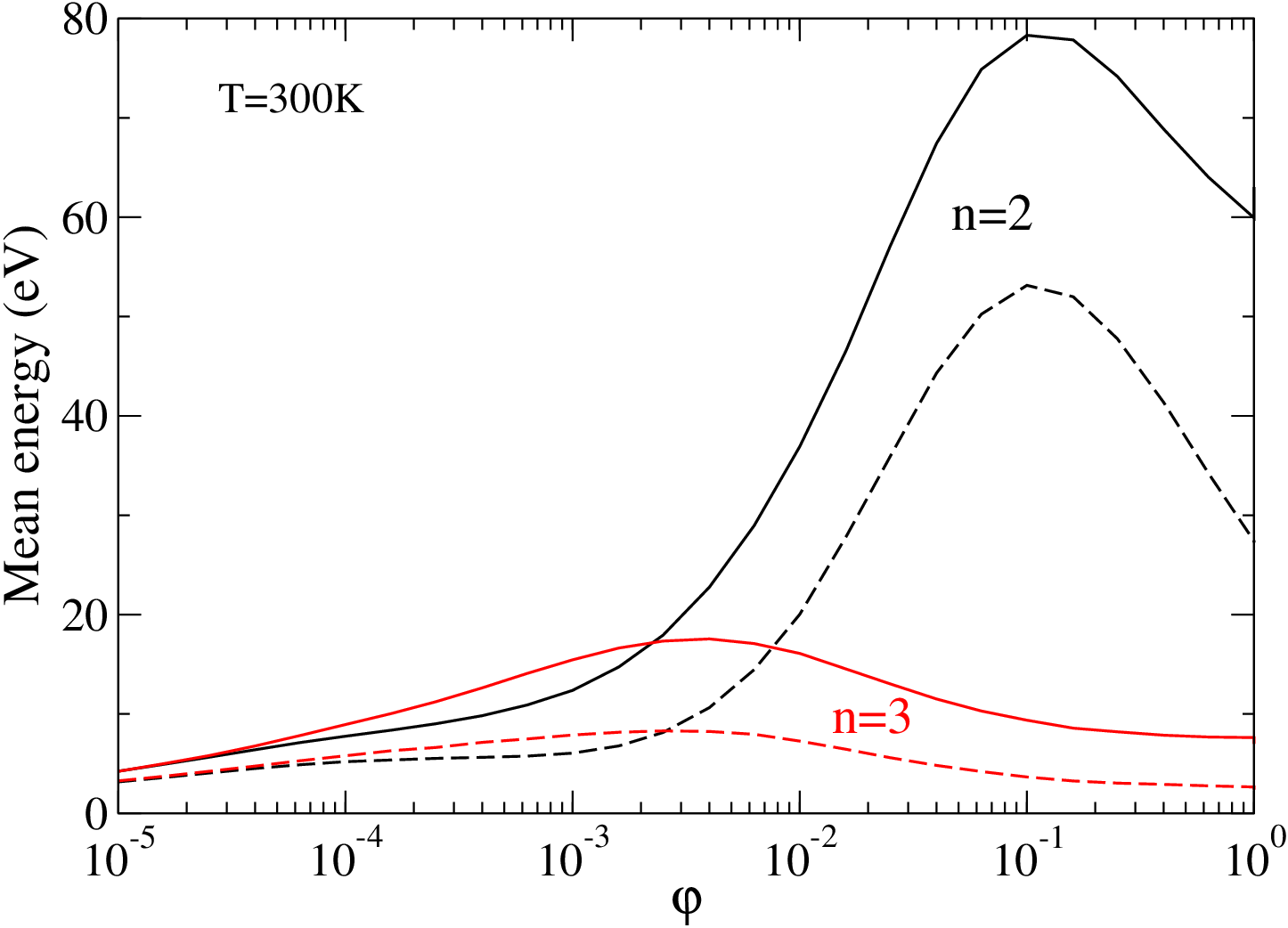}}
 \caption{(Color online) The density dependence of the 
 mean kinetic 
 energy of $\mu^- p$ (solid lines) and $\mu^- d$ 
 (dashed lines) atoms at the instant of their radiative
 transitions $2p\rightarrow 1s$ and $3p\rightarrow 1s$.}
 \label{mean_e_pd}
 \end{figure}

 The acceleration of muonic atoms
during the cascade is also illustrated in
Fig.~\ref{mean_e_pd}. Here, the density 
dependence of the mean kinetic energy
\begin{equation}
\label{meanE}
\overline{E}_{K_i}(\varphi , T)=\int _{0}^{\infty}
E'\omega_{K_i}(E'; \varphi , T)dE'
\end{equation} 
of $\mu^- p$ and $\mu^- d$ atoms, calculated
at the instant 
of their radiative transitions $2p\rightarrow 1s$ and 
$3p\rightarrow 1s$, is shown. The calculations were
performed in the relative density 
range from $10^{-5}$ 
up to 1 at room temperature $T=300$~K.
At a given temperature the mean kinetic energy of the $\mu^- d$
 at the instant of $2p\to1s$ and $3p\to1s$ radiative transitions 
 is always less than the mean kinetic energy of the $\mu^- p$
 at all densities under considerations, excluding the lowest
 density $10^{-8}$, at which the mean energy of both muonic
 hydrogen and deuterium is equal to $1.32$~eV and is in
 perfect agreement with the experimental 
 value $1.3\pm 0.8$~eV \cite{RPD}. 

 In the considered density range the mean kinetic energies
of $\mu^- p$ and $\mu^- d$ atoms increase steadily from 
a few electron volts at density $10^{-5}$ and reach their
maxima (respectively at $3\cdot 10^{-3}$ for $K \beta$ 
lines and at $0.12$ for $K \alpha$ lines 
of $\mu^- p$ and $\mu^- d$ atoms).

In particular, at the density $10^{-5}$ the mean kinetic energies
of both muonic hydrogen and deuterium 
at the instant of their emission of $K \alpha$ and
$K \beta$ lines are the same and equal to 
$4.28$ and 3.30 eV, respectively.
At higher densities, the mean energy of the 
$(\mu^- p)_{3p}$ grows and at the density $3\cdot 10^{-3}$
reaches the maximum value $\simeq 17$ eV and 
about a factor two less, viz.
$\simeq 8$ eV, in the case of the $(\mu^- d)_{3p}$.

The most significant increase in mean energy
is revealed for muonic atoms in the $2p$ state during 
the emission of the
$K \alpha$ line. Here, in the density range from $3\cdot 10^{-3}$ 
to $0.12$ the mean energy of muonic hydrogen and deuterium
rapidly grows from $17$~eV up to $78$ eV for  $\mu^- p$ 
and from $8$~eV up to $53$ eV for $\mu^- d$ atoms, respectively. 
At the liquid hydrogen density the mean kinetic energies of 
$\mu^- p$ and $\mu^- d$ atoms before radiative $2p\to1s$
transition also have a large difference. In our calculations we
obtain $60$~eV and $24$~eV for $(\mu^- p)_{2p}$
and $(\mu^- d)_{2p}$, respectively. 
 This isotopic effect is
 mainly explained by the observed isotopic effect in CD 
 rates of $4\to3$ and $3\to2$ transitions in 
 $\mu^- p$  and $\mu^- d$ atoms 
 (e.g., see Figs. {\ref{rates_n3_av_p} and {\ref{rates_n4_av_p},
 as well as Figs. {\ref{mup_2-1}-{\ref{mu_p_D_3}).
It is noteworthy, that the density dependence of the mean 
kinetic energy calculated for 
$\mu^- p$ and $\mu^- d$ atoms is in qualitative agreement
with the data~\cite{Abbott}.

 \subsection{Cascade time}

The mean time interval between the formation of the muonic atom 
and its arrival in the $1s$ state is determined by
the cascade time. 
In the present study we calculate the so-called prompt cascade time
which does not include the life time of the $2s$ state at kinetic energy 
both above and below $2p$ threshold.  
To find the prompt cascade time, $\tau_{\rm casc}$, 
at low target densities $\varphi \lesssim 10^{-4}$,
we calculate the probability of the $\mu$ decay, $p_{\mu}$,
during the de-excitation cascade (beyond the $2s$ state)
\begin{equation}
\label{prob}
p_{\mu}=1-{\rm e}^{-\tau_{\rm casc}/\tau_{\mu}}.
\end{equation}
Here the probability $p_{\mu}$ is defined as the ratio of the 
number of the $\mu$ decay to the total number of
formed muonic atoms. Then the cascade time is given by
\begin{equation}
\label{tcasc}
\tau_{\rm casc}=\tau_{\mu}\ln\frac{1}{1-p_{\mu}}.
\end{equation}

\begin{table}[ht]
 \caption{The cascade times $\tau_{\rm casc }$ of $\mu p$ and 
  $\mu d$ atoms at various target pressures $p$ ($T=293$ K).}
\tabcolsep=0.25em
\begin{tabular}{c|ccc|ccc} \hline \hline
 $ p $   & &$ \tau_{\rm casc }^{\mu p}$(ns) &
&  & $ \tau_{\rm casc }^{\mu d}$(ns) \\  
 (hPa)  &Theory & Exp.    &Ref.  & Theory & Exp. &Ref.  \\   \hline
$8.55\cdot 10^{-3}$& 37.9        &             &      & 34.3    &        &   \\    
 0.6 & 27.7  &  $26 \pm 5$\footnote{From $K\alpha$ time spectrum} 
           & \cite{Ludh}&28.1   &    & \\   
&  & $39 \pm 5$\footnote{From $K\geqslant \gamma$ time spectrum} 
 &\cite{Ludh}&  &        &    \\  
 1 & 24.1  & $27.7\pm 3^a$ & \cite{Mud} &27.4  & $28.6 \pm 1.9^a$ & \cite{Mud}\\
& &$25.0\pm 1.5^b$ & \cite{Mud} & & $31.6\pm1.7^b$ & \cite{Mud}\\ 
      &   & $28\pm 8$ & \cite{RPD} &    &  & \\         
  4 & 15.5    & $26\pm 6 $&\cite{RPD} & 17.8    &    &   \\
  16 & 8.3       & $11\pm5$&\cite{RPD}  & 8.9       &    &  \\ \hline \hline 
\end{tabular}
 \label{castime}
\end{table} 

At relative target density $\varphi = 10^{-8}$ 
(target pressure about $8.55\cdot 10^{-3}$ hPa), where
the de-excitation cascade is purely radiative, we predict
$\tau_{\rm casc}^{\mu p} = 37.9$ ns and 
$\tau_{\rm casc}^{\mu d} =34.3$ ns. The observed $\approx 10$\% 
difference is explained
by the dependence of the radiative transition rate on the reduced
mass of the muonic atom. We remind, that the reduced mass in 
the muonic
deuterium is about $5$\% more than in the muonic hydrogen
case. 

At higher density the role of collisional processes increases,
that essentially affects the cascade time.  
The cascade time decreases with the increase in target
density owing to collisional processes. 
In our study, at a fixed density, 
the rates of collisional processes in the muonic
hydrogen case are larger than
in the muonic deuterium one 
(e.g., see Figs. \ref{crsec_n4_av} -\ref{rates_n4_av_p}). 
Therefore, the cascade time in muonic
hydrogen should become less than in muonic deuterium 
with the increase in density.

To illustrate this isotopic effect, in Table~I we represent 
cascade times in $\mu p$ and $\mu d$ atoms.
The calculations were performed at a few values of 
the target pressure $p~=~8.55\cdot 10^{-3}, 0.6, 1, 4$, and $16$ hPa 
and temperature $T=293$~K,
corresponding to experimental conditions\cite{RPD,Ludh,Mud}. The 
experimental data \cite{RPD} were obtained in $(\mu^- p)_{1s}$
diffusion experiments,
whereas the measured cascade times \cite{Ludh,Mud}
were extracted by fitting time spectra of the 
muonic hydrogen/deuterium $K\geqslant \gamma$ and $K\alpha$
$x$-rays. These experimental data are also given in Table~1
for comparison with the present results.

As a whole, the cascade times calculated for $\mu p$ and
$\mu d$ atoms
are in very good agreement with experimental data~\cite{RPD,Ludh,Mud}. 
In particular, the experimental fact\cite{Mud} that the cascade time  
in $\mu d$ is about 10\% more than in $\mu p$ 
at a given pressure 1 hPa is in correspondence
with the present theoretical calculations of collisional
cross sections used in the kinetics of the atomic cascade. 
   
\section{Conclusion}

The fully quantum-mechanical description of the scattering 
processes -- elastic scattering, Stark mixing, and CD 
-- has been done in the framework of the close-coupling 
approach for collisions of excited $\mu^- p$ and $\mu^- d$ atoms 
correspodingly with an ordinary hydrogen and deuterium atoms 
in their ground state. 
The energy shifts between $n s$ and
$n l$ states ($n\geqslant 2$, $l\geqslant1$), mainly due to electron
vacuum polarization, have been explicitly taken into account 
in the scattering problem for both muonic hydrogen 
and deuterium atoms. The threshold behavior of all
cross sections was treated in a proper way.
 The explicit analytic expression for the matrix of
interaction potentials has been obtained without
any additional approximations. 
The propagator matrix method \cite{PPForm}
has been used to calculate the scattering matrix defined in 
the subspace of the open channels. 

The influence of the extension of the basis set (by including
the closed channels) on the partial-wave and total cross
sections of the elastic scattering, Stark transitions, and
Coulomb de-excitation has been studied. It was found, that the 
convergence of the close-coupling approach in the scattering
problem under consideration is achieved very slowly due to
a strong coupling between the closed and open channels. 
To our study, the closed channel effect is especially 
significant in the low-energy scattering of exotic atoms 
in low-lying excited states ($n=2-5$). Here the convergence
of the cross sections of different processes has been achieved
by applying the extended basis set including all the exotic atom 
states with the principal quantum number up to 
$n_{\rm{max}}=20$. The cross sections of elastic scattering,
Stark transitions and Coulomb de-excitation have
been calculated reliably both above and below 
$ns - np$ thresholds.
  
The systematic calculations of the 
integrated and differential cross sections of 
elastic scattering, Stark transitions, and CD 
have been performed for the $\mu^- p$ and
$\mu^- d$ atoms in excited states ($n\geqslant2$) and kinetic
energies from 0.001~eV up to $E_{\rm{max}}$ needed for the 
detailed study of the kinetics of the atomic cascade. 
All the cross sections have been calculated by  
applying the extended basis, which 
includes all open and closed channels 
associated with the exotic atom states
with the principal quantum number up to 
$n_{\rm{max}}=20$. 
  
The isotopic effect in the scattering of excited $\mu^- p$ and
$\mu^- d$ atoms has been revealed. At all collision energies the 
cross sections of elastic scattering and Stark transitions in the case 
of the $(\mu^- d)_n +D$ scattering are about (10-15)\%
more than in the $(\mu^- p)_n + H$ one. In contrast, a much
stronger isotopic effect has been discovered in the cross sections
of Coulomb de-excitation. For a given $n$ and fixed 
$\Delta n = n - n'$ the cross sections of the CD 
($n\to n' \leqslant n-1$) in the case of the $\mu^- d$ are,
as a rule, less than in the case of the $\mu^- p$. This
difference grows essentially with the decrease in $n$ and the increase
in $\Delta n$.

The isotopic effect discovered in both cross sections 
and the rates of the CD allow to predict a similar effect in the
pionic deuterium and especially in the muonic tritium.  
This effect should lead to a very strong weakening of the 
high-energy components in the energy spectrum 
of the pionic deuterium and even much stronger in 
the case of the muonic tritium 
compared to the pionic and muonic hydrogen atoms, respectively. 
In particular, the predicted isotopic effect allows for the first 
time to explain the significant suppression of the contributions 
of high-energy components originating from CD 
transitions ($4\to3$) and ($5\to3$) in shift/width 
experiments~\cite{Str2} with the pionic deuterium.

The kinetics of the atomic cascade has been investigated
for both $\mu^- p$ and $\mu^- d$ atoms within the new
version of the cascade model, which has a number of
significant improvements over the previous cascade models:

-- The new results of the fully quantum-mechanical calculations 
of differential and integrated cross sections for collisional processes;  

-- The distributions of exotic atoms in quantum numbers 
$n, l$ and the kinetic energy at the instant of
their formation;   

-- The thermal motion of the target and the exact 
kinematics of binary collisions.

The cascade calculations in the muonic hydrogen and deuterium
have been done in the wide density range
covering 8 orders of magnitude from $10^{-8}$ up to 1.14
(in the units of the liquid hydrogen density, 
$N_{LHD}= 4.25\cdot 10^{22}$~atom/cm$^3$).
A number of different characteristics of the atomic cascade
have been calculated: absolute and relative $x$ ray yields,
kinetic energy distributions, and cascade times. 

The main regularities revealed in the density dependence of the
absolute $x$-ray yields of different $K$ lines have been
discussed. The present cascade calculations predict practically
the same absolute yields of all $K$ lines for both  
$\mu^- p$ and $\mu^- d$ atoms at $\varphi \lesssim 3\cdot 10^{-7}$.
In the density range $3\cdot 10^{-7} < \varphi \lesssim 3\cdot 10^{-3}$
the absolute yields in the case of the $\mu^- d$ atom slightly
differ from the case of the $\mu^- p$ atom. At higher densities
the isotopic effect becomes much stronger in absolute
yields of $K\gamma$, $K\beta$, and especially $K\alpha$ lines.

The calculated relative yields in the muonic hydrogen and
deuterium are in good agreement with the experimental 
data~\cite{HA2,Breg,Laus,Laus1}
practically at all densities. The
observable disagreements between the theoretical and experimental 
relative yields of the $K \beta$ line in the density range 
$\varphi \approx 3\cdot 10^{-5} - 3\cdot10^{-3}$ may be explained
only by experimental reasons: the separation of    
$K \beta$, $K \gamma$, and $K (\geqslant \delta)$ lines, 
the efficiency of their registration, and so on. 
In our opinion, additional measurements 
with much better accuracy and statistics 
in the density region between 
$3\cdot 10^{-5}$ and $3\cdot 10^{-3}$ 
as well as at densities above $\approx5\cdot 10^{-2}$ 
are extremely desirable. In addition to the analysis of experimental data 
it is necessary to include other lines ($K \gamma$ and $K (\geqslant \delta)$)
using our predictions of their absolute yields. 

The calculated kinetic energy distributions of the muonic hydrogen 
and deuterium demonstrate a strong isotopic
effect, which is mainly explained by the strong suppression of
the cross sections of the preceding Coulomb de-excitation 
in the muonic deuterium in comparison with the muonic hydrogen.
The calculated density dependence of the mean kinetic energy of 
$\mu^- p$ and $\mu^- d$ atoms (at the instant of their radiative
($2p\to 1s$) and ($3p\to 1s$) transitions) is in qualitative agreement
with experimental data~\cite{Abbott}. At all densities (above $10^{-5}$)
the mean kinetic energy of the $\mu^- d$ atom is always less than
the mean kinetic energy of the $\mu^- p$ atom. The predicted density
dependence of the mean kinetic energy of 
$\mu^- p$ and $\mu^- d$ atoms (at the instant of their radiative
($2p\to 1s$) and ($3p\to 1s$) transitions) can be used to choose
optimal experimental conditions for experiments similar to 
\cite{Abbott, DSCov} as well as to improve the quality of the
analysis of experimental data.

The isotopic effect in both cross sections and 
the rates of the CD allows to predict a similar effect in 
the pionic deuterium and the muonic tritium.
In particular,  
this effect should lead to a strong weakening of the 
high-energy components in the energy spectrum 
of the pionic deuterium and especially in the case 
of the muonic tritium compared to the pionic and muonic 
hydrogen atoms and could be observed in experiments. 
Besides, the predicted isotopic effect allows for the first time to explain 
the significant suppression of the contributions of high-energy 
components originating from CD 
transitions ($4\to3$) and ($5\to3$) in shift/width 
experiments~\cite{Str2} with the pionic deuterium. 

The cascade times calculated for the $\mu^- p$ and $\mu^- d$ atoms
at low target densities $\varphi \lesssim 10^{-4}$ are in very 
good agreement with available experimental 
data~\cite{RPD,Ludh,Mud}.

We conclude that the present study demonstrates the overall 
reliability of the theoretical approach used for the description
of the scattering processes of the excited muonic hydrogen and
deuterium with ordinary hydrogen isotope atoms and 
a very good understanding of their kinetics. The obtained results serve 
a good foundation to apply the current theory for the realistic study of
hadronic hydrogen atoms.   
 
\section{Acknowledgments} 

The authors thank Detlev Gotta and Randolf Pohl
who had drawn their attention to the importance of the problem 
and e-mail correspondence regarding some aspects of this work.

\appendix

 \section{Potential matrix}
 
Here we present the derivation of the explicit analytic form 
for the matrix elements (\ref{potW}) 
of the interaction potential $V({\bf R},{\boldsymbol{\rho}})$, 
which is the result of 
averaging the four-body interaction potential 
$V({\bf r},{\boldsymbol{\rho}},{\bf R})$ 
over the ground state wave function of hydrogen atom, 
$\frac{1}{\sqrt{4\pi}}R_{1s}(r)$:
 \begin{equation}
 V({\bf R},{\boldsymbol{\rho}})=\frac{1}{4\pi}\int_{0}^{\infty}{\rm
d} {\bf r} R^2_{1s} (r)V({\bf r},{\boldsymbol{\rho}},{\bf R}).
\label{averag}
\end{equation}
The interaction potential, $ V({\bf r},
{\boldsymbol{\rho}},{\bf R})$, includes the two-body 
Coulomb interactions between the
particles from two colliding subsystems:
\begin{equation}
V(\mathbf{r},\boldsymbol{\rho},\mathbf{R})=V_{ab}+V_{\mu
b}+V_{ae}+V_{\mu e}, \label{pots}
\end{equation}
where
\begin{eqnarray}
V_{ab}=|\mathbf{R}+\nu \boldsymbol{\rho}-\nu_e \mathbf{r}|^{-1},\\
V_{\mu b}=-|\mathbf{R}-\xi\boldsymbol{\rho}-\nu_e \mathbf{r}|^{-1}, \\
V_{\mu e}= |\mathbf{R}-\xi \boldsymbol{\rho}+\xi_e \mathbf{r}|^{-1},\\
V_{ae}=-|\mathbf{R}+\nu\boldsymbol{\rho}+\xi_e \mathbf{r}|^{-1}.
\label{potd}
\end{eqnarray}
Here the following notations are used:
\begin{eqnarray}
\nu = m_{\mu}/(m_{\mu} +m_a),\; \xi = 1-\nu, \\ 
\nu_e = m_e /(m_e+m_b), \; \xi_e = 1 - \nu_e,
\end{eqnarray}
($m_a,m_b, m_{\mu}$ and $m_e$ are the masses of 
hydrogen isotopes, muon and electron, respectively).

Averaging in Eq.~(\ref{averag})  results in
\begin{align}
V({\bf R},{\boldsymbol{\rho}})
 = \frac{1}{\xi_e}\{U_{\nu,\xi_e}({\bf
 R},{\boldsymbol{\rho}})-U_{-\xi,\xi_e}({\bf R},{\boldsymbol{\rho}})\}
  \nonumber\\
-\frac{1}{\nu_e}\{U_{\nu,\nu_e}(
{\bf R},{\boldsymbol{\rho}})-U_{-\xi,\xi_e}({\bf R},{\boldsymbol{\rho}})\},
\label{w5}
\end{align}
where  
\begin{multline}
U_{\alpha,\beta}({\bf R}, {\boldsymbol{\rho}})=
\left(1+\frac{\beta}{|{\bf R}+ \alpha\boldsymbol{\rho}|}\right)
{\rm e}^{-\frac{2|{\bf R}+\alpha\boldsymbol{\rho}|}{\beta}}\\
\equiv
 \beta \lim_{x\to 1}\left(1-\frac{1}{2}\frac{\partial}{\partial x}\right)
\frac{{\rm e}^{-\frac{2x|{\bf R}+\alpha\boldsymbol{\rho}|}{\beta}}}
{|{\bf R}+\alpha\boldsymbol{\rho}|}.
\label{w6}
\end{multline}

This representation of 
$U_{\alpha,\beta}({\bf R}, {\boldsymbol{\rho}})$ 
allows to apply the addition theorem for the
spherical Bessel functions~\cite{AbrSt}:
\begin{align}
\frac{{\rm e}^{-\lambda |{\bf R}_1+{\bf r}_1|}}{|{\bf R}_1+{\bf r}_1|}=
\frac{4\pi}{\sqrt{R_1r_1}}\sum_{t\tau}(-1)^t
Y^*_{t\tau}(\hat{\bf R}_1) Y_{t\tau}(\hat{\bf r}_1)\nonumber\\ 
\left\{ K_{t+1/2}(\lambda R_1)\,I_{t+1/2}(\lambda
r_1)\left |_{r_1 <R_1} \right . \right . \nonumber\\ 
\left . + I_{t+1/2}(\lambda R_1)\,K_{t+1/2}(\lambda r_1)
\left |_{r_1 >R_1}\right . \right \},
\label{w7}
\end{align}
where$I_p(x)$ and $K_p(x)$ are the modified  spherical Bessel
functions of the first and third kind, respectively.

The matrix element of interaction
$W^{J {\it p}}_{n'l'L', nlL}(R)$ between the asymptotic initial $(n l
L; J)$ and final $(n' l' L'; J)$  channels is defined by 
\begin{align}
W^{J {\it p}}_{n'l'L,nlL}(R)&=\int{\rm d}{\boldsymbol {\rho}} \,{\rm d} \hat{\bf R} 
R_{nl} (\rho)R_{n'l'}(\rho) \nonumber\\ &\times {\cal
Y}^{JM}_{lL} (\hat{\boldsymbol {\rho}},\hat{\bf R})\, 
V({\boldsymbol{\rho}},{\bf R})\,{\cal Y}^{*JM}_{l'L'}(
\hat{\boldsymbol {\rho}},\hat{\bf R}),
\label{w1}
\end{align}
 where the radial hydrogen-like wave functions
 are given explicitly  by formulas:
\begin{equation}
R_{nl}(\rho)=N_{nl}\left(\frac{2\rho}{n
a}\right)^l\exp(-\rho/na)
 \sum_{q=0}^{n-l-1}S_{q}(n,
 l)\left(\frac{2\rho}{n a}\right)^q \label{w2}
\end{equation}
($a$ is the Bohr' radius of the atom) with
\begin{equation}
N_{nl}=\left(\frac{2}{n a}\right)^{3/2}
\left[\frac{(n+l)!(n-l-1)!}{2n}\right]^{1/2}, \label{w3}
\end{equation}
and
\begin{equation}
S_{q}(n, l) = (-)^q \frac{1}{q!(n-l-1-q)!(2l+1+q)!}.
\label{w4}
\end{equation}

 Furthermore, by
substituting the Eqs.(\ref{w5})-(\ref{w7}) and (\ref{w2})-(\ref{w4}) 
into  Eq.(\ref{w1}) 
one can integrate over the angular variables 
$(\hat{\bf {R}}, \hat{\boldsymbol{\rho}})$.
Finally,  applying the angular momentum algebra and integrating 
over $\rho$, one  obtains:
\begin{align}
&W^{J {\it p}}_{nlL,n'l'L'}(R)=(-1)^{J+l+l'}i^{l'+L'-l-L}\sqrt{\hat{l}
\hat{l'}\hat{L}\hat{L'}} \nonumber \\
&\sum_{t=0}^{t_{\rm max}}(l0l'0|t0)(L0L'0|t0)
\left\{\begin{array}{lll}
l&l'&t\\L'&L&J\end{array}\right\} \nonumber \\
&\left\{\frac{1}{\xi_e}\left[ (-1)^t W_{t;nl,n'l'}(R,\nu
,\xi_e) - W_{t;nl,n'l'} (R,\xi,\xi_e)\right] \right
.- \nonumber \\ &\left .  \frac{1}{\nu_e}\left[ (-1)^{t}
W_{t;nl,n'l'}(R,\nu,\nu_e) - W_{t;nl,n'l'}(R,\xi,\nu_e)\right]
\right\}
\label{w8}
\end{align}
($t_{\rm max}$ is the maximum value of the allowed multipole).
Here the next notations are used:
\begin{align}
W_{t;nl,n'l'}(R,\alpha,\beta)={\cal N}_{nl,n'l'}
\sum_{m_1=0}^{n-l-1}S_{m_1}(n,l)\left(\frac{2n'}{n+n'}\right)^{m_1} 
\nonumber \\
\sum_{m_2=0}^{n'-l'-1} S_{m_2}(n',
l')\left(\frac{2n}{n+n'}\right)^{m_2} 
\left\{H_t(x)J_1^{t,s}(x,\lambda_{n,n'}(\alpha,\beta)) \right . 
\nonumber \\
-h_t(x)J_2^{t,s}(x,\lambda_{n,n'}(\alpha,\beta)) 
+F_t(x)J_3^{t,s}(x,\lambda_{n,n'}(\alpha,\beta)) \nonumber \\
\left . +f_t(x)J_4^{t,s}(x,\lambda_{n,n'}(\alpha,\beta))\right\}, 
\label{w9}
\end{align}
where $x=2R/\beta $, $ s=l+l'+m_1+m_2 $, $ \hat{L}\equiv
2L+1$;

\begin{align}
&{\cal N}_{nl,n'l'} =
\frac{1}{n+n'}\!\left(\frac{2n'}{n+n'}\right)^{l+1}\!\left(\frac{2n}
{n+n'}\right)^{l'+1}\! \nonumber \\
&\sqrt{(n+l)!(n-l-1)!(n'+l')!(n'-l'-1)!}; 
\label{ww1}
\end{align}

\begin{equation}
\lambda_{n,n'}(\alpha,\beta)=\frac{2nn'}{n+n'}\frac{a\alpha}{\beta};
\label{ww2}
\end{equation}
\begin{equation}
H_t(x)=(1-2t)h_t(x)+xh_{t+1}(x); \label{ww3}
\end{equation}
\begin{equation} F_t(x)=(1-2t)f_t(x)-xf_{t+1}(x). \label{ww4}
\end{equation}
The functions $h_t(x)$ and $f_t(x)$ are given by
\begin{equation}
h_t(x)\equiv\sqrt{\frac{2}{\pi x}}K_{t+1/2}(x) \label{ww5}
\end{equation}
and
\begin{equation}
f_t(x)\equiv\sqrt{\frac{\pi}{2 x}}I_{t+1/2}(x). \label{ww6}
\end{equation}

The radial integrals $J_i^{t,s}(x,\lambda)$ are defined as
follows:
\begin{equation}
J_1^{t,s}(x,\lambda)=\int_{0}^{x/\lambda}
y^{s+2}e^{-y}f_t(\lambda y)\,{\rm d}y,
\label{ww7}
\end{equation}

\begin{equation}
J_2^{t,s}(x,\lambda)=\lambda J_1^{t+1,s+1}(x,\lambda),
\label{ww8}
\end{equation}

\begin{equation}
J_3^{t,s}(x,\lambda)=\int_{x/\lambda}^{\infty}
y^{s+2}e^{-y}h_t(\lambda y)\,{\rm d}y,
\label{ww9}
\end{equation}

\begin{equation}
J_4^{t,s}(x,\lambda)=\lambda J_3^{t+1,s+1}(x,\lambda)
\label{w12}
\end{equation}
and can be calculated analytically using  the power series
for the modified Bessel functions
or expressed via incomplete gamma functions.

\end{document}